\documentclass[aps,pre,amsmath,amsfonts,floatfix,superscriptaddress,nofootinbib]{revtex4}

\usepackage{t1enc,latexsym,amsfonts, amssymb, amsmath,empheq,amsthm,bm}
\usepackage{graphicx}
\usepackage{stmaryrd}

\usepackage{mathrsfs}
\usepackage[usenames,dvipsnames]{xcolor}
\usepackage{color}
\usepackage{multirow}
\usepackage{bbm}
\usepackage{commath}

\usepackage{tikz}
\usetikzlibrary{shapes,arrows,positioning,calc}
\usepackage{hyperref}

\usepackage[french,english]{babel}
\usepackage[utf8x]{inputenc}
\usepackage[T1]{fontenc}
\usepackage{times}
\usepackage{url}

\definecolor{darkgreen}{rgb}{0,0.6,0}
\definecolor{darkblue}{rgb}{0,0,0.6}
\definecolor{darkred}{rgb}{0.6,0,0}
\definecolor{darkpurple}{rgb}{0.5,0,0.5}

\hypersetup{
bookmarksopen=true,
pdftitle="Out-of-equilibrium dynamical equations of infinite-dimensional particle systems -- I. The isotropic case",
pdfauthor="Agoritsas-Maimbourg-Zamponi", 
pdftoolbar=false, 
pdfstartview={FitH},		
pdfmenubar=true,			
pdfhighlight=/O,			
colorlinks=true,			
urlcolor=darkblue,
citecolor=darkblue,		
linkcolor=darkpurple,	
}

\let\a=\alpha \let\b=\beta \let\g=\gamma \let\d=\delta
 \let\z=\zeta \let\h=\eta \let\k=\kappa
\let\l=\lambda \let\m=\mu \let\n=\nu \let\x=\xi \let\p=\pi
\let\s=\sigma \let\t=\tau \let\f=\varphi 
   \let\G=\Gamma
\let\D=\Delta  \let\X=\Xi 
   
 \let\r=\rho \let\th=\theta \let\io=\infty

\def\ie{{\textit{i.e.} }}\def\eg{{\textit{e.g.} }}

\def\PP{{\cal P}}\def\MM{{\cal M}} 
\def\CC{{\cal C}}\def\FF{{\cal F}} 
 \def\BB{{\cal B}}
\def\cR{{\cal R}}  \def\OO{{\cal O}}
\def\DD{{\cal D}}\def\AA{{\cal A}}\def\GG{{\cal G}} 
  
\def\ZZ{{\cal Z}}

\def\xx{{\bf x}} \def\yy{{\bf y}} 

\def\uu{{\bf u}}
\def\vv{{\bf w}}

\def\ul{\underline}

\def\redv{\bar v}

\def\bth{\bar\th}

\def\rr{\mathbf{r}}
\def\RR{\mathbf{R}}
\def\xx{\mathbf{x}}
\def\de{\mathrm d}
\def\Lap{\nabla^2}
\def\Ti{T_0}
\def\Bi{\b_0}

\def\to{\rightarrow} \def\la{\left\langle} \def\ra{\right\rangle}

\newcommand{\beq}{\begin{equation}} \newcommand{\eeq}{\end{equation}}
\newcommand{\wh}{\widehat} \newcommand{\wt}{\widetilde}
\newcommand{\Tr}{\text{Tr}}

\newcommand{\afunc}[1]{\operatorname{\mathsf{#1}}}
\def\DE{\afunc{D}}
\def\AE{\afunc{A}}
\def\ff{\mathrm{f}}

\newcommand{\argc}[1]{\left[#1\right]}
\newcommand{\arga}[1]{\left\lbrace #1\right\rbrace }
\newcommand{\argp}[1]{\left(#1\right)}
\newcommand{\valabs}[1]{\vert #1\vert}
\newcommand{\moy}[1]{\left\langle  #1 \right\rangle }

\newcommand{\ththbar}[1]{\theta_{#1} \bar{\theta}_{#1}}

\begin{document}

\title{
Out-of-equilibrium dynamical equations of infinite-dimensional particle systems
\\
I.~The isotropic case
} 


\author{Elisabeth Agoritsas}
\affiliation{Institute of Physics, EPFL, CH-1015 Lausanne, Switzerland}

\author{Thibaud Maimbourg}
\affiliation{The Abdus Salam International Centre for Theoretical Physics, Strada Costiera 11, 34151 Trieste, Italy}

\author{Francesco Zamponi}
\affiliation{Laboratoire de Physique de l’Ecole normale sup\'erieure, ENS, Universit\'e PSL, CNRS, Sorbonne Universit\'e, Universit\'e Paris-Diderot, Sorbonne Paris Cit\'e, Paris, France}


\begin{abstract}

We consider the Langevin dynamics of a many-body system of interacting particles in $d$ dimensions, 
in a very general setting suitable 
to model several out-of-equilibrium situations, such as liquid and glass rheology, active self-propelled particles, and glassy aging dynamics.
The pair interaction potential is generic, and can be chosen to model colloids, atomic liquids, and granular materials.
In the limit ${d\to\io}$, we show that the dynamics can be exactly reduced to a single one-dimensional effective stochastic equation,
with an effective thermal bath described by kernels that have to be determined self-consistently.
%
We present two complementary derivations,
via a dynamical cavity method and via a path-integral approach.
From the effective stochastic equation, one can compute dynamical observables such as pressure, shear stress, particle mean-square displacement, and the associated response function.
As an application of our results, we derive dynamically the `state-following' equations that describe the response of a glass to quasistatic perturbations, thus bypassing the use of replicas.
The article is written in a modular way, that allows the reader to skip the details of the derivations and focus on the physical setting and the main results.

\end{abstract}


\maketitle

\begin{center}
\today
\end{center}


\begin{center}
\rule{200pt}{0.5pt}
\end{center}

\tableofcontents

\begin{center}
\rule{200pt}{0.5pt}
\end{center}

\section{Introduction}
\label{sec-introduction}

Solving the dynamics of classical many-body systems of interacting particles in the dense liquid regime is
an extremely difficult task~\cite{hansen}.
Yet, this problem is relevant to describe many physical situations, such as
{\it (i)} the rheology of pastes, emulsions, colloids, and other soft materials~\cite{BDBDM17},
{\it (ii)} the behaviour of dense assemblies of active self-propelled particles, such as bacteria, tissues, and flocks~\cite{MJRL13},
{\it (iii)} the response of glasses to perturbations such as compression and shear strain, which can induce plasticity, yielding, and flow~\cite{RTV11}.
Nowadays, most of these problems are studied via numerical simulations, while a general microscopic theory encompassing all these phenomena is still lacking.

A classical approach to study the dynamics of dense liquids is the Mode-Coupling Theory (MCT)~\cite{hansen,Go09}.
In a nutshell, MCT considers a set of relevant dynamical correlation functions, such as the intermediate scattering function, and derives closed equations for their evolution using a series of approximations of the exact dynamics.
Several alternative derivations of MCT have been reported in the literature, see Refs.~\cite{Go09,reichman2005mode,janssen2018mode} for examples.
In its standard form, MCT describes the equilibrium time-translationally invariant (TTI) liquid dynamics, and has been particularly successful in describing its slowing down upon approaching the glass transition and the associated dynamical arrest~\cite{Ca09,Go99},
in a variety of model systems ranging from Lennard-Jones potentials for structural glasses~\cite{KA95a,KA95b} to square-well potentials for attractive colloids~\cite{fabbian1999,dawson2000,bergenholtz1999}.
In addition, several extensions of MCT have been attempted in order to deal with out-of-equilibrium situations.
For instance, MCT equations for rheology are reported in Refs.~\cite{miyazaki2002molecular,FC02,miyazaki2006nonlinear,Br09,BCF12}
and for active matter in Refs.~\cite{SFB15,janssen2018mode,NG17},
and a MCT analysis of yield stress and jamming has been described in Ref.~\cite{IB13}.
Also, while already the classical MCT is often quantitatively accurate, many attemps to improve it have been reported, a literature that has become so large that an exhaustive account would be a challenge, see Refs.~\cite{schweizer1989microscopic,schweizer2003entropic,BBW08,MK11,szamel2013mode,kim2014equilibrium,rizzo2015qualitative,rizzo2016dynamical,janssen2015microscopic,janssen2018mode} for a few examples.
In fact, it is fair to say that, to date, MCT is essentially the unique microscopic theory capable of describing a variety of out-of-equilibrium properties of strongly-interacting particle systems.

A very important step in understanding the foundations of MCT has been 
the recognition that the `schematic' MCT equations~\cite{L84,BGS84} correspond to an exact description of the dynamics of a wide class of mean-field spin-glass models~\cite{KT87,KT87b,BCKM96}, belonging to the so-called `Random-First-Order-Transition' (RFOT) universality class.
RFOT models have thus provided a lot of inspiration for the construction of dynamical theories of glasses~\cite{KTW89,LW07,WL12,BBBCS11}, providing in particular schematic MCT-like 
equations to describe aging~\cite{CK93}, rheology~\cite{BBK00}, and active matter~\cite{BK13}.
At the core of the RFOT theory is the idea that the long-time limit of the dynamics can be described in terms of a restricted, metastable equilibrium (of varying nature depending on the problem under consideration), which allows one to obtain dynamical informations without fully solving the dynamics, 
typically using the replica~\cite{KT87,KW87,Mo95,FP95,CK93,BBM96,RUYZ15} or cavity~\cite{KZ10,KZ10b} methods.
However, spin-glass models are microscopically very different from liquids, and  
the connection between the two classes of systems is not {\it a priori} clear.

The origin of the RFOT universality was first understood by studying the MCT equations for $d$-dimensional liquids in the infinite-dimensional limit, 
in order to show that they become mathematically very similar to the ones describing spin glasses~\cite{KW87}.
However, while the idea of Ref.~\cite{KW87} is essentially correct, later studies~\cite{IM10,SS10,CIPZ11} have shown that the ${d\to\io}$ limit of MCT equations themselves is ill-defined because the theory develops several pathologies.
It took a few more years, and the use of ideas coming from the static replica approach~\cite{PZ06a,PZ10,KPZ12,KPUZ13,CKPUZ17}
in addition to a general mapping of replicas onto the dynamics~\cite{Ku92,Ku03,PR12,CFLPRR12},
to derive a set of equations that describe exactly the dynamics of particle systems in the high-dimensional limit~\cite{KMZ16,MKZ16,Sz17}.
These equations are, however, restricted to the \emph{equilibrium} case.
They are mathematically distinct from MCT, but have a similar structure, and in particular they display
a dynamical ergodicity-breaking glass transition and a dynamical criticality around it that are precisely identical to those predicted by MCT (only non-universal numerical values differ) and, more generally, fall into the RFOT universality class~\cite{CKPUZ17}.

In this work, we generalise the derivation of Refs.~\cite{KMZ16,MKZ16,Sz17} in order to obtain an exact solution of the dynamics of a system of interacting particles in the ${d\to\io}$ limit, in a fully general setting which includes essentially any kind of non-equilibrium situation, including
rheology, active matter, aging, response to static perturbations, etc.
Our equations obviously reduce to those of Refs.~\cite{KMZ16,MKZ16,Sz17} when assuming an equilibrium dynamics.
Furthermore, in the long-time limit within a metastable glass phase, they reproduce the replica equations derived in Refs.~\cite{PZ10,RUYZ15}, 
thus providing an alternative derivation that does not require the use of replicas.

The main challenge from there is that the dynamical equations that we will obtain have an involved self-consistent structure, similar to the ones of the Dynamical Mean Field Theory (DMFT) of strongly correlated electrons~\cite{GKKR96} which also holds in the ${d\to\io}$ limit.
They will therefore not be easy to solve numerically, in order to extract the observables beyond the equilibrium  case or the long-time limit in a metastable state.
Here we restrict ourselves to establish the equations for the ${d \to \io}$ dynamics
--~which is already a non-trivial task~--
leaving  for future work their numerical solution in cases of interest.
While our equations should not be expected to provide quantitatively accurate predictions in the physically relevant dimensions ${d=2,3}$, they have nevertheless a number of interesting features that we highlight thereafter:
\begin{itemize}

\item
Their derivation is based on a series of clear and intuitive physical assumptions (mostly identified in Ref.~\cite{Sz17}), that all become exact in the limit ${d\to\io}$.
This allows one to have an idea of why and how the theory can fail in lower dimension and what are the most crucial hypotheses for its validity range.

\item
They can provide qualitatively accurate results, as it has been shown in a number of applications, in particular for hard spheres~\cite{PZ10,RUYZ15,CKPUZ17,UZ17} and for attractive colloids~\cite{sellitto2013thermodynamic,altieri2018microscopic}.
This is especially true in the vicinity of the jamming transition, where they even become \emph{quantitatively} accurate~\cite{nature,CKPUZ17}.

\item 
Being exact equations for a microscopically well-defined model (namely interacting particles in ${d\to\io}$) they cannot, by definition, display any pathological behaviour: they are bound to remain correct in all the possible regimes.

\item
Hence, they can be handled as a `black box': one can choose the desired inter-particle potential and physical setting, and from there run the numerical solution of the equations to output all the relevant dynamical observables.

\item
These equations can serve as a starting point for a controlled perturbative expansion in ${1/d}$, as well as uncontrolled resummations to improve the quantitative accuracy~\cite{PZ10,MZ16}.

\item 
Many phase transitions that are just crossovers in finite $d$ become sharp when ${d\to\io}$.
This may of course be a drawback, possibly introducing `spurious' mean-field transitions that might not exist in reality, but at least these transitions can be studied in a controlled framework in ${d\to\io}$, and one can then try to understand any remnant behaviour in low $d$ by either perturbative expansions or numerical simulations where $d$ would be varied systematically~\cite{CIPZ11,CIPZ12,CCPZ15}.

\end{itemize}
Our equations can thus provide a useful complement to MCT or to atomistic numerical simulations, in order to understand the complex behaviour of non-equilibrium dense liquids and glasses.
Note that in principle it should be possible to derive them from, \textit{e.g.}, generalised MCT~\cite{janssen2015microscopic,janssen2018mode} by a proper treatment of the expansion in the limit ${d\to\io}$, see Refs.~\cite{Sz17,CCS18} for preliminary attempts in this direction.


We would like to stress that, while this article is long and very detailed, it has a modular structure.
The reader can thus navigate through the sections to identify the parts she is mostly interested in.
In particular, the setting of the problem is given in Sec.~\ref{sec:setting} and one
can then directly skip the derivations to Sec.~\ref{sec:summary-results} where our main results for the dynamical equations are summarised, and to Sec.~\ref{sec:limit-cases} where some applications are presented.

More precisely, in the following, we start by introducing in Sec.~\ref{sec:setting} the generic setting of the problem we are considering.
Then we present two complementary derivations
--~first via a dynamical `cavity' method in Sec.~\ref{sec:cavity}
and secondly via a path-integral approach in its supersymmetric formulation in Sec.~\ref{sec:path}~--
of two effective stochastic processes, respectively for the individual displacements of particles and for the inter-particle distances, in their high-dimensional \emph{vectorial} form.
Exploiting further the high-dimensional limit, we show in Sec.~\ref{secC3:largedM} how we can eventually simplify the effective dynamics into a \emph{scalar} stochastic process for the rescaled inter-particle distance (or `gap'), 
involving three distinct time-dependent kernels self-consistently defined by statistical averages over the effective process itself.
In order to close our mean-field dynamics, we then present in Sec.~\ref{sec:dynamical-equations-correlation-response} the corresponding evolution equations for the correlation and response functions, and in particular for the mean-square displacement.
A summary of our mean-field out-of-equilibrium dynamics is given in Sec.~\ref{sec:summary-results}, as consistent as possible with our most general initial setting.
Then in Sec.~\ref{sec:limit-cases} we check that we can recover from there both the equilibrium and the quasistatic-protocol results,
and in particular we discuss some direct applications to the case of static random forces.
We finally conclude in Sec.~\ref{sec-conclusion} and discuss some immediate and long-term perspectives.


We emphasise that our two parallel derivations --~cavity \textit{versus} path-integral~-- are in the same spirit as the study presented in Ref.~\cite{ABUZ18} for the continuous random perceptron in its thermodynamic limit.
Such combined approaches are indeed quite generic for coupled (generalised) Langevin dynamics in infinite dimension, and in particular for our system of interacting particles:
the first derivation via the cavity is more intuitive although it relies on a few physical but not-fully controlled assumptions, whereas the second one is more involved technically but properly under control, justifying \textit{a posteriori} the cavity hypotheses.
Thus, throughout the paper, we will highlight the relations between the physical assumptions and their technical implementation in the path-integral formalism.


In the initial approach presented in Refs.~\cite{MKZ16,KMZ16}, the equilibrium counterpart of the high-dimensional mean-field dynamics has been studied through path integrals   on a hypersphere of diverging radius, 
hence physically equivalent to an infinite Euclidean space.
We have checked that we recover --~as we should~-- the same out-of-equilibrium results in this formulation as well, but here we have chosen to work directly in a flat Euclidean space, as we managed on the one hand to bypass the technical need of the hyperspherical setting, and on the other hand to encode in a straightforward way the initial condition, a necessary ingredient for generic out-of-equilibrium situations. 
The present derivations are considerably simplified in comparison.

\section{Setting of the problem}
\label{sec:setting}

We consider a system of $N$ interacting particles labeled by ${i=1, \dots , N}$,  
in $d$ spatial dimensions labeled%
\footnote{Throughout the article we will denote:
\textit{(i)}~$\bm a$ for a vector with components $a_\m$,
\textit{(ii)}~$\hat{\bm a}=\bm a/|\bm a|$ for a unit vector with components $\hat a_\m$,
and~\textit{(iii)}~$\hat a$ for a matrix with components~$a_{\m\n}$.}
by ${\mu =1, \dots ,d}$,
with positions ${\ul{X}(t) = \arga{\xx_i(t) \in \Omega \subset \mathbb{R}^d}_{i=1,\dots,N}}$ at time $t$ 
belonging to a region $\Omega$ of volume $|\Omega|$, hence at number density ${\r= N/|\Omega|}$.
For simplicity we consider $\Omega$ to be a cubic region with periodic boundary conditions.
In our most general setting, the interacting particles are supposed to be embedded in a fluid, and
to evolve under the following coupled Langevin dynamics:
\beq
\label{eqC3:GENLang}
\begin{split}
 & m \ddot \xx_i(t) +  \z [\dot \xx_i(t)- \bm v_f(\xx_i,t) ]
 	+ \int_0^t \!\! \de s \, \G_R(t,s) \, [\dot\xx_i(s) - \bm v_f(\xx_i,s)] 
 	=  \bm F_i(t)  + \bm{\x}_i(t) +\bm{\l}_i(t)  \ ,
 \\
 &	
 	\text{Gaussian noise } \arga{\bm{\xi}_i(t)}_{i=1,\dots,N}: \, \quad\quad
 	\moy{\x_{i\m}(t)}_{\bm\xi} =0 , \qquad
 	\moy{\x_{i\m}(t) \x_{j\n}(s)}_{\bm\xi}
		= \delta_{ij}  \delta_{\m\n} [2 T\z  \delta(t-s) + \G_C(t,s)] \ ,
 \\
 &
 	\text{Interaction force and potential}: \, \quad\quad \bm F_i(t)  = -\frac{\partial V(\ul X(t))}{\partial \xx_i(t)} \ , \qquad
 	V(\ul X) = \sum_{i<j} v(\xx_i - \xx_j) \ .
\end{split}
\eeq
On the left side of the generalised Langevin equation, the first term describes inertia, with
$m$ being the individual mass of particles.
The second term describes the frictional force exerted by the fluid;
here, ${\bm v_f(\xx,t)}$ is the velocity field of the fluid in which the particles are embedded,
${\dot \xx_i(t)- \bm v_f(\xx_i,t)}$ is the relative velocity of particle $i$ with respect to the fluid,
and the frictional force is proportional to the relative velocity via the friction coefficient $\zeta$.
The third term describes the same frictional force, but taking into account possible retarded effects via a friction
kernel ${\G_R(t,s)}$.
On the right side, the first term ${\bm F_i(t)}$ is the conservative force due to the interaction with the other
particles via the potential ${V(\ul X)}$, which is a sum of pair interactions with radial pair potential ${v(\xx)}$, assumed to be a standard potential between atoms or colloids, such as the Lennard-Jones or hard sphere potential~\cite{hansen}.
The second term ${\bm\x_i(t)}$ is a Gaussian noise, assumed to describe the noisy force due to collisions with
particles in the embedding fluid: it has a zero mean, and time-correlation given by a white noise term proportional
to ${\d(t-s)}$, with coefficient ${T=\beta^{-1}}$ which is the temperature of the fluid, and a colored noise term
with kernel ${\Gamma_C(t,s)}$ which can describe several physical effects, see below.
The third term ${\bm{\l}_i}$ is an external field which is only included to define the linear response of the system, and otherwise set to zero.
Note that the brackets ${\moy{\cdots}_{\bm\xi}}$ denote the statistical average over the microscopic noise~${\bm\xi}$,
that hydrodynamic interactions between particles are not considered,
and that the Boltzmann constant is set to $k_B=1$, thus fixing the units of temperature and entropy.

We assume that at time $t=0$ the particles start from an equilibrium configuration at a temperature ${\Ti=\beta_0^{-1}}$:
we define ${\xx_i(0) = \RR_i}$, with $\RR_i$ sampled from a Boltzmann distribution ${\propto e^{-\beta_0 V(\ul{X}(0))}}$.
The initial velocities, if needed, are also sampled from a Maxwell distribution at temperature $\Ti$.
The bare brackets ${\moy{\cdots}}$ correspond in the following to the statistical average over both the noise and this stochastic initial condition.
Note that, unless the potential has a hard core, a randomly uniform initial condition corresponds to the particular case of ${\Ti \to\io}$,
whereas at finite $\Ti$ we assume that we start in the equilibrium (possibly supercooled) liquid phase, \textit{i.e.}~${\Ti > T_{\rm K}}$ where $T_{\rm K}$ is the Kauzmann temperature below which the equilibrium liquid does not exist%
\footnote{Technically, this assumption is required in order to have a simple
thermodynamical description of the initial condition in the limit ${d\to\io}$: we assume that at temperature $\Ti$ there is a single thermodynamic equilibrium state. Note that it might turn out that ${T_{\rm K}=0}$~\cite{BB11}.}.
We emphasise that the key property of this initial stochastic distribution, regarding our whole derivations, is that it displays a \emph{statistical isotropy}, at equilibrium in particular the Boltzmann distribution being defined with respect to a \emph{radial} pair potential ${v(\xx)}$.
Besides, we introduce the following definitions:
\begin{equation}
\label{eq-def-wrt-initial-condition}
\begin{split}
 \uu_i(t) = \xx_i(t) - \RR_i
 \, , \quad
 \rr_{ij}(t) = \xx_i(t) - \xx_j(t)
 \, , \quad
 \vv_{ij}(t) = \uu_i(t) - \uu_j(t) = \rr_{ij}(t) - \rr_{ij}(0)
 \, ,
\end{split}
\end{equation}
where ${\uu_i(t)}$ is the individual displacement of particle~$i$ with respect to its initial position,
${\rr_{ij}(t)}$ the inter-particle distance of the pair ${(i,j)}$,
and ${\vv_{ij}(t)}$ the difference with respect to its initial value.

The general setting of Eq.~\eqref{eqC3:GENLang} can be specialised to describe several physically interesting situations:
\begin{itemize}

\item {\it Equilibrium:}
The standard equilibrium white-noise dynamics corresponds to setting ${\bm v_f}=\bm 0$, ${\G_R = \G_C =0}$ and ${T=\Ti}$, in which case Eq.~\eqref{eqC3:GENLang} simply reads:
\beq
\label{eqC3:GENLang-state-following}
 m \ddot \xx_i(t) +  \z \dot \xx_i(t)
 	=  \bm F_i(t) + \bm{\x}_i(t)  \ , \qquad \moy{\x_{i\m}(t) \x_{j\n}(s)}_{\bm\xi}
		=2 T\z \delta_{ij}  \delta_{\m\n}   \delta(t-s)
 \ .
\eeq
In this case%
\footnote{Note that our notations are fully equivalent to considering a friction coefficient $\z_R$ in the derivative term ${\z_R \dot\xx_i(t)}$, and a distinct coefficient in the noise term
${\moy{\x_{i\m}(t) \x_{j\n}(s)} =  \d_{ij}  \d_{\m\n} 2 \z_C  \d(t-s)}$.
This choice corresponds to a white noise with an effective temperature ${T = \z_C/\z_R}$.
Our notations thus correspond to ${\z_R = \z}$ and ${\z_C = T\z}$.},
we will recover the results previously obtained in Refs.~\cite{MKZ16,KMZ16,Sz17,CKPUZ17}.
An equilibrium colored noise dynamics can also be implemented by using time-translationally invariant (TTI) friction and noise kernels ${\G_R(t-s)}$ and ${\G_C(t-s)}$ related by ${\G_R(t) = \b \th(t)\G_C(t)}$; this last relation is a necessary condition to be at equilibrium, \textit{i.e.}~for the fluctuation-dissipation theorem (FDT) to be valid at asymptotically large times~\cite{Ku66,Cu02,Ha97,ZBCK05,MSVW13}.

\end{itemize}
Here we want to address more generally \emph{the dynamics when the system is driven out of equilibrium}.
This can be due to several reasons, as for instance because the FDT relation between the kernels is not satisfied, or because the system is non-ergodic.
In all cases, we will consider for simplicity the overdamped case, formally corresponding to ${m=0}$, as inertia is not an essential feature in any of these applications.
Also we consider ${\bm v_f=\bm 0}$ unless otherwise specified.
We are interested in particular in the following cases:
\begin{itemize}

\item {\it State-following protocol:}
While the system is prepared at equilibrium at~$\Ti$, 
an equilibrium dynamics is run but at a different temperature ${T \neq \Ti}$~\cite{FP95,BBM96,BFP97}.
In this case, one can consider ${\G_C=\G_R=0}$ for simplicity (since the noise being colored is qualitatively not the relevant feature) and the dynamical equation coincides with Eq.~\eqref{eqC3:GENLang-state-following}.
At high temperatures, the system equilibrates at large times, in other words the memory of the initial condition is lost and equilibrium dynamics is recovered~\cite{Cu02}.
At sufficiently low temperatures, however, ergodicity might be broken.
In this case the system eventually reaches an equilibrium-like steady state but in a restricted phase space, corresponding to a `glass' basin, randomly sampled by the initial condition~\cite{FP95,BBM96,BFP97}.
Our dynamical equations allow one to describe this restricted equilibrium, recovering in particular the results obtained via the replica method~\cite{RUYZ15}, as we will explicitly check in Sec.~\ref{sec:limit-cases-state-following}.

\item {\it Active matter}:
An active self-propulsion of the particles can be modelled by choosing for instance ${\G_R=0}$ and ${\G_C \neq 0}$, the latter being the correlation of the self-propulsion.
A common choice consists in removing the white noise by setting ${T=0}$ and ${\G_C(t,s) =\G_C(t-s) = \G_a e^{-|t-s|/\t_p}/\t_p}$~\cite{BK13}, where $\t_p$ is the persistence time of the self-propulsion and $\G_a$ its amplitude.
In this case one can introduce an auxiliary noise $\bm\h_i(t)$ via an Ornstein-Uhlenbeck process, and write
Eq.~\eqref{eqC3:GENLang} as
\beq
\label{eqC3:GENLang-active}
\begin{split}
 \z \dot \xx_i(t)
 	&=  \bm F_i(t)  + \bm{f}_i(t)
 \ , \\
\t_p  \dot{\bm f}_i(t)
	&= - \bm f_i(t)  + \bm{\h}_i(t)
 \, , \qquad  
 	\moy{\h_{i\m}(t) \h_{j\n}(s)}_{\bm\h}
		=  2 \G_a \delta_{ij}  \delta_{\m\n} \,  \delta(t-s) 
 \ .
\end{split}
\eeq
Here ${\bm{f}_i(t)}$ is a Gaussian random noise with zero mean and autocorrelation ${\delta_{ij}  \delta_{\m\n}\G_C(t-s)}$, which describes the self-propulsion~\cite{SFB15,BFS17}.
Similar models have been used for instance in Refs.~\cite{FM12,YMM14}.

\item {\it Micro-rheology}:
In this setting, an external Gaussian random force $\bm f_i$ with zero mean and a variance $f_0^2$ (independently for each component) is applied on each particle, see for instance Refs.~\cite{GPVF09,LX18}.
This formally corresponds to setting ${\G_R=0}$ and ${\G_C(t,s) = f_0^2}$:
\beq
\label{eqC3:GENLang-micro-rheol}
\begin{split}
 \z \dot \xx_i(t)
 	=  \bm F_i(t) + \bm{\x}_i(t)
 \, , \quad \text{with} \,
 	\moy{\x_{i\m}(t) \x_{j\n}(s)}_{\bm\xi}
		= \delta_{ij}  \delta_{\m\n} \, \argc{ 2 T\z  \delta(t-s) + f_0^2}
 \ .
\end{split}
\eeq
In this way one can probe the local rheological response of the liquid.
Note that this setting corresponds to a stress- (or force-) controlled experiment.
Formally, this also corresponds to the limit of self-propelled particles when the persistence time diverges (${\tau_p \to\io}$).

\item {\it Shear strain}:
A shear strain is applied to the fluid in which the particle are immersed, which is assumed to result in a laminar flow of the form ${\bm v_f(\xx,t) =\dot\g(t) x_2 \hat \xx_1}$, where the flow is along direction ${\m=1}$ (${\hat \xx_1}$ is the unit vector along this direction) and its gradient is along direction ${\mu=2}$.
Setting for simplicity ${\G_C=\G_R=0}$, the resulting equation of motion is~\cite{FC02,Br09,BCF12,IBS13,KCIB15}:
\beq
\label{eqC3:GENLang-shear}
 \z  [ \dot \xx_i(t) - \dot\g(t) x_{i2} \hat \xx_1 ]
 	=  \bm F_i(t) + \bm{\x}_i(t)  \ , \qquad \moy{\x_{i\m}(t) \x_{j\n}(s)}_{\bm\xi}
		=2 T\z \delta_{ij}  \delta_{\m\n}   \delta(t-s) \ .
\eeq
Here, ${\dot\g(t)}$ is the {\it shear strain rate}, which can also be assumed to be time-dependent.
This setting models a strain-controlled experiment and can be used to study the rheology of liquids and glasses.
However, in this case the statistical isotropy is broken because of the external shear, demanding a generalization that will be treated separately in the companion paper~\cite{AMZ19bis}.

\end{itemize}

Solving the dynamics consists by definition in being able to compute the dynamical observables and their time dependence.
Typical examples of one-time observables are the average potential energy and pressure tensor, respectively~\cite{hansen}:
\beq
\label{eq:ePi}
 e(t)
 	=\frac1N \la V[\ul X(t)] \ra 
 \ , \qquad 
 \Pi_{\m\n}(t)
 	= \frac{\r}{ N} \moy{ \sum_{i=1}^N m \dot x_{i\m}(t) \dot x_{i\n}(t)
	 -\frac12 \sum_{i\neq j}^{1,N} \frac{r_{ij\m}(t) r_{ij\n}(t)}{r_{ij}(t)^2} \rr_{ij}(t) \cdot \nabla v(\rr_{ij}(t))}
 \ .
\eeq
Note that the pressure is given by ${P(t) = \Tr \,\hat\Pi(t) / d}$, and the shear stress is defined as the ${\s(t) = -\Pi_{12}(t)}$ component of the stress tensor (the negative of
the pressure tensor), \textit{i.e.}~the observable conjugated  to the shear strain $\dot\gamma(t)$ defined above.
For the micro-rheology setting, an interesting quantity to compute is also the average velocity (or current) in the direction of the force,
\beq
 J(t) = \frac1{N d} \la \dot\xx_i(t) \cdot \bm\x_i(0) \ra 
 \ .
\eeq
Typical two-time observables are the mean-square displacement (MSD), and the correlation and response functions, respectively given by
\beq
\label{eq:CRdef}
\begin{split}
\DE(t,t')
 	&=\frac{1}{Nd}\sum_{i=1}^N\la [ \xx_i(t) - \xx_i(t')]^2 \ra =\frac{1}{Nd}\sum_{i=1}^N\la [ \uu_i(t) - \uu_i(t')]^2 \ra 
 \, ,  \\
C(t,t')
 	&=\frac{1}{Nd}\sum_{i=1}^N\la\uu_i(t)\cdot\uu_i(t')\ra 
 \, , \quad
 R(t,t')
 	=\frac{1}{Nd}\sum_{i,\m}\left.\frac{\d\la u_{i\m}(t)\ra}{\d \l_{i\m}(t')}\right\vert_{\{\bm{\l}_i\}=\{\bm0\}}
 \ .
\end{split}
\eeq
One could also be interested to the auto-correlation of the stress tensor defined above, which is related to the viscosity~\cite{hansen}.

The above discussion shows that essentially all the interesting physical situations can be modeled by particular cases of Eq.~\eqref{eqC3:GENLang}, hence the interest of studying this equation in the most general case. However, a full solution of Eq.~\eqref{eqC3:GENLang} for interacting particles in the physically interesting dimensions $d=2,3$ is impossible.
Our goal is to study Eq.~\eqref{eqC3:GENLang} in the limit of infinite dimensions (${d\to\io}$), taken after the thermodynamic limit (${N\to\io}$).
In this limit, the mean-field description is exact~\cite{CKPUZ17} and the solution of the dynamics is reduced to
two effective single-particle stochastic processes, 
describing the fluctuations of the individual displacement of a representative particle ${\uu(t)}$ and the representative inter-particle fluctuating distance ${\vv(t)}$;
they can be also simplified further thanks to the high-dimensional scaling of the interaction potential ${v(\rr)}$
and eventually yield the dynamical equations for the correlation and response functions~\eqref{eq:CRdef} in terms of two effective one-dimensional
equations.

For simplicity we are going to set ${m=0}$ and ${\G_R=0}$ in our following derivations, 
corresponding to an overdamped dynamics without retarded friction,
as both terms do not play a very important role in any of the relevant examples discussed above.
However, both the inertial term and the non-local friction kernel can be reinserted at any time, since they are single-particle terms in the Langevin dynamics~\eqref{eqC3:GENLang};
the corresponding explicit formulae will thus be given directly in the summary of our infinite-dimensional mean-field dynamics, in Sec.~\ref{sec:summary-results}.
We will also set the fluid velocity $\bm v_f(\xx_i,t)=\bm 0$ throughout this paper.
Reinserting it requires a little bit more work because one has to take care of the assumption of statistical isotropy that can be broken.
So here we focus on the isotropic case --~starting from an isotropic initial condition (equilibrium) and with a rotational-invariant dynamics~-- whereas we will discuss the anistropic case under shear strain in the companion paper~\cite{AMZ19bis}.

\section{Derivation via a dynamical `cavity' method}
\label{sec:cavity}

We first present a derivation inspired by the cavity method,
based on an extension --~with some key differences~-- of the equilibrium derivation of Ref.~\cite{Sz17},
and in the same spirit as what has been done in Ref.~\cite{ABUZ18} for the continuous random perceptron.
It relies on the following two main ideas:
\begin{enumerate}

\item
Since we know that in ${d\to\io}$ the interesting dynamics happens on a length scale of order ${1/d}$ with respect to the inter-particle distances~\cite{MKZ16}, 
we are going to assume that the individual displacements with respect to the initial position,
${\uu_i(t) = \xx_i(t) - \RR_i }$, 
are small (more precisely ${\uu_i(t)=\mathcal{O}(1/d)}$) and treat them perturbatively.

\item
In ${d\to\io}$ each particle has so many neighbours --~their number growing proportionally to $d$~-- that we can assume that they are uncorrelated,
allowing us to invoke the central limit theorem in order to assume Gaussian fluctuations.

\end{enumerate}
Moreover, along the way we will make a few additional assumptions that are neither fully justified for the time being.
For this reason, we will present in Sec.~\ref{sec:path} a second derivation, based on dynamical path integrals and a supersymmetric formulation
--~ fully under control in the infinite-dimensional limit, although technically more involved and less intuitive~--
that reproduces the same results and justifies \textit{a posteriori} all our assumptions for the cavity derivation.


In this section, we start by showing in Sec.~\ref{secC3:onep} how to rewrite the original Langevin dynamics~\eqref{eqC3:GENLang} for one given particle, isolating it and treating perturbatively its interaction with the others as a mean-field isotropic `liquid'.
We then write similarly in Sec.~\ref{secC3:twop} a two-particle effective stochastic process, for the inter-particle distances which control the pairwise interactions.
In doing so, we obtain self-consistent definitions of the three time-dependent kernels ${\arga{k(t),M_C(t,s),M_R(t,s)}}$ controlling this mean-field dynamics.

\subsection{Single-particle effective stochastic process}
\label{secC3:onep}

\subsubsection{Isolating one particle in the `liquid'}
\label{secC3:onep-isolating-1particle}

In a nutshell, the cavity method consists in isolating one particle (our `cavity') and rewriting its dynamics coupled to the rest of the system as an effective Langevin equation, with kernels encoding the interactions in a mean-field way~\cite{MPV87}.
In the thermodynamic limit that we consider, we can equivalently add or isolate one particle, and call it $0$.
The other particles can be seen as a `liquid' acting on it%
\footnote{This terminology is inspired by the usual picture for the Brownian motion, of a large colloidal particle in a suspension. However, the system might not be in a true liquid state, in the sense of equilibrium with a single thermodynamic state, hence the quotation marks `liquid' in the whole section.}.

First, the Langevin equation~\eqref{eqC3:GENLang} for particle $0$ has the form
\beq
\label{eq:dyn0}
 \z \dot \xx_0(t) = \bm{F}_0(t) + \bm{\x}_0(t)
 \, , \quad \text{with} \quad
 \bm{F}_0(t) =  -\sum_{j (\neq 0)} \nabla v( \xx_0(t) - \xx_j(t))
\ .
\eeq
We can expand the force at small displacement ${\uu_0(t)=\xx_0(t)-\RR_0}$, truncating it after first order:
\beq
\label{eq:forcesplit}
\bm{F}_0(t)
 \approx 
 \underbrace{-\sum_{j (\neq 0)} \nabla v( \RR_0 - \xx_j(t))}_{= \wt{\bm{F}}_0(t)}
 - \underbrace{\sum_{j (\neq 0)} \nabla \nabla^T v( \RR_0 - \xx_j(t))}_{= \hat k(t)} \, \uu_0(t)
 = \wt{\bm{F}}_0(t)  - \hat k(t) \, \uu_0(t)
 \ .
\eeq
Physically, the first term ${\wt{\bm{F}}_0(t)}$ is the force produced in the point $\RR_0$ by the other particles, while the second term ${-\hat k(t) \uu_0(t)}$ is an effective restoring force due to the local confinement of particle $0$ in the `liquid'%
\footnote{Note that the notation in coordinates for the matrix ${\hat k(t)}$ is ${k^{\mu \nu}(t) = \sum_{j (\neq 0)} \nabla_\mu \nabla_\nu v (\RR_0 - \xx_j(t))}$, \textit{i.e.}~we indicated by $\nabla \nabla^T$ the matrix
$\nabla_\mu \nabla_\nu$.}.
However, we emphasise that both ${\wt{\bm{F}}_0(t)}$ and the matrix ${\hat k(t)}$ are fluctuating quantities which still depend implicitly on ${\uu_0(t)}$, via the other trajectories ${\ul X (t)=\lbrace \xx_j(t) \rbrace _{j (\neq 0)}}$.
We thus need to treat this implicit dependence perturbatively --~at first order in ${\uu_0(t)}$ in order to be consistent~-- and we will do it statistically by averaging on the `liquid' fluctuations.

The complete trajectories ${\ul X (t)}$ can thus be decomposed into two contributions:
on the one hand the trajectories ${\ul X^{(0)} (t)=\lbrace \xx_j^{(0)}(t) \rbrace}$ that the particles would have in the fictitious `liquid' for $\uu_0(t)=\bm 0$; 
and on the other hand their correction due to the individual displacement of particle $0$ itself, which we truncate at first order in ${\uu_0(t)}$.
Once the particle $0$ has been treated separately, we have to characterise the `liquid' formed by the other particles, which follows the dynamics given by Eq.~\eqref{eqC3:GENLang} but with a modified potential
\beq
\label{eq:pert}
 V(\ul X;\xx_0)
 = \sum_{i<j (\neq 0)} v(\xx_i - \xx_j)
	+ \sum_{i (\neq 0)} v(\xx_i - \RR_0 - \uu_0)
\approx V(\ul X;\RR_0)
		- \bm{F}_0(\ul X; \RR_0) \cdot \uu_0
\ .
\eeq
Here ${V(\ul X;\RR_0)}$ is the potential of the particles in the `liquid' with a fixed particle in $\RR_0$,
and ${\bm{F}_0(\ul X; \RR_0)=-\nabla_{\RR_0} V(\ul X;\RR_0)}$ is the corresponding force%
\footnote{We used ${\bm{F}_0(\ul X; \RR_0) = \sum_{i (\neq 0)} \nabla v( \xx_i - \RR_0)=-\sum_{i (\neq 0)} \nabla v(\RR_0 - \xx_i)}$.}
exterted by the `liquid' on the fixed particle.
So along the original dynamics we have ${\wt{\bm{F}}_0(t)=\bm{F}_0(\ul X(t); \RR_0)\neq \bm{F}_0(\ul X^{(0)}(t); \RR_0)}$: in other words the particles in the `liquid' feel an obstacle in $\RR_0$ and an external time-dependent perturbation ${\uu_0(t)}$ acting on their trajectories ${\ul X (t)}$.

From now on, the brackets $\moy{\bullet}_{0}$ will denote the dynamical average over this fictitious system composed of the `liquid' of particles ${i=1,\dots,N}$ and of the particle $0$ fixed in ${\xx_0(t)=\RR_0}$, with the Langevin dynamics~\eqref{eqC3:GENLang} but with the modified potential~\eqref{eq:pert} in absence of an external field ${\uu_0}$.
Similarly, ${\moy{\bullet}_{\uu_0}}$ will denote the average in presence of a time-dependent external field ${\uu_0(t)}$ in Eq.~\eqref{eq:pert}.
As for the bare brackets $\moy{\bullet}$, we recall that they denote the dynamical average on the original dynamics defined in Sec.~\ref{sec:setting}, 
where ${\uu_0(t)}$ is not an arbitrary function but the true trajectory of the particle $0$  interacting with its surroundings,
and we do not truncate at first order in ${\uu_0(t)}$.

In the limit ${d\to\io}$, the particle 0 has many neighbours and this allows us to simplify further the fluctuating ${\wt{\bm{F}}_0(t)}$ and ${\hat k(t)}$ in Eq.~\eqref{eq:forcesplit}, by introducing the above dynamical averages.
First, the matrix ${\hat k(t)}$ concentrates around its average as a result of its many uncorrelated components:
\beq
\label{eq-k0-average0-v1}
\hat k(t)
 \stackrel{\eqref{eq:forcesplit}}{=}
	\sum_{j (\neq 0)} { \nabla \nabla^T v( \RR_0 - \xx_j(t))}
 =  \sum_{j (\neq 0)} { \nabla \nabla^T v( \RR_0 - \xx_j^{(0)}(t))} + \mathcal{O}(\uu_0)
 \approx \sum_{j (\neq 0)} \moy{ \nabla \nabla^T v( \RR_0 - \xx_j^{(0)}(t))}_0
\ .
\eeq
Here the perturbation in $\uu_0$ can be neglected because the restoring force ${\hat k(t) \, \uu_0(t)}$ in Eq.~\eqref{eq:forcesplit} is already of order ${\mathcal{O}(\uu_0)}$,
whereas we need to take this perturbation into account for the force ${\wt{\bm{F}}_0(t)}$.
Indeed, as a direct consequence of Eq.~\eqref{eq:pert}, we can decompose this quantity in linear response with respect to a small $\uu_0(t)$ as
\beq
\label{eq:F0lin}
\begin{split}
 \wt{\bm{F}}_0(t)
 \stackrel{\eqref{eq:forcesplit}}{=} \bm{F}_0(\ul X(t); \RR_0)
 & \approx \wt{\bm{F}}^f_0(t) + \int_0^t \!\! \de s\, \hat M_R(t,s) \uu_0(s)
 \ ,
 \\
 & \text{with} \quad
 \wt{\bm{F}}^f_0(t)=\bm{F}_0(\ul X ^{(0)}(t); \RR_0) \, , \quad
 M^{\m\n}_R(t,s) = \left. \frac{\delta \moy{F_{0\mu}(\ul X(t); \RR_0)}_{\uu_0}}{\delta u_{0\nu}(s)} \right\vert_{\uu_0=\bm 0}
 \ ,
\end{split}
\eeq
where the first term in Eq.~\eqref{eq:F0lin} is the fluctuating part at fixed ${\xx_0(t) = \RR_0}$, 
and the second term is the shift of the force due to the perturbation ${\uu_0(t)}$, treated as an external field.
The latter is a sum over a large number of uncorrelated particles, and therefore it gets concentrated on its non-zero average value noted ${\hat M_R(t,s)}$. 
Similarly, ${\wt{\bm{F}}^f_{0}(t)}$ is a sum of a large number of uncorrelated fluctuating forces, and it is then a Gaussian random variable by the central limit theorem;
since the average of ${\wt{\bm{F}}_0(t)}$ in absence of perturbation ${\uu_0(t)}$ must be zero by isotropy of the `liquid', it is thus fully characterised by a zero mean and its autocorrelation~${\hat M_C(t,s)}$.

Gathering all the above expressions, we can  rewrite the Langevin dynamics~\eqref{eq:dyn0} for the particle $0$ as follows, directly for its individual displacement ${\uu_0(t)=\xx_0(t)-\RR_0}$:
\beq
\label{eq:part0intermed}
\begin{split}
 \z \dot \uu_0(t)
 	&=  -\hat k(t) \uu_0(t) + \int_0^t \de s\, \hat M_R(t,s) \uu_0(s) + \wt{\bm{F}}_0^f(t) + \bm{\x}_0(t)
 \ , \\
 & \moy{\wt{\bm{F}}^f_{0}(t)}_0=0
 \, , \quad
 \moy{\wt{F}^f_{0\m}(t) \wt{F}^f_{0\n}(s) }_0
 =  M^{\m\n}_C(t,s)
 \ ,
\end{split}
\eeq
and we recall that ${\bm{\xi}_0(t)}$ is the Gaussian noise defined in the original dynamics~\eqref{eqC3:GENLang}.
This equation of motion contains three unknown matricial kernels: ${\lbrace \hat k(t),\hat M_C(t,s),\hat M_R(t,s)\rbrace}$ that we need to determine self-consistently, in order to close our equations.
Note that in this whole subsection we only had to assume small displacements for the particle 0, not for the others yet.

\subsubsection{Isotropy of the `liquid': diagonal kernels and their initialisation values}
\label{secC3:onep-isotropy}

The statistical isotropy of the `liquid' implies that the kernels are proportional to the identity, allowing one to further simplify their expression:
\beq
 k^{\m\n}(t) = k(t) \d_{\m\n}
 \, , \qquad
 M_{C}^{\m\n}(t,s) = M_C(t,s) \d_{\m\n}
 \ , \qquad
 M_{R}^{\m\n}(t,s) =M_R(t,s) \d_{\m\n}
 \ .
\eeq
This implies first for Eq.~\eqref{eq-k0-average0-v1}:
\beq
\label{eq:kF}
 k(t) = \frac1d  \sum_{j (\neq 0)} \moy{ \nabla^2 v( \RR_0 - \xx_j^{(0)}(t))}_0 
 \ .
\eeq
Secondly one obtains from Eq.~\eqref{eq:part0intermed}:
\beq
\label{eqC3:MvsF}
\begin{split}
 M_C(t,s)
	& =\frac1d \sum_{\m=1}^d  \moy{\wt{F}^f_{0\m}(t) \wt{F}^f_{0\m}(s) }_0
 	\stackrel{\eqref{eq:F0lin}}{=}
 	\frac1d  \moy{ \sum_{j (\neq 0)} \nabla v( \RR_0 - \xx_j^{(0)}(t)) \cdot \sum_{k (\neq 0)}\nabla v( \RR_0 - \xx_k^{(0)}(s)) }_0 \\
	&\approx 	\frac1d \sum_{j (\neq 0)} \moy{ \nabla v( \RR_0 - \xx_j^{(0)}(t)) \cdot \nabla v( \RR_0 - \xx_j^{(0)}(s)) }_0
\ ,
\end{split}
\eeq
where to obtain the second line we used the assumption that different particles have uncorrelated contributions, 
which implies that the double sum over particles ${j,k}$ contains only terms with ${j=k}$.
Similarly for the response memory kernel, from Eq.~\eqref{eq:F0lin}:
\beq
\label{eq:MRsim}
\begin{split}
 M_R(t,s)
 	= \left.\frac1d \sum_{\m=1}^d \frac{\delta \moy{F_{0\mu}(\ul X(t); \RR_0)}_{\uu_0}}{\delta u_{0\mu}(s)} \right\vert_{\uu_0=\bm{0}}
	=-\frac{1}{d} \sum_{\mu=1}^d \sum_{j(\neq 0)}\left.\frac{\d \moy{  \nabla_\m v( \RR_0 - \xx_j(t))}_{\uu_0}}{\d u_{0\mu}(s)}\right\vert_{\uu_0=\bm 0} 
 \ .
\end{split}
\eeq
Note that here the same perturbation $\uu_0(t)$ affects all the distances between particle 0 and its neighbours labeled by $j$.
Under the assumption that the neighbours are uncorrelated, we can replace this perturbation by an independent perturbation on each distance.
This amounts to replace the perturbation in Eq.~\eqref{eq:pert}, by ${\sum_{i (\neq 0)} v(\xx_i - \RR_0 - \uu_{0}) \to \sum_{i (\neq 0)} v(\xx_i - \RR_0 - \bm P_{i0})}$ and write
\beq
\label{eq:MRsim2}
\begin{split}
 M_R(t,s)=-\frac{1}{d} \sum_{\mu=1}^d \sum_{j(\neq 0)}\left.\frac{\d \moy{  \nabla_\m v( \RR_0 - \xx_j(t))}_{\bm P}}{\d P_{j0,\mu}(s)}\right\vert_{\bm P=\bm 0} 
\ .
\end{split}
\eeq

These three kernels are explicitly defined as statistical averages over a fictitious `liquid' where one single particle is kept fixed at its initial position $\RR_0$, so we still need to get rid of this constraint.
In fact, in the thermodynamic limit ${N\to\io}$ and at high dimension, we can reasonably assume that in these dynamical averages we can simply remove the constraint of this obstacle, and replace in these average the fictitious `liquid' by the normal system in which all particles can move equivalently.
This is an assumption for the time being, and its proper justification will be postponed to the path-integral formulation in Sec.~\ref{sec:path}.

In practice, this means replacing ${\moy{\bullet}_0}$ by $\moy{\bullet}$, and the unperturbed trajectories ${\xx_j^{(0)}(t)}$ by the full trajectories ${\xx_j(t)}$ (note that this is consistent with all our perturbative expansion truncated at first order in~$\uu_0$).
In addition, since the particle $0$ is not special anymore, we can as well replace ${\xx_0(t)}$ by ${\xx_i(t)}$ and sum over all the pairs ${(i,j)}$, \textit{i.e.}~replacing ${\sum_{j(\neq 0)}}$ by ${\frac{1}{N} \sum_{j \neq i}}$.
Consequently, we can rewrite the three kernels as follows:
\beq
\label{eq-diagonial-kernels-sum-ij}
\begin{split}
 k(t)
 	&= \frac1{Nd} \sum_{i \neq j}  \moy{ \nabla^2 v( \xx_i(t) - \xx_j(t))} 
 \ , \\
 M_C(t,s)
 	&= \frac1{Nd} \sum_{i \neq j} \moy{ \nabla v( \xx_i(t) - \xx_j(t)) \cdot \nabla v( \xx_i(s) - \xx_j(s))}
 \ ,\\
 M_R(t,s)
 	&= \frac1{Nd} \sum_{\m=1}^d \sum_{i\neq j} \left.\frac{\d \moy{  \nabla_\m v( \xx_i(t) - \xx_j(t) )}_{\bm P}}{\d P_{ij,\mu}(s)}\right\vert_{\bm P=\bm 0} 
 \ .
\end{split}
\eeq
In the last expression, we defined the response similarly to Eq.~\eqref{eq:MRsim2}
by shifting the potential independently for each pair
${\sum_{j\neq i} v(\xx_i-\xx_j) \mapsto \sum_{j\neq i} v(\xx_i-\xx_j -\bm P_{ij}(t))}$ in the Langevin dynamics.
Note that we must impose for consistency that ${\bm P_{ij}(t) = -\bm P_{ji}(t)}$, which is why the minus sign in Eq.~\eqref{eq:MRsim2} has disappeared.

At time ${t=0}$, we have ${\xx_i(0)=\RR_i}$, and the ${t=0}$ values for the kernels are simple thermodynamical averages 
over the `liquid' in equilibrium at ${\Ti=\beta_0^{-1}}$.
These can be expressed as
\beq
\label{eqC3:kF0}
\begin{split}
 k(0)
 &= \frac1{Nd} \sum_{j \neq i} \moy{ \nabla^2 v( \RR_i - \RR_j) }_{\text{eq}, T_0}
 = \frac{\r}d \int \de\rr_0\, g_{\text{in}}(\valabs{\rr_0}) \Lap v( \rr_0)
 \ , \\
 M_C(0,0)
 &=  \frac1{Nd} \sum_{i \neq j} \moy{ | \nabla v(\RR_i-\RR_j)|^2 }_{\text{eq}, T_0}
 =  \frac{\r}d \int \de \rr_0 \, g_{\text{in}}(\valabs{\rr_0}) \left|\nabla v(\rr_0) \right|^2
 \ , 
\end{split}
\eeq
where $\r$ is the number density and we introduced the radial distribution function ${g_{\text{in}}(\valabs{\rr_0})}$~\cite{hansen}.
This function characterises the initial stochastic condition, and we need it to be radial in order to initially have a statistical isotropy.
In particular, under our assumption that we start at equilibrium at temperature $T_0$, it is given by its first virial contribution in the high-dimensional limit~\cite{FRW85,WRF87,FP99}:
\beq
\label{eq-radial-distribution-Ti}
 g_{\text{eq}}(\valabs{\rr_0}|\Ti) = e^{-\Bi v(\rr_0) }
\ .
\eeq
Note that we do not write explicitly the expression of $M_R(0,0)$ because while it is formally similar to Eq.~\eqref{eqC3:kF0}, the response at equal time is ill-defined and depends on the discretisation of the stochastic equation as well as on the presence (or absence) of inertia.

\subsubsection{Effective stochastic process for the individual displacements}
\label{secC3:onep-eff-stoch-process}

Because the particle $0$ has nothing special, the above derivation can be applied to any particle by extending the assumption of small individual displacements ${\uu_j(t)=\xx_j(t) - \RR_j}$, generalising the stochastic process given in Eq.~\eqref{eq:part0intermed} to any particle in the `liquid' as well.
We define moreover the effective noise%
\footnote{We include the factor $\sqrt{2}$ for later convenience.} ${\sqrt{2}\, \bm{\X}_i = \bm{\x}_i + \wt{\bm{F}}_i^f}$, which is a sum of Gaussian processes and thus Gaussian itself, and its associated average ${\moy{\bullet}_{\bm{\Xi}}}$.
Substituting the diagonal kernels~\eqref{eq-diagonial-kernels-sum-ij}, we thus obtain:
\beq
\label{eqC3:effproc}
\begin{split}
 & \z \dot \uu_i(t)
 = - k(t) \uu_i(t) + \int_0^t \!\! \de s\, M_R(t,s) \uu_i(s) + \sqrt{2}\, \bm{\X}_i(t)
 \ , \\
 & \moy{\X_{i\m}(t)}_{\bm{\Xi}} =0
 \ , \qquad
 \moy{\X_{i\m}(t) \X_{i\n}(s)}_{\bm{\Xi}}
 = \d_{\m\n} \left[ T \z \d(t-s) +\frac12 \G_C(t,s) +\frac12 M_C(t,s) \right]
 \ ,
\end{split}
\eeq
which is a \textit{single-particle} effective process with the kernels ${\lbrace k(t), M_C(t,s),M_R(t,s) \rbrace}$, now scalar functions that remain to be determined.
Remember that by definition the initial condition for $\uu_i(t)$ is ${\uu_i(0)=\bm 0}$.
Note at last that if we remove the indices $i$, this single effective process accounts for the typical fluctuations of the individual displacements of any particle.


\subsection{Two-particles effective stochastic process}
\label{secC3:twop}

\subsubsection{Effective stochastic process for the inter-particle distances}
\label{secC3:twop-eff-stoch-process}

To determine self-consistently the kernels ${\lbrace k(t), M_C(t,s), M_R(t,s) \rbrace}$, 
we need to write the effective process for the dynamics of two particles~\cite{Sz17}.
Physically, this is because the dynamics is driven by pairwise interactions, controlled by the inter-particle distance as it can be read directly from the kernel expressions in Eq.~\eqref{eq-diagonial-kernels-sum-ij}.

The original Langevin equation~\eqref{eqC3:GENLang} for particles $i$ and $j$ reads
\beq
\begin{split}
 &\z \dot \xx_i(t)
 	= \bm{F}_{j\to i}(t) + \bm{F}_{i}^{(j)} + \bm{\x}_i(t)
 \ , \\
 &\z \dot \xx_j(t)
 	= - \bm{F}_{j\to i}(t) + \bm{F}_{j}^{(i)} + \bm{\x}_j(t)
 \ ,
\end{split}
\eeq
where ${\bm{F}_i^{(j)}}$ denotes the total force on the particle $i$, to which the contribution ${\bm{F}_{j\to i}(t)}$ coming from the particle $j$ has been substracted.
In the limit ${d\to\io}$, the number of terms in~${\bm{F}_i^{(j)}}$ is proportional to $d$ and removing one contribution is a small correction%
\footnote{Note that, as such, ${\bm{F}_{j\to i}(t)}$ is also a small correction, as can be seen from power counting in $d$ from the ${d\to\io}$ scalings presented in Sec.~\ref{secC3:largedM}.
One therefore recovers Eq.~\eqref{eqC3:effproc} for both particles $i$ and $j$, as it should. 
Nevertheless, it is important to keep the additive contribution ${\bm{F}_{j\to i}(t)}$, as we will see that in the relative motion of the particle pair ${(i,j)}$ its projection along the dipole they form sums coherently and gives a contribution which cannot be neglected.}.
We can thus apply to ${\bm{F}_i^{(j)}}$ the same treatment applied to the total force in Sec.~\ref{secC3:onep}, and invoking Eq.~\eqref{eqC3:effproc} we obtain
\beq
\label{eqC3:app1}
\begin{split}
 & \z \dot \uu_i(t)
	=- k(t) \uu_i(t) + \int_0^t \!\! \de s\, M_R(t,s) \uu_i(s) + \sqrt{2}\bm{\X}_i(t) + \bm{F}_{j\to i}(t)
 \ , \\
 & \z \dot \uu_j(t)
 	= - k(t) \uu_j(t) +\int_0^t \!\! \de s\, M_R(t,s) \uu_j(s) + \sqrt{2}\bm{\X}_j(t) - \bm{F}_{j\to i}(t)
 \ , \\
\end{split}
\eeq
where the noises ${\bm{\X}_i(t)}$ and ${\bm{\X}_j(t)}$ are independent and have the same statistics as in Eq.~\eqref{eqC3:effproc}.

We recall the notations for the relative displacements that we have defined in Eq.~\eqref{eq-def-wrt-initial-condition}, though removing the indices designating the pair ${(ij)}$ since we will obtain a stochastic process describing the typical fluctuations of these quantities:
\begin{equation}
 \rr(t) = \xx_i(t) - \xx_j(t)
 \, , \quad
 \rr_{0} = \RR_i - \RR_j
 \, , \quad
 \vv(t) = \uu_i(t) - \uu_j(t) = \rr(t) - \rr_0
 \, .
\end{equation}
The fluctuating inter-particle distance $\vv(t)$ satisfies an equation obtained by taking half the difference of the equations in Eq.~\eqref{eqC3:app1}, and using that the force is given by ${\bm{F}(\rr)=-\nabla v(\rr)}$:
\beq
\label{eqC3:twop}
 \begin{split}
 &\frac\z2 \dot \vv(t)
 	=- \frac{k(t)}2 \vv(t) + \frac12 \int_0^t \!\! \de s\, M_R(t,s) \vv(s) - \nabla v(\rr_0 + \vv(t) - \bm P(t)) + \bm{\X}(t)
 	\ , \\
 & \moy{\X_{\m}(t)}_{\bm{\Xi}}=0
 \, , \quad
 \moy{\X_{\m}(t) \X_{\n}(s)}_{\bm{\Xi}}
 	= \d_{\m\n} \left[ T \z \d(t-s) + \frac12 \G_C(t,s) + \frac12 M_C(t,s) \right]
 \ .
\end{split}
\eeq
Here, the noise ${\bm{\X}(t) = ( \bm{\X}_i(t) - \bm{\X}_j(t) )/\sqrt{2}}$ has the same statistics as in Eq.~\eqref{eqC3:effproc}.
By definition the initial condition for ${\vv(t)}$ is ${\vv(0) =\bm 0}$,
and $\rr_0$ is a fixed parameter, whose role is explained in the next section. The vector $\bm P(t)$
is only used to compute the response function, as in Eq.~\eqref{eq-diagonial-kernels-sum-ij}, and otherwise set to zero.

\subsubsection{Self-consistent definition of the kernels}
\label{secC3:twop-self-consist-def-kernels}

We can finally express the three kernels ${\lbrace k(t), M_C(t,s), M_R(t,s) \rbrace}$ as dynamical averages over the effective process given by Eq.~\eqref{eqC3:twop}.
The normalisation constant of the averages is determined by continuity with the values at ${t=0}$ given by Eq.~\eqref{eqC3:kF0}.
We can similarly rewrite Eq.~\eqref{eq-diagonial-kernels-sum-ij}, denoting ${\moy{\bullet}_{\vv}}$ the dynamical average over the stochastic process ${\vv(t)=\rr(t)-\rr_0}$:
\beq
\label{eqC3:Mself}
\begin{split}
 k(t)
 	&= \frac{\r}d \int \de\rr_0 \, g_{\text{in}}(\valabs{\rr_0}) \moy{ \Lap v( \rr_0 + \vv(t))}_{\vv}
 \ , \\
 M_C(t,s)
 	&=  \frac{\r}d \int \de \rr_0 \, g_{\text{in}}(\valabs{\rr_0)} \moy{ \nabla v(\rr_0 + \vv(t))  \cdot \nabla v(\rr_0 + \vv(s)) }_{\vv}  
 \ , \\
 M_R(t,s)
 &= \frac{\rho}d \sum_{\m=1}^d \int \de \rr_0 \, g_{\text{in}}(\valabs{\rr_0)} \,\left. \frac{\d \moy{  \nabla_\m v(\rr_0 + \vv(t)) }_{\vv,\bm{P}}}{\d P_{\mu}(s)}\right\vert_{\bm P=\bm0}
 \ .
\end{split}
\eeq
One should then choose a $\rr_0$, for that value of $\rr_0$ compute the dynamical averages over $\vv(t)$ using the effective process~\eqref{eqC3:twop} with initial condition ${\vv(0)=\bm 0}$, and finally integrate over $\rr_0$.
Since we assume that we start from equilibrium, the integration over $\rr_0$ involves the equilibrium distribution of the inter-particle distances at the initial temperature~$\Ti$, that is proportional to the radial distribution function ${g_{\text{in}}(\valabs{\rr_0})=g_{\text{eq}}(\valabs{\rr_0} | \Ti) = e^{-\Bi v(\rr_0)}}$, defined in Eq.~\eqref{eq-radial-distribution-Ti}~\cite{FRW85,WRF87,FP99}.
Note that this quantity is not normalisable because ${g_{\text{eq}}(\valabs{\rr_0} | \Ti)\to 1}$ when ${r_0 \to\io}$, and therefore it does not define a probability distribution over $\rr_0$.
Yet, the integration over $\rr_0$ is convergent because all the dynamical averages vanish fast enough at large $\rr_0$.

Physically,
${k(t)}$ is simply the mean divergence of local forces,
${M_C(t,s)}$ the force-force correlator,
and ${M_R(t,s)}$ the mean linear response of the local force.
They are obtained as statistical averages over the initial condition ${\rr_0}$
and dynamically over the trajectories ${\vv(t)}$.
Note that, at equilibrium, the two memory kernels would be related by a FDT relation, whereas in a more generic out-of-equilibrium setting they must be treated as two distinct functions.

This finally closes our mean-field dynamics, in its high-dimensional vectorial form:
the two effective stochastic processes for the individual displacement of each particle in Eq.~\eqref{eqC3:effproc},
the inter-particle distance of each pair of particles in Eq.~\eqref{eqC3:twop},
and the self-consistent equations for the three kernels in Eq.~\eqref{eqC3:Mself}.
A complex many-body problem with ${N\to\io}$ particles 
has thus been reduced to the self-consistent solution of a single $d$-dimensional stochastic differential equation.
These expressions are recalled in the summary section~\ref{sec:summary-results}, reinstating there a finite inertia (${m \neq 0}$) and a retarded friction kernel (${\Gamma_R \neq 0})$.
All our derivation here relied essentially on three assumptions:
small displacements of particles around their initial position (${\uu_i(t) = \mathcal{O}(1/d)}$),
uncorrelated numerous neighbours,
and statistical isotropy of the system.
We will show in the next section that we can obtain exactly the same effective dynamics via a path-integral description in high dimension, where these assumptions can be properly implemented and justified.

\section{Derivation via a path integral in Euclidean space}
\label{sec:path}

In this section, we present an alternative derivation of the self-consistent effective dynamics obtained in Sec.~\ref{sec:cavity}, based on a path-integral representation of the dynamics and specifically its high-dimensional saddle-point simplification, along the lines of Refs.~\cite{MKZ16,KMZ16}.
We emphasise that it is complementary to the cavity derivation, in the sense that it allows to properly implement the physical assumptions that we have used until now, though it is more technical and thus less intuitive; those assumptions will however be fully justified only in~Sec.~\ref{secC3:largedM-path-integral}.

We first compute in Sec.~\ref{sec:path-MSRDJ-SUSY} the dynamical action in a supersymmetric (SUSY) form.
Then in Sec.~\ref{sec:path-Gaussian-ansatz} we rewrite it assuming a \emph{Gaussian} ansatz for the high-dimensional fluctuations of the individual displacements ${\lbrace \uu_i(t) \rbrace}$, and we derive the self-consistent equation for their variance  by using an extremalisation criterion.
We finally extract in Sec.~\ref{sec:path-eff-stoch-process} the resulting \emph{effective} stochastic processes for $\vv$  and for $\uu$ with their corresponding kernels,
recovering the same equations as in the previous section.

\subsection{Martin-Siggia-Rose-De Dominicis-Janssen action and supersymmetry}
\label{sec:path-MSRDJ-SUSY}

We now consider the same Langevin dynamics given in Eq.~\eqref{eqC3:GENLang}, once again setting ${m=0}$ and ${\G_R=0}$ for simplicity (they will be restored in the summary in Sec.~\ref{sec:summary-results}).
Within the Martin-Siggia-Rose-De Dominicis-Janssen (MSRDDJ) formalism~\cite{MSR,Ja76,DD78,Cu02,CC05,Kamenev} we introduce a dynamical partition function of the form%
\footnote{Note that we did not normalise the Gibbs-Boltzmann distribution for the initial condition, and as a result $\ZZ_N$ coincides with the thermodynamic partition function at temperature ${1/\Bi}$.}
\beq\label{eq:ZNdyndef}
\begin{split}
 &\mathcal{Z}_N
 	= \int \mathcal{D}\ul{\x}  \int \mathcal{D} \ul X  \  e^{-\Bi V(\ul X(0))} \prod_{i=1}^{N} \d( \z \dot \xx_i(t)  -\bm F_i(t) - \bm\x_i(t) )
 \\
  &=
	\int \mathcal{D} \ul{X}  \int \mathcal{D} \ul{\tilde X}  \
	e^{ \int \de t \sum_i \argc{ T \z i \tilde \xx_i(t) \cdot i \tilde \xx_i(t) - i \tilde \xx_i(t) \cdot \z\dot \xx_i (t) }
	+ \frac12 \int \de t \de s \sum_i i \tilde\xx_i(t) \cdot \Gamma_C(t,s) i \tilde\xx_i(s)
 	+\int \de t \sum_i  i\tilde \xx_i(t) \cdot \bm F_i(t)  -\Bi V(\ul X(0))}
 \ .
\end{split}
\eeq
Here ${\mathcal{D} \ul X}$ denotes a functional integral over the trajectories ${\ul X(t) = \{ \xx_i(t) \}}$,
and similarly ${\mathcal{D}  \ul \x}$ over the noises ${\ul \xi (t)= \lbrace \bm{\x}_i(t) \rbrace}$.
The trajectories $\ul{\tilde X}(t)  = \{\tilde\xx_i(t)\}$ are introduced to construct the path-integral action via an integral representation
of the functional $\delta$ function in the first line of Eq.~\eqref{eq:ZNdyndef}. They are usually called `response fields' because 
the response function can be written as a correlation between $\xx_i(t)$ and $\tilde\xx_i(t)$, see
Refs.~\cite{MSR,Ja76,DD78,Cu02,CC05,Kamenev} for more details\footnote{The response fields are often
denoted by $\hat\xx$, but here we use the less standard notation $\tilde\xx$ to avoid confusion with unit vectors, which are denoted
by $\hat \xx = \xx/|\xx|$ in our conventions.}.
Note that because we assume that we start in an equilibrium configuration at time ${t=0}$, all integrals
over time extend from~$0$~to~$\io$ and the path integral includes the Boltzmann factor ${e^{-\Bi V(\ul X(0))}}$.

As it is well-known from previous studies on equilibrium dynamics~\cite{MKZ16,KMZ16}, this partition function can be simplified further in the high-dimensional limit, and in addition it can conveniently be put in a SUSY form which has a similar structure to the static case.
Thereafter we start by adapting these simplifications to our out-of-equilibrium setting with an equilibrium initial condition at time ${t=0}$.

\subsubsection{Dynamical generating functional, virial expansion, and initial condition}
\label{sec:path-MSRDJ-SUSY-formalism-part1}

We can rewrite the dynamical partition function as:
\begin{equation}
 \mathcal{Z}_N
 	=\int \argp{ \prod_{i=1}^N \mathcal{D} \xx_i(t) \mathcal{D} \tilde{\xx}_i (t) } \,e^{-\AA[\{\xx_i,\tilde{\xx}_i \}]}
\label{eq:Z}
\end{equation}
where the dynamical action~$\mathcal{A}$ is the sum of a one-particle (friction and noise) and a two-particle (interaction) terms:
\begin{equation}
\label{eq:action}
\begin{split}
 \AA[\{\xx_i,\tilde{\xx}_i\}]
	&=\sum_{i=1}^N \Phi[\xx_i,\tilde{\xx}_i]
		+ \frac12 \sum_{i,j=1}^{N}W[\xx_i-\xx_j,\tilde{\xx}_i-\tilde{\xx}_j]
 \, ,
 \\
 \Phi[\xx,\tilde{\xx}]
 	&=\z\int \de t \,\left( T\tilde{\xx}(t)^2 + i\tilde{\xx}(t)\cdot\dot{\xx}(t) \right)
		- \frac12 \int \de t \de s\, i \tilde\xx(t) \cdot \Gamma_C(t,s) \, i \tilde\xx(s)
 \, , \\  
 W[\xx-\yy,\tilde{\xx}-\tilde{\yy}]
 	&=\Bi v(\xx(0)-\yy(0)) + \int \de t \,[i \tilde{\xx}(t)- i \tilde{\yy}(t)] \cdot \nabla v(\xx(t)-\yy(t)) \ .
\end{split}
\end{equation}
One can therefore apply the standard virial expansion to the calculation of the dynamical equivalent of the `free energy' associated to ${\ZZ_N\sim e^{\FF}}$~\cite{MKZ16}.
In high dimension, this expansion can be truncated to the second virial coefficient~\cite{FRW85,WRF87,FP99,MKZ16,KMZ16};
physically, this amounts to assuming uncorrelated neighbours, or equivalently neglecting any interaction loop.
This truncated expansion results in the effective action
\beq
\label{eq-truncated-virial-expansion}
 \FF
	= -\int \mathcal{D} \xx \mathcal{D} \tilde{\xx} \,\rho[\xx,\tilde{\xx}](\Phi[\xx,\tilde{\xx}]+\log \rho[\xx,\tilde{\xx}])
		+ \frac12 \int \mathcal{D} \xx_1 \mathcal{D}  \tilde{\xx}_1 \mathcal{D} \xx_2 \mathcal{D} \tilde{\xx}_2 \,\rho[\xx_1,\tilde{\xx}_1]\rho[\xx_2,\tilde{\xx}_2] f[\xx_1-\xx_2,\tilde \xx_1-\tilde \xx_2]
 \ ,
\eeq
where ${f=e^{-W} -1}$ is the so-called `dynamical Mayer function'
and all the time-dependence of the trajectories are implicit.
More importantly, ${\rho[\xx,\tilde{\xx}]}$  is the probability density of trajectories whose actual profile is determined through the extremalisation/Legendre condition%
\footnote{In this expansion, the probability density of trajectories is conjugated to the kinetic one-particle term ${\Phi[\xx,\tilde{\xx}]}$ through a Legendre transformation, and the usual definition of conjugated Legendre variables translates here into this extremalisation condition on the dynamical action $\FF$.}
${\d\FF/\d\rho[\xx,\tilde{\xx}]=0}$.
Its definition and normalisation are:
\beq
\label{eq-def-normalisation-Pi}
\begin{split}
 \rho[\xx,\tilde{\xx}]
	=\la\sum_{i=1}^N\d[\xx-\xx_i]\d[\tilde \xx-\tilde\xx_i] \ra 
 \, , \qquad
 \int \mathcal{D} \xx \mathcal{D} \tilde \xx\, \r[\xx,\tilde \xx] = N
 \, .
\end{split}
\eeq
The Dirac~$\delta$ functions act at the level of trajectories, \textit{i.e.}~they are product of Dirac~$\delta$ over time.
Of course this definition of the probability density is implicit at this stage, since ${\moy{\bullet}}$ is the statistical average corresponding to the dynamical partition function ${\ZZ_N\sim e^{\FF}}$, while $\FF$ depends on ${\rho[\xx,\tilde{\xx}]}$ and the latter is fixed by the extremalisation condition on the dynamical action itself.

We now change variables in order to absorb the initial condition~${\ul X (0)}$.
We recall the definitions~Eq.~\eqref{eq-def-wrt-initial-condition}, removing the specific particle indices which are not relevant in our path-integral formulation:
${\RR = \xx(0)}$, ${\uu(t) = \xx(t) - \xx(0)}$, and ${\tilde \uu(t) = \tilde \xx(t)}$.
Note that the initial condition for $\uu(t)$ is ${\uu(0)=\bm 0}$, while there is no initial condition for $\tilde \uu(0)$.
By translational invariance, in a cubic region $\Omega$ with periodic boundary conditions and volume $|\Omega|$, we can write
${\r[\xx,\tilde \xx] = \r [\RR + \uu,\tilde{\uu}] = \r \, \Pi[\uu,\tilde \uu]}$
where ${\rho=N/|\Omega|}$ is now the actual number density of particles in the system (not to be confused with the functional ${\r[\xx,\tilde \xx]}$).
By definition, the new probability density ${\Pi[\uu,\tilde \uu]}$ does not depend on $\RR$,
and its normalisation is given by 
\beq
\label{eq-def-normalisation-Pi-bis}
 \int \de \RR \, \mathcal{D} \uu \mathcal{D} \tilde \uu \, \r[\xx,\tilde \xx]
	= |\Omega| \r \int \mathcal{D} \uu \mathcal{D} \tilde \uu \, \Pi[\uu,\tilde \uu]
	= N 
 \quad \Rightarrow \quad
 \int \mathcal{D} \uu \, \mathcal{D} \tilde \uu \, \Pi[\uu,\tilde \uu]
 	=1
 \ .
\eeq
Similarly, the one-particle term of the action in Eq.~\eqref{eq:action} does not depend on $\RR$ so we have simply ${\Phi[\xx,\tilde \xx] = \Phi[\uu,\tilde \uu]}$.
Therefore, neglecting additive constants, we obtain a free-energy density that splits into its `ideal gas' component ${\ff^{\text{id}}}$ and its excess
component ${\ff^{\text{ex}}}$:
\beq
 \label{eq:freePI}
 \begin{split}
 \ff & = \frac{\FF}N = \ff^{\text{id}} + \ff^{\text{ex}} \ ,
 \\
 \ff^{\text{id}} &= - \int \mathcal{D} \Pi[\uu,\tilde{\uu}](\Phi[\uu,\tilde{\uu}]+\log \Pi[\uu,\tilde{\uu}]) \ ,
 \\
 \ff^{\text{ex}} &= \frac\r2 \int \de \rr_0 \left\{ e^{-\Bi v(\rr_0)} \int \mathcal{D} \Pi[\uu_1,\tilde{\uu}_1] \mathcal{D} \Pi[\uu_2,\tilde{\uu}_2] \,
e^{ - \int \de t\,[i \tilde{\uu}_1(t)-i \tilde{\uu}_2(t)] \cdot \nabla v(\rr_0+ \uu_1(t)-\uu_2(t))} 
 -1 
\right\}
 \ ,
\end{split}
\eeq
where ${\rr_0 = \RR_1 - \RR_2}$, and 
${\mathcal{D} \Pi[\uu,\tilde{\uu}] = \mathcal{D} \uu \mathcal{D} \tilde{\uu}\, \Pi[\uu,\tilde{\uu}]}$ is the measure with respect to the (normalised) distribution $\Pi$.

We emphasise that, in order to derive the free-energy density $\ff$ of Eq.~\eqref{eq:freePI}, we have essentially considered its standard virial expansion truncated to the second order  (Eq.~\eqref{eq:action}), and then we used the statistical translation invariance of the system in order to rewrite it as a function of the probability density ${\Pi[\uu,\tilde{\uu}]=\rho[\xx,\tilde{\xx}]/\rho}$, with the normalisation given in Eq.~\eqref{eq-def-normalisation-Pi};
the latter is then self-consistently given by the extremalisation condition ${\d\FF/\d\Pi[\uu,\tilde{\uu}]=0}$.
Under these assumptions, the initial stochastic condition only appears via the integration over $\rr_0$ in the interaction term, and we recall that at equilibrium the initial radial distribution is simply given by ${g_{\text{in}}(\valabs{\rr_0}) = g_{\text{eq}} (\valabs{\rr_0} \vert T_0) = e^{-\Bi v(\rr_0)}}$.

\subsubsection{Supersymmetric formulation}
\label{sec:path-MSRDJ-SUSY-formalism-part2}

The free-energy density $\ff$ can be put in a SUSY form, allowing us to formulate the dynamics in a compact manner, advantageous for the subsequent calculations of the path-integral saddle point in high dimension.
In order to do so, we introduce SUSY fields defined in terms of Grassmann variables $\th$ and $\bth$~\cite{Ku92,Ku03,Cu02,Zinn-Justin}:
\beq
\begin{split}
\label{eq:defSUSY-part1}
 & a = \{ t_a, \th_a, \bth_a \}
 \ , \\
 & \uu(a) = \uu(t_a)  + i \tilde \uu_a(t) \ththbar{a}
 \ , \\
 & \d(a,b)
 	= \d(t_a - t_b) (\ththbar{a} + \ththbar{b})
 \ , \\
 & \Rightarrow \quad
 	\int \de a \,f[ \uu(a) ]
 	= \int \de t_a \de \bth_a \de \th_a \, f[\uu(t_a) + i \tilde \uu(t_a) \ththbar{a}]
	= \int \de t \,  i\tilde \uu(t) \cdot \nabla f[\uu(t)]  \ ,
\end{split}
\eeq
and specifically for the SUSY representation%
\footnote{Note that we do not introduce the fermions by following It\^o convention~\cite{Cu02}.} 
of our dynamics (with ${m =0}$) we define:
\beq
\begin{split}
\label{eq:defSUSY-part2}
 & \partial^2_a
 	= 2T\z \frac{\partial^2}{\partial\th \partial\bth} -2 \z\th \frac{\partial^2}{\partial\th\partial t}+2\z\frac{\partial}{\partial t} 
 \ , \\
 & \G(a,b) = \G_C(t_a,t_b) + \ththbar{a} \, \partial_{t_a} \Gamma_R(t_b,t_a) + \ththbar{b} \, \partial_{t_b} \Gamma_R(t_a,t_b)
 \ , \\
 & \Phi[\uu(a)]
 	= \frac12 \int \de a \de b \, \uu(a)  \cdot \left[\partial^2_a \d(a,b) -  \G(a,b) + \Gamma_R(t_a,t_a)  \, \delta(a,b) \right]  \uu(b)
 \ .
\end{split}
\eeq
We choose as an overall convention that the argument indicates if we are considering a superfield or a scalar,  for instance $\uu(a)$ or ${\uu(t_a)}$.
Note that these expressions include the general case with ${\Gamma_R \neq 0}$, following Ref.~\cite{ABUZ18}, but from now on we set for simplicity ${\Gamma_R =0}$ and we will reinstate it only in the summary in Sec.~\ref{sec:summary-results}.

In terms of these variables, we have for the free-energy density ${\ff = \ff^{\text{id}} + \ff^{\text{ex}}}$ of Eq.~\eqref{eq:freePI}:
\beq
 \label{eq:freePISUSY}
\begin{split}
 \ff^{\text{id}}
 	&= - \int {\mathcal{D}}\Pi[\uu(a)](\Phi[\uu(a)]+\log \Pi[\uu(a)]) \ ,
 \\
  \ff^{\text{ex}}
  	&= \frac\r2 \int \de \rr_0 \left\{ e^{-\Bi v(\rr_0)}  \int {\mathcal{D}}\Pi[\uu_1(a)] {\mathcal{D}}\Pi[\uu_2(a)] \,
e^{ - \int \de a\, v(\rr_0 + \uu_1(a)-\uu_2(a))} 
 -1 
\right\} \ .
\end{split}
\eeq

\subsection{Gaussian approximation in the supersymmetric formulation}
\label{sec:path-Gaussian-ansatz}

We will now assume that the probability density ${\Pi[\uu (a)]}$ is Gaussian,
and thus fully determined by its mean and variance:
\beq
\label{eq-Gaussian-meas-part1}
 \moy{\uu(a)} =0
 \ , \qquad
 \moy{\uu(a) \cdot \uu(b)} = d \, A(a,b)
 \ ,
\eeq
hence we introduce the compact notation ${\Pi[\uu(a)] = G_A[\uu(a)] \propto e^{- \frac12 \int \de b \de c \, \uu(b) A^{-1}(b,c) \uu(c)}}$.
The immediate consequence of this assumption is that the inter-particle distance ${\vv(a)= \uu_1(a)-\uu_2(a)}$ is also a Gaussian variable, with variance ${2A(a,b)}$.
This implies for the path-integral measures:
\beq
\label{eq-Gaussian-meas-part2}
 \mathcal{D} \Pi[\uu(a)] \to \mathcal{D} \uu \, G_A[\uu(a)]
 \, , \qquad
 \mathcal{D} \Pi[\uu_1(a)] \, \mathcal{D} \Pi [\uu_2(a)] \to \mathcal{D} \vv \, G_{2A}[\vv(a)] \ ,
\eeq
where the second replacement is possible whenever one is averaging a function of $\uu_1(a)-\uu_2(a)$.
The idea of using a Gaussian assumption for the single-particle trajectory distribution has been proposed in Ref.~\cite{MK11}, in a similar context, but technical aspects of the calculations are different here.
Even in the limit ${d\to\io}$, this assumption is, strictly speaking, not exact~\cite{LCDBB10,CIPZ12}.
However, it becomes exact at the ${d\to\io}$ saddle-point level of the free-energy path integral of Eq.~\eqref{eq:freePISUSY}, which is all that we need in order to compute averaged quantities or the free energy itself~\cite{KPZ12,KPUZ13}.
This assumption will be better justified in Sec.~\ref{secC3:largedM-path-integral}, but for the time being it can be seen as the explicit implementation of the assumption of single-particle Gaussian fluctuations --~based on the central limit theorem, thanks to the numerous uncorrelated neighbours of each particle~-- that we invoked several times in the cavity derivation of Sec.~\ref{sec:cavity}.
%
We emphasise that although the measure for $\vv$ in Eq.~\eqref{eq-Gaussian-meas-part2} is Gaussian, it is an assumption at the level of the path-integral measure and equivalently on the \textit{single-particle} dynamics, \textit{not} on the full dynamics of the inter-particle distance encoded in Eq.~\eqref{eq:freePISUSY}.
We shall therefore see --~as in the cavity approach~-- that the effective stochastic process for the inter-particle distance $\vv$ is non-Gaussian.

Using the Gaussian measures~\eqref{eq-Gaussian-meas-part2} to explicitly average over the fluctuations of $\uu(a)$, we can further simplify the free-energy density, and express it as a function of the superkernel ${A(a,b)}$.
First, ${\Phi[\uu(a)]}$ is by definition quadratic in the displacements ${\uu(a)}$ and under the Gaussian assumption so is ${\log\Pi[\uu(a)]}$;
this implies that the `ideal gas' single-particle term $\ff^{\text{id}}$ in Eq.~\eqref{eq:freePISUSY} becomes a pure Gaussian integral, and its direct evaluation reads (up to irrelevant additive constants independent of $A$)%
\footnote{
For the term ${\log\Pi[\uu(a)]}$ we explicitly computed the standard integral, introducing the `determinant' of the superkernel ${A(a,b)}$:
\beq
\nonumber
\begin{split}
 &\int \mathcal{D} \uu \, G_A[\uu(a)] \, \log G_A[\uu(a)]
 = - \frac12  \int \mathcal{D} \uu \, G_A[\uu(a)] \, \int \de b \de c \, \uu(b) A^{-1}(b,c) \uu(c)
 	- \log \arga{\int \mathcal{D} \uu \, e^{-\frac12 \int \de b \de c \, \uu(b) A^{-1}(b,c) \uu(c)}}
 \\
 &= - \frac12 \int \de b \de c \, \moy{ \uu(b) A^{-1}(b,c) \uu(c)}_\uu
 	- \frac{d}{2} \log \det A
 	= 	- \frac{d}{2} \log \det A + \text{constant}
\ . 
\end{split}
\eeq}:
\begin{equation}
\label{eq:Gaussian_id}
 \ff^{\text{id}}
 	=- \int \mathcal{D} \uu \, G_A[\uu(a)] \argp{\Phi[\uu(a)]+\log G_A[\uu(a)]}
 	=- \frac{d}2 \int \de a \de b \,  \left[\partial^2_a \d(a,b) - \G(a,b) \right]   A(a,b) +\frac{d}2 \log\det A
\ .
\end{equation}
Focusing on the two-particle interaction term, we can use the general identity for Gaussian convolutions
\beq
\label{eq:idGA}
 \int \mathcal{D} \vv \, G_{2A}[\vv(a)] \, e^{ - \int \de a\, v(\rr_0 + \vv(a))}
 	= \left. e^{\sum_\m \int\de a\de b\, A(a,b) \frac{\delta^2}{\d w_\m(a) \d w_\m(b)}}
			e^{ - \int \de a\, v(\rr_0 + \vv(a))} \right|_{\vv=0} \ ,
\eeq
and rewrite the excess free-energy density in Eq.~\eqref{eq:freePISUSY}:
\beq
\label{eq:Gaussian_exchange}
 \ff^{\text{ex}}
 	= \frac\r2 \int \de \rr_0 \left\{ e^{-\Bi v(\rr_0)} \left[ 
e^{\sum_\m \int\de a\de b \,A(a,b) \frac{\delta^2}{\d w_\m(a) \d w_\m(b)}}
e^{ - \int \de a\, v(\rr_0 + \vv(a))} \right]_{\vv=0} -1 \right\}
\ .
\eeq
Combining Eqs.~\eqref{eq:Gaussian_id} and~\eqref{eq:Gaussian_exchange} we recover the total dynamical action ${\mathcal{F}=N \ff=N(\ff^{\text{id}}+\ff^{\text{ex}})}$.

We recall that in the truncated virial expansion of Eq.~\eqref{eq-truncated-virial-expansion}, valid in high dimension, the probability density ${\Pi[\uu,\tilde{\uu}]}$ was defined by the extremalisation condition ${\d\FF/\d\Pi[\uu,\tilde{\uu}]=0}$.
Within our Gaussian assumption~\eqref{eq-Gaussian-meas-part1},
the extremalization of the dynamical action $\ff$ in order to determine the probability density of trajectories amounts now to take the derivative with respect to $A(a,b)$, \textit{i.e.}~to compute ${\d \ff/\d A(a,b)=0}$.
The variance ${A(a,b)}$ being symmetric in its arguments and using the identity ${\d [\log\det A]/\d A(a,b)= A^{-1} (a,b)}$, the extremalisation condition yields the following self-consistent equation:
\beq
\label{eq:dynA}
 A^{-1}(a,b)
 	= \frac12 [ \partial^2_a \d(a,b)+ \partial^2_b \d(b,a)] - \argc{ \G(a,b) + M(a,b)} + k(a) \d(a,b)
  \ ,
\eeq
with the superkernel definitions:
\beq
\label{eqC3:dyn2}
\begin{split}
 M(a,b)
 	&= \frac{\r}d\int \de \rr_0 \, e^{-\Bi v( \rr_0 )}  \moy{ \nabla v(\rr_0 + \vv(a) ) \cdot \nabla v(\rr_0 + \vv(b) ) }_\vv
 \ , \\
 k(a)
 	&= \frac{\r}d\int \de \rr_0 \, e^{-\Bi v( \rr_0  )}   \moy{ \Lap v(\rr_0+ \vv(a)) }_\vv
 \ ,
\end{split}
\eeq
and the SUSY operator $\partial^2_a$ defined in Eq.~\eqref{eq:defSUSY-part2}.
The averages ${\moy{\bullet}_\vv}$ are defined with respect to the measure
\beq
\label{eq:Pv}
 P[\vv(a)]
 \, \propto \, G_{2A}[\vv(a)] \, e^{- \int \de a\, v(\rr_0 + \vv(a))}
 \, \propto \, e^{ -\frac12 \int \de a\de b \,\vv(a) \cdot (2A)^{-1}(a,b) \vv(b) - \int \de a\, v(\rr_0 + \vv(a) )}
 \ ,
\eeq
\textit{i.e.}~the Gaussian measure~\eqref{eq-Gaussian-meas-part2} modified by the potential term for a given initial condition~$\rr_0$ 
and normalised such that ${\int \mathcal{D} \vv(a) \, P[\vv(a)]=1}$.
Note that this is the first occurrence of the effective stochastic process for the inter-particle distance in this path-integral derivation, and the measure $P[\vv]$ in Eq.~\eqref{eq:Pv} corresponds to the inter-particle dynamics and should not be confused with the Gaussian measure in Eq.~\eqref{eq-Gaussian-meas-part2}.

The extremalisation equation in its SUSY form~\eqref{eq:dynA} has actually two complementary implications.
First it allows us to rewrite more explicitly the measure~\eqref{eq:Pv}:
\beq
\label{eq:dynA-stochastic-process}
 P[\vv(a)]
 \, \propto \,
 e^{ -\frac14 \int \de a\de b \,\vv(a) \cdot \arga{ \partial^2_a \d(a,b) - \argc{ \G(a,b) + M(a,b)} + k(a) \d(a,b)} \vv(b) - \int \de a\, v(\rr_0 + \vv(a) )}
 \ ,
\eeq
which can be taken as a definition of the effective stochastic process~${\vv(a)}$.
Secondly it provides the following closure relation for the variance ${A(a,b)}$:
\beq
\label{eq:dynA-closure-relation}
\begin{split}
 \delta(a,b)
 	&= \int \de c \, A^{-1}(a,c) A(c,b)
 \\ 	
 	&= - \int \de c \, \argc{ \G(a,c) + M(a,c)} A(c,b)
 		+ k(a) \, A(a,b) 
		+ \int \de c \, \frac12 [ \partial^2_a \d(a,c)+ \partial^2_c \d(c,a)] \, A(c,b)
\end{split}
\eeq
from which one can derive the dynamical equations for the correlation and response functions that we will discuss in Sec.~\ref{sec:dynamical-equations-correlation-response}.

All that we have assumed since the beginning of this section is that the fluctuations of the SUSY field ${\uu(a)}$ can be considered
as Gaussian for the calculation of the extremal value of the free energy.
Coming back to the definition of ${A(a,b)}$ in Eq.~\eqref{eq-Gaussian-meas-part1}, its components encode both the correlation and response functions; besides, causality implies that $\moy{ \tilde \uu(t_a) \cdot \tilde \uu(t_b)}=0$, hence the SUSY correlators have no `${\ththbar{a}\ththbar{b}}$' component and
 have the form~\cite{Cu02,ABC10}:
\beq
\label{eq:SUSYnoneq}
\begin{split}
 & A(a,b)
 	= \frac{1}{d} \moy{ \uu(a) \cdot \uu(b) }
 	= C(t_a,t_b) + \th_a\bth_a R(t_b,t_a) + \th_b\bth_b R(t_a,t_b)
 \ , \\
 & C(t_a,t_b)
 	=\frac{1}{Nd}\sum_{i=1}^N\la\uu_i(t_a)\cdot\uu_i(t_b)\ra 
 \, , \quad
 R(t_a,t_b)
 =\frac{1}{Nd}\sum_{i=1}^N\la\uu_i(t_a)\cdot i \tilde\uu_i(t_b)\ra 
 	=\frac{1}{Nd}\sum_{i,\m}\left.\frac{\d\la u_{i\m}(t_a)\ra}{\d \l_{i\m}(t_b)}\right\vert_{\{\bm{\l}_i\}=\{\bm0\}} \ ,
\end{split}
\eeq
and similarly for the superkernels defined in Eq.~\eqref{eqC3:dyn2}:
\beq
\label{eq:SUSYnoneq-Mk}
 M(a,b)
 	= M_C(t_a,t_b) + \th_a\bth_a M_R(t_b,t_a) + \th_b\bth_b M_R(t_a ,t_b)
 \quad \Rightarrow \quad k(a) = k(t_a)
 \ .
\eeq
The last implication, that ${k(a)}$ is a real scalar function, derives from the following reasoning:
if both ${M(a,b)}$ and ${A(a,b)}$ do not have a ${\ththbar{a}\ththbar{b}}$ component, then neither does ${\int \de c\, M(a,c) A(c,b)}$;
moreover, ${\d(a,b)}$ has no ${\ththbar{a}\ththbar{b}}$ component either;
thus, in order to fulfill Eq.~\eqref{eq:dynA}, $k(a)$ must have no components with Grassmann variables.
Note also that for a causal superfield $A(a,b)$, the measure in Eq.~\eqref{eq:dynA-stochastic-process} is already normalised, because the partition function
of a dynamical generating functional with fixed initial condition (here ${\vv=0}$) is equal to 1.

\subsection{Single- and two-particle effective stochastic processes}
\label{sec:path-eff-stoch-process}

At last we can write down the two effective stochastic processes corresponding to the high-dimensional Gaussian approximation of the previous section
--~respectively for the inter-particle distances $\vv(a)$ and for the individual displacements $\uu(a)$~--
and see that we recover indeed the same results as via the cavity method in Sec.~\ref{sec:cavity}.

The former process is straightforward to obtain from the measure ${P[\vv(a)]}$ given in Eq.~\eqref{eq:dynA-stochastic-process}, once we write explicitly the scalar and Grassmann variables components.
It corresponds to a free particle moving in the potential ${v(\rr_0+\vv(t))}$, with a noise having a variance ${[\G_C(t,s) + M_C(t,s)]}$,
a retarded friction kernel related%
\footnote{A friction kernel in the Langevin equation is conventionally defined by a convolution with the velocity and not the position, so that by integration by parts it is related  to a time-derivative of $M_R$ (with extra boundary terms).
Incidentally, this is why at equilibrium the FDT `of the second kind'~\cite{Ku66,Cu02,Ha97,ZBCK05,MSVW13} for the pair of TTI kernels $M_{C,R}$ has the `first kind' form ${M_R(t) = -\b \th(t) \dot M_C(t)}$
with a time derivative, unlike the pair of TTI kernels $\G_{C,R}$: ${\Gamma_R(t)=\beta \theta(t) \Gamma_C(t)}$.
Yet this is at the same time natural since one can alternatively view them as (force-force) correlation and response functions over a stochastic process as in Eq.~\eqref{eqC3:Mself-Path}.
These FDT relations are discussed in Sec.~\ref{sec:limit-cases-equilibrium}.} to ${M_R(t,s)}$, and a harmonic constant ${k(t)}$,
which is exactly the same equation as the one obtained via the cavity method in Eq.~\eqref{eqC3:twop}. One can explicitly check that a SUSY path integral representation of Eq.~\eqref{eqC3:twop} gives the measure in Eq.~\eqref{eq:dynA-stochastic-process}.

The kernels are self-consistently given by the combination of Eqs.~\eqref{eqC3:dyn2} and~\eqref{eq:SUSYnoneq-Mk}:
\beq
\label{eqC3:Mself-Path}
\begin{split}
 k(t)
 	&= \frac{\r}d\int \de \rr_0 \, e^{-\Bi v( \rr_0  )}   \moy{ \Lap v(\rr_0+ \vv(t)) }_{\vv}
 \ , \\
 M_C(t,s)
 	&= \frac{\r}d\int \de \rr_0 \, e^{-\Bi v( \rr_0 )}  \moy{ \nabla v(\rr_0 + \vv(t) ) \cdot \nabla v(\rr_0 + \vv(s) ) }_{\vv}
 \ , \\
 M_R(t,s)
 	&=  \frac{\r}d\int \de \rr_0 \, e^{-\Bi v( \rr_0 )}  \moy{ \nabla v(\rr_0 + \vv(t)) \cdot \nabla \nabla^T v(\rr_0 + \vv(s) ) \cdot i \tilde \vv(s) }_{\vv}
 \ ,
\end{split}
\eeq
which are again the same as in Eq.~\eqref{eqC3:Mself}, since the definition for ${M_R(t,s)}$ here is simply the MSRDDJ translation of the linear response in Eq.~\eqref{eqC3:Mself}.

As for the effective stochastic process for the individual displacements $\uu(a)$, by definition of the Gaussian approximation~\eqref{eq-Gaussian-meas-part1}, it is also straightforward to obtain from the Gaussian measure ${G_A[\uu(a)]}$ under the extremalisation constraint~\eqref{eq:dynA} for $A$,
\beq
 P[\uu(a)]
	\propto e^{ -\frac12 \int \de a\de b \,\uu(a) \cdot \arga{ \partial^2_a \d(a,b) - \argc{ \G(a,b) + M(a,b)} + k(a) \d(a,b)} \uu(b)}
	\, ,
\eeq
which is the probability density of single-particle trajectories and corresponds to the path integral representation of the same Langevin dynamical equation as in Eq.~\eqref{eqC3:effproc}.
As for Eq.~\eqref{eq:dynA-stochastic-process}, if ${M(a,b)}$ is causal, this measure is normalised to~$1$.

This concludes our alternative derivation of the mean-field dynamics in its high-dimensional vectorial form.
Via a high-dimensional path-integral derivation, we have obtained strictly the same equations as in Sec.~\ref{sec:cavity}, also recalled in the summary section~\ref{sec:summary-results}:
the two effective stochastic processes Eqs.~\eqref{eqC3:effproc} and~\eqref{eqC3:twop},
and the self-consistent equations for the three kernels~\eqref{eqC3:Mself-Path}.
Besides, in this SUSY formulation, we have the general closure relation~\eqref{eq:dynA-closure-relation} which encodes the dynamical equations for the correlation and response functions
that we will examine in Sec.~\ref{sec:dynamical-equations-correlation-response}.
We emphasise that our derivation here relied on four ingredients:
the standard virial expansion truncated to the second order of the dynamical free energy,
a statistical translation invariance of the particle trajectories,
a Gaussian approximation for the fluctuations of the individual displacements,
and the causal structure of the supercorrelator and superkernels in Eqs.~\eqref{eq:SUSYnoneq}-\eqref{eq:SUSYnoneq-Mk}.
In fact the only assumption we are making is the Gaussian approximation, and it will be fully justified in~Sec.~\ref{secC3:largedM-path-integral}.
Otherwise, the cavity assumptions of many uncorrelated neighbours in a statistically isotropic system are properly implemented and accounted for, in this high-dimensional effective dynamics.

\section{Infinite-dimensional scaling of the effective dynamics}
\label{secC3:largedM}

In both Sec.~\ref{sec:cavity} and Sec.~\ref{sec:path}, we have obtained two effective stochastic processes for the individual displacements of particles ${\uu(t)}$ and for the inter-particle distance ${\vv(t)}$, in their high-dimensional \textit{vectorial} form. 
Now, exploiting rotational invariance under a proper high-dimensional scaling of the pairwise potential ${v(\rr)}$, we will show how we can further simplify the effective dynamics into a \textit{scalar} stochastic process, 
with its corresponding self-consistent kernels, following Refs.~\cite{MKZ16,KMZ16,Sz17}.
Note that from now on, we will explicitly write that the potential is radial, ${v(\rr)=v(|\rr|)}$.

In the limit ${d\to\io}$, typical displacements are of order ${\mathcal{O}(1/d)}$ with respect to the initial condition, and this controls the dynamics of the system.
Consistently, in this limit, in order to obtain a non-trivial result one has to scale the potential as
\beq
\label{eq:scalpotd}
 v(r) = \redv(h)
 \ , \qquad
 h = d(r/\ell -1)
 \quad \Leftrightarrow \quad
 r = \ell (1+h/d)
 \ ,
\eeq
$\ell$ being the typical interaction scale and $h$ the fluctuating gap~\cite{CKPUZ17,KMZ16}.
Under the appropriate rescaling of all quantities with respect to the dimension ${d}$ (and using $\ell$ as the unit of length), and in particular with the definition of the rescaled potential ${\redv(h)}$,
the effective stochastic process for $\vv(t)$ (see Eq.~\eqref{eqC3:twop}) can be written in terms of a single one-dimensional
variable. 
Here, we present this derivation, first in Sec.~\ref{sec:Langlarged} directly from the effective Langevin dynamics, and then in Sec.~\ref{secC3:largedM-path-integral} in the path-integral setting.
In the same spirit as in the two previous sections, the former derivation relies on a few but not fully-controlled assumptions, whereas the latter provides a proper but more technical treatment of the ${d\to\infty}$ limit.
As usual, for simplicity we set ${m=0}$ and ${\G_R=0}$ in our following derivations.

\subsection{Derivation from the effective Langevin dynamics}
\label{sec:Langlarged}

We start from the effective stochastic process for $\vv(t)$ given in Eq.~\eqref{eqC3:twop}.
The main idea is that the non-trivial fluctuations of ${\vv(t)=\rr(t)-\rr_0}$, which are assumed to be small compared to $\rr_0$, 
essentially happen along the direction given by $\rr_0$ (the `longitudinal' motion in the following):
the projection of the motion along $\rr_0$ yields an effective one-dimensional equation for a scalar ${y(t) \propto \hat{\rr}_0 \cdot \vv(t)}$, while the `transverse' motion is essentially diffusive.
This is a direct consequence of the assumption that the individual displacements of particles with respect to their initial positions are small, \textit{i.e.}~${\uu(t) = \mathcal{O}(1/d)}$, as first mentioned in Sec.~\ref{sec:cavity}, and of the scaling~\eqref{eq:scalpotd} of the potential.
From there, the fluctuating gap ${h(t)= d(r(t)/\ell -1)}$ will have contributions of both the longitudinal and transverse motions, and will essentially take the form ${h(t)=h_0 + y(t) + \Delta_r(t)}$ with the MSD ${\Delta_r(t) \propto \moy{\uu(t)^2}}$.

\subsubsection{`Longitudinal' \textit{versus} `transverse' motions}
\label{sec:Langlarged-projection}

We make a first assumption about the scaling of the equilibrium distribution
--~which derives from the thermodynamics of the system~\cite{KPZ12,KPUZ13,MK16,KMZ16}~--
namely that ${r_0=|\rr_0|}$ has the form ${r_0 = \ell (1+h_0/d)}$, with the initial gap ${h_0 = \mathcal{O}(1)}$.
Next, we can approximate, at the leading order in $\vv(t)$, the force deriving explicitly from the interaction potential:
\beq
\label{eq:approxFr0}
 \nabla v(| \rr_0 + \vv(t) | )
 	= v'(| \rr_0 + \vv(t) |) \, \frac{\rr_0 + \vv(t)}{| \rr_0 + \vv(t) |}
	\approx v'(| \rr_0 + \vv(t) |) \,  \hat \rr_0
\ ,
\eeq
where ${\hat \rr_0=\rr_0/|\rr_0|}$ is the longitudinal unit vector.
The potential term ${v'(| \rr_0 + \vv(t) |)}$ is of ${\mathcal{O}(d)}$ at all orders in perturbation, due to the scaling~\eqref{eq:scalpotd} such that ${v'(r(t))=\redv'(h(t)) \, d/\ell}$, and therefore must not be approximated perturbatively.
Eq.~\eqref{eq:approxFr0} essentially states that the force is concentrated along the longitudinal direction.
Conversely, if one projects the stochastic process of Eq.~\eqref{eqC3:twop} on the transverse directions, the force contribution is of ${\mathcal{O}(1)}$, and thus subdominant with respect to the other terms; this implies that the transverse motion is 
essentially diffusive (with colored noise and friction), and on the transverse directions the equations for $\vv(t)$ and $\uu(t)$ have
the same structure.
In addition, there is no coupling between the different transverse directions, which are therefore  mutually uncorrelated.
On the contrary, along the longitudinal motion, the force~\eqref{eq:approxFr0} is of ${\mathcal{O}(d)}$ and contributes as a nonlinear term in the projected Langevin equation, at the same order as its other terms.

So let us now focus on the longitudinal equation of motion of the stochastic process~\eqref{eqC3:twop},
and examine the force term~\eqref{eq:approxFr0}.
We need to approximate, at the leading order in $\vv(t)$, the argument in the interaction potential:
\beq
\label{eq:expnorm0}
 | \rr_0 + \vv(t) |
 = r_0 \sqrt{ 1+ 2 \frac{\hat \rr_0 \cdot \vv(t)}{r_0} + \frac{|\vv(t)|^2}{r_0^2} } 
 \approx r_0 + \hat \rr_0 \cdot \vv(t) + \frac{|\vv(t)|^2}{2r_0}
 \ ,
\eeq
and inspect specifically the averages and fluctuations of  the two dynamical variables in the latter expression. 
First, since ${|\vv(t)|^2}$ is dominated by the ${d-1}$ transverse components of $\vv(t)$
--~which are independent identically distributed (i.i.d.) variables~--
it has an average of ${\mathcal{O}(1/d)}$ with subdominant fluctuations of ${\mathcal{O}(1/d^{3/2})}$.
So we will assume that we can simply replace ${|\vv(t)|^2}$ by its dynamical average, which in turn is dominated by the rescaled MSD:
\beq
\label{eq-proj-term1}
\begin{split}
 \moy{ |\vv(t)|^2 }
	& \stackrel{\eqref{eq-def-wrt-initial-condition}}{=}
		\moy{ \uu_1(t)^2 } + \moy{ \uu_2(t)^2 }- 2 \moy{ \uu_1(t) \cdot \uu_2(t )}
	\approx 2\frac{\ell^2}{d} \D_r(t)
 \ ,\\
 & \text{with} \quad \D_r(t)
 	=\frac{d}{\ell^2} \moy{ | \xx(t) - \xx(0) |^2}
 	=\frac{d}{\ell^2} \moy{\uu(t)^2} 
 \ .
\end{split}
\eeq
The average of the scalar product ${\langle  \uu_1 \cdot \uu_2 \rangle = \langle \uu_1 \rangle \cdot \langle \uu_2 \rangle =0}$, 
because $\uu_1$ and $\uu_2$ are independent and have zero average.
Secondly, we assume that the projection ${\hat \rr_0 \cdot \vv(t) \sim \OO(1/d)}$ and we then 
define its rescaled counterpart
\beq
\label{eq-def-y-proj-longitud}
 y(t) = \frac{d}{\ell} \hat \rr_0 \cdot \vv(t)
 \ , \qquad
 y(0) =0 \ ,
\eeq
assuming that $y(t)$ remains finite in the limit $d\to\io$, which will be justified {\it a posteriori} in the following.
We then rewrite the expansion in Eq.~\eqref{eq:expnorm0} as follows:
\beq
\label{eq:rtoy}
 | \rr_0 + \vv(t) |
 	\approx \ell \argp{1+\frac{h_0}{d}} + \frac{\ell}{d} y(t) + \frac{\moy{|\vv(t)|^2}}{2\ell(1+h_0/d)}
	\approx \ell \argp{ 1+ \frac{h_0+y(t)+\D_r(t)}{d}}
 \ ,
\eeq
where the leading order is of ${\mathcal{O}(1/d)}$ and we neglect higher-order corrections.
We can finally use the potential rescaling~\eqref{eq:scalpotd} to express the force projected on $\rr_0$, from Eq.~\eqref{eq:approxFr0},
as
\beq
\label{eq:app985}
 \hat \rr_0 \cdot \nabla v(| \rr_0 + \vv(t) | )
 	\approx v'(| \rr_0 + \vv(t) |)
 	\approx \frac{d}{\ell} \redv'(h_0+y(t)+\D_r(t))
 \ ,
\eeq
which provides the leading order in the limit ${d\to\infty}$.

\subsubsection{Effective Langevin process for the longitudinal motion}
\label{sub:EffLangevin}

Projecting the effective stochastic process for $\vv(t)$ given in Eq.~\eqref{eqC3:twop} along its initial condition $\hat\rr_0$ and using the above results, we obtain the new scalar effective stochastic process valid in the high-dimensional limit for ${y(t)}$ defined in Eq.~\eqref{eq-def-y-proj-longitud}:
\beq
\label{eqC3:app56}
 \frac{\z \ell}{2d} \dot y(t)
 = - \frac{k(t)\ell}{2d}  y(t) 
	+\frac{\ell}{2d} \int_0^t \!\! \de s \, M_R(t,s) y(s)
	- \frac{d}\ell  \redv'(h_0+y(t)+\D_r(t))
	+ \frac{d}\ell\, \X(t)
\ ,
\eeq
where ${\X(t) = (\ell/d) \hat\rr_0 \cdot \bm{\X}(t)}$ is the longitudinal projection of the noise, rescaled by ${\ell/d}$ for a reason that will become immediately clear.
Its correlation is deduced from Eq.~\eqref{eqC3:twop}.
It is manifest from Eq.~\eqref{eqC3:app56} that, in order to obtain a finite result when taking the limit ${d\to\io}$, one needs to rescale the friction and noise contributions as follows~\cite{KMZ16,MKZ16}:
\beq
 \label{eqC3:fricscal}
 \wh\z = \frac{ \ell^2}{2d^2} \z 
 \ , \qquad
 \GG_C(t,s) = \frac{ \ell^2}{2d^2} \G_C(t,s)
 \ .
\eeq
Furthermore, we will see that the self-consistent kernels are naturally scaled as
\beq
\label{eqC3:fricscal2}
 \k(t) = \frac{ \ell^2}{2d^2} k(t)
 \ , \qquad
 \MM_C(t,s) = \frac{ \ell^2}{2d^2} M_C(t,s)
 \ , \qquad
 \MM_R(t,s) = \frac{ \ell^2}{2d^2} M_R(t,s)
 \ .
\eeq
With these rescalings, all terms in Eq.~\eqref{eqC3:app56} are of the same order with respect to $d$, this was actually our very goal with these specific rescalings%
\footnote{We see here that the force between the pair of particles gives a contribution of ${\mathcal{O}(1)}$ to the equation of the longitudinal motion, 
while it does not contribute at the level of the dynamical equation for each one of the particles (nor for the transverse motion), cf.~the discussion following Eq.~\eqref{eq:approxFr0}.
A similar effect occurs in the dynamics of the perceptron model~\cite{ABUZ18}.}.
Then we get
\beq
\label{eqC3:effproch}
\begin{split}
 & \wh\z \dot y(t)
 	=	- \k(t) y(t)
 		+ \int_0^t \!\! \de s \, \MM_R(t,s) y(s)
 		- \redv'(h_0+y(t)+\D_r(t)) +  \X(t)
 \ , \\
 & \moy{ \X(t)}_{\X}=0
 	\ , \quad
 	\moy{ \X(t) \X(s)}_{\X}
 	=  2 T \wh\z \d(t-s) + \GG_C(t,s) + \MM_C(t,s)
 \ , \\
\end{split}
\eeq
recalling the initial condition ${y(0)=0}$, by its definition~\eqref{eq-def-y-proj-longitud}.
The dynamics of the variable $y(t)$ is then  governed by Eq.~\eqref{eqC3:effproch}, which shows that $y(t)$ remains finite for ${d\to\io}$, as initially assumed.
We still need to specify the initial distribution of $h_0$ and the kernels ${\lbrace \kappa(t), \MM_C(t,s), \MM_R(t,s) \rbrace}$, from the initial distribution of $\rr_0$ and the kernels expressions~\eqref{eqC3:Mself},
as well as the MSD~${\Delta_r(t)}$.

\subsubsection{Initial condition and self-consistent equations for the kernels}

The longitudinal motion and the kernels themselves turn out to depend on the fluctuating quantity
\beq
 h(t) = h_0+y(t)+\D_r(t)
\ .
\eeq
According to the potential rescaling ${v(r) = \redv(h)}$ of Eq.~\eqref{eq:scalpotd}, the perturbative expansion~\eqref{eq:rtoy} of a representative inter-particle distance becomes
\beq
 r(t) = |\rr(t)| = | \rr_0 + \vv(t) |
	\approx \ell \argp{ 1+ \frac{h_0+y(t)+\D_r(t)}{d}}
	= \ell \argp{ 1+ \frac{h(t)}{d}}
 \ ,
\eeq
which allows one to identify $h(t)$ with the rescaled inter-particle gap.
Introducing $\Omega_d$ and $V_d$ as respectively the solid angle and the volume of the unit radius sphere in dimension $d$, with ${V_d=\Omega_d/d=\p^{d/2}/\G(d/2+1)}$,
we can rewrite the measure for the vectorial initial condition, in high dimension and under rotational invariance, as
\beq
\label{eq-def-measure-rr0-h0}
 \de\rr_0
	=\Omega_d \,\de r_0\,r_0^{d-1}
	= \Omega_d \, \de h_0 \frac{\ell^d}{d} \argp{1 + \frac{h_0}{d}}^{d-1}
	\approx V_d \,\ell^d\,\de h_0\, e^{h_0}
 \ ,
\eeq
where the last approximation is the exponential limit formula.
Besides we define the bare and rescaled packing fractions, respectively $\f$ and $\wh\f$, as
\beq
\label{eq:pf}
 \f
 	=\r V_d\left(\frac\ell2\right)^d=\frac{d}{2^d}\wh\f 
 \hskip15pt\Leftrightarrow\hskip15pt
 \wh\f
 	=\frac{\r V_d\ell^d}{d}
 \ ,
\eeq
which is the usual density scaling for  liquids and glasses in high dimension~\cite{PZ10,CKPUZ17}.
Gathering all the last expressions, 
we have that, in the high-dimensional limit of the three kernels~\eqref{eqC3:Mself}, one can replace:
\beq
\label{eq-high-dim-averages-gap}
 \frac{\r}{d} \int \de \rr_0 \, e^{-\beta_0 v(|\rr_0|)}  \to  \wh\f \int_{-\infty}^{\infty} \!\! \de h_0 e^{h_0 - \beta_0 \redv (h_0)} 
  \ ,
\qquad
 v'(|\rr_0 + \vv(t)|) \to \redv'(h(t))\, d/\ell \ ,
 \qquad
\moy{\bullet}_\vv \to \moy{\bullet}_h \ .
\eeq
With these replacements, the kernels~\eqref{eqC3:Mself} and more specifically their rescaled counterparts ${\lbrace \kappa(t), \MM_C(t,s), \MM_R(t,s) \rbrace}$ defined by Eq.~\eqref{eqC3:fricscal2} have more explicit expressions in the high-dimensional limit.
First, using the expression of the Laplacian for a rotationally invariant function, we have for the average of the local force divergence:
\beq
\label{eqC3:divergence-forces-self}
\begin{split}
 \k(t)
 & \stackrel{\eqref{eqC3:Mself}}{=}
 	\frac{\r\ell^2}{2d^3} \int \de\rr_0 \, e^{-\Bi v(|\rr_0|)} \moy{ \Lap v(| \rr_0 + \vv(t)|)}_{\vv}
 = \frac{\r\ell^2}{2d^3} \int \de\rr_0 \, e^{-\Bi v(|\rr_0|)} \moy{ v''(r(t)) + \frac{d-1}{r(t)} v'(r(t)) }_\vv
 \\
 & \stackrel{\eqref{eq-high-dim-averages-gap}}{\approx}
 	\frac{ \wh\f}{2}  \int_{-\io}^\io \!\! \de h_0\, e^{h_0 -\Bi \redv(h_0)} \moy{ \redv''(h(t)) + \redv'(h(t))}_h
 \ ,
\end{split}
\eeq
where by self-consistency we sticked again to the leading order in $d$.
Secondly, under the assumption that ${\hat\rr(t)\approx \hat\rr_0}$ is constant at leading order in $d$ for every pair of particles
--~the very same assumption we used to focus on the longitudinal motion ~-- so is the angular direction of the interaction force ${\nabla v (\rr(t))}$, which is parallel to $\hat\rr(t)$.
This implies for the force-force correlation, encoded in the rescaled memory kernel~\eqref{eq:app985}, that
\beq
\label{eqC3:MMself}
 \MM_C(t,s)
 	\stackrel{\eqref{eqC3:Mself}}{=}
		\frac{\r\ell^2}{2d^3} \int \de \rr_0 \, e^{-\Bi v(|\rr_0|)} \moy{ \nabla v(|\rr_0+\vv(t)|)  \cdot \nabla v(|\rr_0+\vv(s)|)}_\vv 
 	\stackrel{\eqref{eq-high-dim-averages-gap}}{\approx}
 		\frac{ \wh\f}{2}  \int_{-\io}^\io \!\! \de h_0\, e^{h_0 -\Bi \redv(h_0)} \moy{ \redv'(h(t))  \redv'(h(s))}_h
\ .
\eeq
Similarly, for the response kernel in Eq.~\eqref{eqC3:Mself}, the average over all directions ${\m=1, \dots,d}$ 
is dominated by the longitudinal direction, 
so we can choose ${\bm{P}(t) = \hat\rr_0 \, P(t)}$ and introduce the rescaled perturbation ${P(t) =  (\ell/d) \PP(t)}$:
\beq
\label{eqC3:MMself-response}
\begin{split}
\MM_R(t,s)
 & \stackrel{\eqref{eqC3:Mself}}{=}
 	\frac{\r\ell^2}{2d^3} \int \de \rr_0 \, e^{-\Bi v(|\rr_0|)} \left.\frac{\d \moy{ \hat\rr_0 \cdot \nabla v(|\rr_0+\vv(t)|}_{\vv,\bm{P}}}{\d P(s)} \right\vert_{P=0}
 \\
 & \stackrel{\eqref{eq:app985}}{=}
 	 \frac{\r\ell^2}{2d^3} \int \de \rr_0 \, e^{-\Bi v(|\rr_0|)} \left.\frac{\d \moy{v'(|\rr_0+\vv(t)|}_{\vv,\bm{P}}}{\d P(s)} \right\vert_{P=0}
 \stackrel{\eqref{eq-high-dim-averages-gap}}{\approx}
 	 \frac{ \wh\f}{2}  \int_{-\io}^\io \!\! \de h_0\, e^{h_0 -\Bi \redv(h_0)}\left. \frac{\d \moy{ \redv'[h(t)] }_{h,\PP}}{\d \PP(s)} \right\vert_{\PP=0}
\ ,
\end{split}
\eeq
where the field $\PP(t)$ amounts to shifting ${\redv'(h(t))  \to \redv'(h(t) - \PP(t))}$ in the dynamical Eq.~\eqref{eqC3:effproch}.

\subsubsection{Effective stochastic process for the gaps}
\label{sec:effprocgap}

So far in this section, we first have argued that in high dimension the only non-trivial motion of each pair of particles is longitudinal to its initial condition, hence we have derived the effective stochastic process for the rescaled projection ${y(t) = (d/\ell) \hat{\rr}_0 \cdot \vv(t)}$ given in Eq.~\eqref{eqC3:effproch},
and rescaled all the parameters and quantities with respect to ${d/\ell}$.
However, we have just shown that the rescaled kernels 
${\lbrace \kappa(t), \MM_C(t,s), \MM_R(t,s) \rbrace}$ defined by Eq.~\eqref{eqC3:fricscal2}
are more naturally defined as dynamical averages over the fluctuating gap ${ h(t) = h_0+y(t)+\D_r(t)}$.
So we conclude our derivation by rewriting the scalar effective dynamics in an alternative form, directly for the gap ${h(t)}$:
\beq
\label{eq:yeff}
\begin{split}
\wh\z \dot h(t) 
	&= \BB_{\text{MSD}}(t)
		- \k(t) \argp{h(t)-h_0}
		+ \int_{0}^t \!\! \de s\, \MM_R(t,s)  \argp{h(s)-h_0}
		- \redv'(h(t)) + \Xi(t) 
 \\
 h(0)
 	&=h_0 
 \ , \hskip15pt
 \langle \Xi(t) \Xi(s) \rangle  = 2 T \wh \z \d(t-s)+\GG_C(t,s) + \MM_C(t,s)
 \ , \\
  \BB_{\text{MSD}}(t)
 	&:=\wh\z \dot \D_r(t) + \k(t) \D_r(t) - \int_{0}^t \de s\, \MM_R(t,s)\D_r(s)
 \ ,
\end{split}
\eeq
and for completeness we gather here the three self-consistent definition of the kernels given by Eqs.~\eqref{eqC3:divergence-forces-self}-\eqref{eqC3:MMself}-\eqref {eqC3:MMself-response}:
\beq
\label{eq:yeff-kernels-trio}
\begin{split}
 \k(t)
 	&= \frac{\wh \f}2 \int^{\infty}_{-\infty} \!\! \de h_0 \, e^{h_0 -\Bi \redv( h_0  )}   \la \redv''(h(t)) + \redv'(h(t)) \ra_{h}
 \ , \\
 \MM_C(t,s)
 	&=  \frac{\wh\f}2 \int^{\infty}_{-\infty} \!\! \de h_0 \, e^{h_0 -\Bi  \redv(h_0)}  
 		\langle \redv'(h(t)) \redv'(h(s)) \rangle_{h} 
 \ , \\
 \MM_R(t,s)
 	&=  \frac{\wh\f}2 \int^{\infty}_{-\infty} \!\! \de h_0 \, e^{h_0 -\Bi  \redv(h_0)}  
		\left. \frac{\d \langle \redv'(h(t))  \rangle_{h,\PP}}{\d \PP(s)}\right\vert_{\PP=0}
 \ .
\end{split}
\eeq
In order to reconnect these equations to the initial setting of Sec.~\ref{sec:setting}, we recall that one should use the rescalings in Eqs.~\eqref{eqC3:fricscal}-\eqref{eqC3:fricscal2}.

The quantity that still needs to be elucidated is the rescaled MSD ${\Delta_r(t)}$,  defined in Eq.~\eqref{eq-proj-term1} as ${\D_r(t)=\frac{d}{\ell^2} \moy{ | \xx(t) - \xx(0) |^2}=\frac{d}{\ell^2} \moy{\uu(t)^2}}$.
Physically, it first appeared in Eqs.~\eqref{eq:expnorm0} and~\eqref{eq:rtoy} as the contribution to the gap fluctuations due to the transverse motion, in the high-dimensional limit.
But for now we want to emphasise that, in the stochastic process for the longitudinal motion, ${\Delta_r(t)}$ only intervenes in the rescaled force ${\redv(h_0+y(t)+\Delta_r(t))}$, whereas for the gap itself it adds a whole new additive term $\BB_{\text{MSD}}(t)$.
We will see in Sec.~\ref{sec:limit-cases-equilibrium} that at equilibrium we have simply ${\BB_{\text{MSD}}(t)=T}$, but out of equilibrium it is a complicated term unknown beforehand, 
encoding a time-dependent tilt of the effective potential  `seen' by degree of freedom~${h(t)}$.
So, although in high dimension we can nicely decompose the vectorial fluctuations of $\vv(t)$ into the longitudinal and transverse motions, their respective contributions to the gap $h(t)$ and thus to the kernels are nonlinearly intertwined.
We will actually be able to characterise $\Delta_r(t)$ further only in Sec.~\ref{sec:dynamical-equations-correlation-response-section3-high-dim}.

\subsection{Derivation via the Euclidean path integral}
\label{secC3:largedM-path-integral}

In the previous section, our starting point was the vectorial effective stochastic process for the inter-particle distance fluctuations ${\vv(t)=\rr(t)-\rr_0}$, given in Eq.~\eqref{eqC3:twop}, that we had consistently derived either by the dynamical cavity method in Sec.~\ref{sec:cavity}, either by the SUSY path-integral approach in Sec.~\ref{sec:path}.
Our main assumption was a continuation of the one invoked at the beginning of the cavity derivation, namely that the individual displacements of particles remain of ${\mathcal{O}(1/d)}$ with respect of their initial position, so that the fluctuations of ${\vv(t)}$ happen mostly in the `longitudinal' direction of $\rr_0$.

Here we present an alternative derivation of the same results, based on a SUSY path-integral approach, exploiting again the high-dimensional scaling~\eqref{eq:scalpotd} of the interaction potential ${v(r)=\bar{v}(h)}$ and the other parameters and quantities of the model via Eqs.~\eqref{eqC3:fricscal}-\eqref{eqC3:fricscal2}.
This will essentially allow us to get rid of the Gaussian ansatz used in Sec.~\ref{sec:path-Gaussian-ansatz} for the fluctuations of the superfield ${\uu(a)}$,
and to derive the decomposition into longitudinal and transverse motion presented in Sec.~\ref{sec:Langlarged-projection} directly as a feature of the ${d \to \infty}$ saddle point.

\subsubsection{High-dimensional effective dynamical action}
\label{secC3:largedM-path-integral-effective-action}

We start just one step before the Gaussian ansatz in Eq.~\eqref{eq-Gaussian-meas-part1}, and consider the free-energy density ${\ff = \ff^{\text{id}} + \ff^{\text{ex}}}$ given in Eq.~\eqref{eq:freePISUSY}.
Instead of assuming that the fluctuations of $\uu(a)$ are Gaussian with zero mean and a variance ${\moy{\uu(a) \cdot \uu(b)} = d \, A(a,b)}$, we will first express everything in terms of the scalar product ${\uu(a)\cdot \uu(b)}$ and show that in the forthcoming saddle point it does not fluctuate, and thus gets concentrated on its average value.
In the same spirit as the rescalings of the parameters and kernels in Eqs.~\eqref{eqC3:fricscal}-\eqref{eqC3:fricscal2}, we introduce the rescaled superkernel
\beq
\label{eq:defalpha}
 \a(a,b)
 	= \frac{d}{\ell^2} \moy{ q(a,b) }
 	= \frac{d^2}{\ell^2} \, A(a,b)
 \ , \qquad
 \text{with} \quad q(a,b)= \uu(a) \cdot \uu(b)
 \, ,
\eeq
and we will see later that ${\a(a,b)}$ remains finite for ${d\to\io}$,
with in particular ${\langle (\uu(a)-\uu(b))^2 \rangle \propto \alpha(a,a)+\alpha(b,b) - 2 \alpha(a,b)}$.

The `ideal gas' term is independent of the initial condition --~as emphasised by the trajectory probability ${\Pi[\uu,\hat{\uu}]=\rho[\xx,\hat{\xx}]/\rho}$, normalised according to Eq.~\eqref{eq-def-normalisation-Pi-bis}~--
and it can be treated as in Ref.~\cite[App. II.C]{MKZ16}.
We briefly recall the argument.
One can use the isotropy to write ${\ff^{\rm id}}$ in terms of ${q(a,b)}$.
The kinetic term given by Eqs.~\eqref{eq:defSUSY-part1}-\eqref{eq:defSUSY-part2} is readily expressed as ${\Phi[\uu(a)]=\Phi[q(a,b)]}$, while the density of trajectories is naturally scaled as ${\Pi[\uu(a)]=\Pi[q(a,b)]=e^{d \, \pi[q(a,b)]}}$ in order to counterbalance the Jacobian $\exp\left[\frac d2\log\det q\right]$ coming from the change of variables ${\uu(a)\leftrightarrow q(a,b)}$~\cite{MKZ16,KMZ16}.
Indeed, one has
\begin{equation}
\begin{split}
 1
 	= \int \mathcal{D} \Pi[\uu(a)]
 	&= \int \mathcal{D}\uu \int \mathcal{D} q \, \d[q(a,b)-\uu(a)\cdot \uu(b)] \,\Pi[q(a,b)]
 \propto \int \mathcal{D} q(a,b) \,e^{d \, \pi[q(a,b)]+\frac d2\log\det q}
 \ .
 \end{split}
\end{equation}
Because of the overall factor $d$ in the exponential argument of the normalisation, 
when ${d\to\io}$ the integration over ${q(a,b)}$ gets concentrated on its saddle point $q^*(a,b)$, 
which, as we will see below, turns out to be ${q^*(a,b) = {\moy{\uu(a) \cdot \uu(b)}=(\ell^2/d)\a(a,b)}}$.
Note that the scaling ${\Pi[q(a,b)]=e^{d \, \pi[q(a,b)]}}$ guarantees that the exponential argument satisfies 
the relation ${d \, \pi[\a \ell^2/d]+\frac d2 \log\det \a=0}$ at leading order ${\mathcal{O}(d)}$, 
consistently with the normalisation condition ${\int \mathcal{D} \Pi[\uu(a)]=1}$.
Consequently, the `ideal gas' free-energy density in Eq.~\eqref{eq:freePISUSY} is given, up to irrelevant constants (independent of ${\a(a,b)}$), by its evaluation at the saddle-point value in high dimension:
\begin{equation}
 \ff^{\rm id}
 =-\left(\Phi[\a(a,b)]+d \, \pi[\a(a,b)\ell^2/d]\right)=-\Phi[\a(a,b)]+\frac d2 \log\det \a
 \ .
\end{equation}
Keeping track of the factors in the definitions~\eqref{eq:defSUSY-part1}-\eqref{eq:defSUSY-part2} and~\eqref{eq:defalpha}, we obtain
\beq
\label{eq:ffidinf}
 \frac{\ff^{\rm id}}{d}
	= - \int \de a \de b\, [\wh \partial^2_a \d(a,b) - \GG(a,b)] \a (a,b) + \frac12 \log\det\a
 \, , \quad
 \text{with} \quad \wh \partial^2_a = \frac{d^2}{2 \ell^2} \partial^2_a
\eeq
where the operator ${\wh \partial^2_a}$ corresponds to the original $\partial_a^2$, see Eq.~\eqref{eq:defSUSY-part2}, 
with $\wh\z$ and $\GG$ rescaled as in Eq.~\eqref{eqC3:fricscal}.
Note that, as in Sec.~\ref{sub:EffLangevin}, these rescalings in $d$ have been chosen in such a way that both contributions, the kinetic term and the Jacobian ${\propto \log\det\a}$, scale in the same way with $d$.
Moreover, we emphasise that the exact result of Eq.~\eqref{eq:ffidinf} gives the same expression for $\ff^{\rm id}$ as the Gaussian approximation in Eq.~\eqref{eq:Gaussian_id}~\cite{KPZ12,KPUZ13}.

Next we can focus on the `excess' free-energy density, \textit{i.e.}~the second term in Eq.~\eqref{eq:freePISUSY} which is due to the interaction.
Using explicitly the definition of the rescaled potential ${v(|\rr|)=\redv (d (r/\ell -1))}$ given in Eq.~\eqref{eq:scalpotd} ,
it can be rewritten as
\beq
\label{eqC2:fcm33}
\ff^{\text{ex}}
	=\frac\r2 \int \de \rr_0  \argc{ e^{-\Bi \redv( d ( r_0/\ell - 1 ))} \moy{ e^{-\int \de a \redv( d ( |\rr_0 + \vv(a)|/\ell - 1 )}} - 1}
\ ,
\eeq
where we have defined as usual ${\vv(a) = \uu_1(a) - \uu_2(a)}$, and both $\uu_1(a), \uu_2(a)$ are independently distributed with the probability density ${\Pi[\uu(a)]}$.
We recall from Sec.~\ref{sec:path-MSRDJ-SUSY-formalism-part1} that the bare ${\moy{\bullet}}$ denotes the statistical average corresponding to the dynamical partition function, \textit{i.e.}~over the trajectory ${\vv(a)}$.
From the rotational invariance of the distribution ${\Pi[\uu(a)]}$ and from the definition of ${\alpha(a,b)}$ in Eq.~\eqref{eq:defalpha}, we moreover have that
\beq
\label{eq:app399}
 \moy{w_\m(a) w_\n(b)}
 	= \moy{ u_{1\mu}(a) u_{1\nu}(b)} + \moy{u_{2\mu}(a) u_{2\nu}(b)}
 	= \d_{\m\n} \, 2\a(a,b) \frac{\ell^2}{d^2}
 \ .
\eeq
The latter result implies the independence of the variables ${(w_\m,w_\n)}$ describing two different directions ${\m\neq\n}$. 
In Eq.~\eqref{eqC2:fcm33}, the argument of the rescaled potential depends on the random variable
\beq
\label{eq:app400}
 |\rr_0 + \vv(a) |
 	= \sqrt{r_0^2 + 2 \, \rr_0 \cdot \vv(a) + |\vv(a)|^2}
 \ ,
\eeq
whose three components are actually fluctuating independently, and we want to identify the dominant contributions which survives in the infinite-dimensional limit.
First, the statistics of $r_0^2$ is known from the initial stochastic condition (assumed to be at equilibrium, but we need it only to be statistically isotropic), and is independent of the distribution of $\vv(a)$.
It is thus independent of ${|\vv(a)|^2}$ and also of ${\rr_0 \cdot \vv(a)}$, since by rotational invariance of $\rr_0$ we have 
${\moy{ r_0^2 \rr_0 }=\bm 0}$ (here the average is over the initial condition). 
For the same reason, we have that ${\moy{|\vv(a)|^2 \vv(b)} =0}$, and so ${|\vv(a)|^2}$ and ${\rr_0 \cdot \vv(a)}$ are independent.
Secondly, as a direct consequence of Eq.~\eqref{eq:app399}, the term ${|\vv(a)|^2 = \sum_{\m=1}^d |w_\m(a)|^2}$ is the sum of $d$ i.i.d. terms, each having a finite average of order ${1/d^2}$, therefore the average of ${|\vv(a)|^2}$ is of order $1/d$ and its fluctuations are of order $1/d^{3/2}$ (and can be neglected).
Thirdly, the term ${\rr_0 \cdot \vv(a)  = \sum_{\m=1}^d r_{0\m} w_\m(a)}$ is as well a sum of $d$ i.i.d. terms (by isotropy) which
has zero average.
Its fluctuations are of order ${1/d}$, and as such they cannot be neglected;
due to the central limit theorem in the limit ${d \to \infty}$, these fluctuations are in fact Gaussian with the variance
\beq
\label{eqC2:zvar}
\moy{ |\rr_0 \cdot \vv(a) |^2  }
	=  \sum_{\m\n} r_{0\m} r_{0\n} \moy{ w_\m(a) w_\n(b) }
	\stackrel{\eqref{eq:app399} }{=} r_0^2 \, 2 \a(a,b) \frac{\ell^2}{d^2}
 \ . 
\eeq
So we can naturally define the Gaussian superfield ${Y(a)=\rr_0 \cdot \vv(a)}$ with its mean and variance
\beq
 \moy{Y(a)}=0
 \, , \quad
 \moy{Y(a)Y(b)}= r_0^2 \, 2 \a(a,b) \ell^2 / d^2
 \ .
\eeq
Fourthly, collecting these results and defining the change of variable ${r_0 = \ell (1+ h_0/d)}$,
we can write at leading order in ${1/d}$:
\beq
\label{eq-SUSY-superfield-y-h-part1}
 |\rr_0 + \vv(a) |
 	\approx \sqrt{ r_0^2 + 2Y(a) + \moy{ |\vv(a)|^2 }}
	\approx \ell (1 + h_0/d)  + Y(a) + \a(a,a) \, \ell^2/d
 \ .
\eeq
Equivalently, introducing the rescaled Gaussian superfield:
\beq
\label{eq-SUSY-superfield-y-h-part2}
 y(a) = \frac{d}{\ell} \frac{Y(a)}{r_0} = \frac{d}{\ell} \hat{\rr}_0 \cdot \vv(a)
 \quad \Rightarrow \quad
 \moy{y(a)}=0
 \, , \quad
 \moy{y(a) y(b)} = 2 \a (a,b)
\eeq
we have the following decomposition of the fluctuating `super'-gap ${h(a)}$ which appears in the exponential in Eq.~\eqref{eqC2:fcm33}:
\beq
\label{eq-SUSY-superfield-y-h-part3}
 h(a) = d \left( \frac{|\rr_0 + \vv(a)|}{\ell} - 1 \right)
 	\approx h_0 + \a(a,a) + y(a)
 \ .
\eeq
In addition, the measure for the initial condition  ${\rho \int \de \rr_0 \, e^{- \Bi v(\rr_0)}}$ can be replaced by ${d \, \wh\f   \int \de h_0 \, e^{h_0 - \Bi \redv(h_0)}}$ exactly as in Eq.~\eqref{eq-high-dim-averages-gap}.
Therefore the interaction term of Eq.~\eqref{eqC2:fcm33} can be rewritten as:
\beq
\label{eq:ffexinf-final}
\begin{split}
 \frac{\ff^{\rm ex}}{d}
 	&= \frac{\wh\f}2 \int \de h_0 \, e^{h_0} \,
	\left\{ e^{-\b_0 \redv( h_0  )} \moy{e^{-\int \de a\, \redv( h_0 + \a(a,a) + y(a) )}} - 1 \right\}
 \ .
\end{split}
\eeq
Note that by definition, we have as initial condition ${y(t=0)=0}$ and ${h(t=0)=h_0}$;
we also have that ${\moy{y(t) y(s)} = 2 \a(t,s)= (2d/\ell^2) \moy{ \uu(t) \cdot \uu(s)}}$, hence ${\moy{y(t)^2}=2 \Delta_r(t)}$ and in particular ${\moy{y(0)^2}=0}$.

We emphasise that Eqs.~\eqref{eq-SUSY-superfield-y-h-part1}-\eqref{eq-SUSY-superfield-y-h-part2}-\eqref{eq-SUSY-superfield-y-h-part3} are simply the SUSY counterpart of the argument presented in Sec.~\ref{sec:Langlarged-projection}, and in particular of Eqs.~\eqref{eq:expnorm0}-\eqref{eq-def-y-proj-longitud}-\eqref{eq:rtoy}.
Here the decomposition of the gap ${h(a)}$ into three contributions
--~its initial condition $h_0$, its transverse motion contribution ${\a (a,a)}$ and its longitudinal motion contribution ${y(a) \propto \hat{\rr}_0 \cdot \vv(a)}$, with ${y(a)}$ a Gaussian stochastic process~--
is an exact feature in ${d \to \infty}$.
In other words, the effective Gaussianity of ${\Pi[\uu(a)]}$ is not an assumption here, but a direct consequence of the infinite-dimensional limit and rotational invariance, and as such it fully justifies \textit{a posteriori} our Gaussian ansatz in Sec.~\ref{sec:path-Gaussian-ansatz}.

Collecting the results for $\ff^{\rm id}$ and $\ff^{\rm ex}$ of Eqs.~\eqref{eq:ffidinf} and~\eqref {eq:ffexinf-final} we finally obtain for the total free-energy density at its infinite-dimensional saddle point:
\beq
\label{eq:fftotalinf-final}
 \frac{\ff}{d} 
 	= \frac12 \log\det\a - \int \de a \de b\, [\wh \partial^2_a \d(a,b) - \GG(a,b)] \a(a,b)
+ \frac{\wh\f}2 \int \de h_0 \, e^{h_0} \,
\arga{e^{-\b_0 \redv( h_0  )} \moy{ e^{-\int \de a\, \redv( h_0 + \a(a,a) + y(a) )}} - 1 }
 \ .
\eeq

\subsubsection{Dynamical equations in supersymmetric form}
\label{secC3:largedM-path-integral-dynam-equations}

From the explicit high-dimensional free-energy density given in Eq.~\eqref{eq:fftotalinf-final},
the remaining steps are exactly the same as in Sec.~\ref{sec:path-Gaussian-ansatz}.
The functional derivative with respect to ${\a(a,b)}$ provides the self-consistent equation
\beq
\label{eqC3:dyn1}
 (2\a)^{-1}(a,b)
 	-\frac{\wh \partial^2_a \d(a,b)
 	+\wh \partial^2_b \d(b,a)}2
 	+\GG(a,b)+ \MM(a,b) - \k(a) \d(a,b)
 	=0
 \ ,
\eeq
with the superkernels
\beq
\label{eqC3:dyn2bis}
\begin{split}
 \MM(a,b)
	&= \frac{\wh \f}2 \int_{-\infty}^{\infty} \!\! \de h_0 \, e^{h_0 -\Bi \redv( h_0  )}  \moy{ \redv'(h_0 + \a(a,a) + y(a) ) \redv'(h_0 + \a(b,b) + y(b)) }_y
 \ , \\
 \k(a)
 	&= \frac{\wh \f}2 \int_{-\infty}^{\infty} \!\! \de h_0 \, e^{h_0 -\Bi \redv( h_0  )}   \moy{ \redv''(h_0 + \a(a,a) + y(a) ) + \redv'(h_0 + \a(a,a) + y(a))}_y	
 \ ,
\end{split}
\eeq
where we have defined the effective (non-Gaussian) stochastic process by including the rescaled potential into the dynamical average, exactly as we had done in Sec.~\ref{sec:path-Gaussian-ansatz} and specifically Eq.~\eqref{eq:Pv}:
\beq
\label{eq:defGauss}
 \moy{ \bullet }_y
 	=\moy{ \bullet \,  e^{-\int \de a\, \redv( h_0 + \a(a,a) + y(a) )}}
 \ ,
\end{equation}
normalised to 1 (\textit{i.e.}~${\moy{ 1}=1}$) by probability conservation, since we sum over all the possible trajectories.
Equivalently we can define the effective stochastic process on the gaps, for a given initial condition~$ h_0$, and from there update the expressions for the kernels:
\beq
\label{eq:defGauss-gaps}
 \moy{ \bullet }_h
 	=\moy{ \bullet \,  e^{-\int \de a\, \redv( h(a) )}}
 \ , \quad
 \left\lbrace \begin{array}{l}
	\MM(a,b)
	= \frac{\wh \f}2 \int_{-\infty}^{\infty} \!\! \de h_0 \, e^{h_0 -\Bi \redv( h_0  )}  \moy{ \redv'(h(a) ) \redv'(h(b)) }_h
	\ , \\ \\
	\k(a)
	= \frac{\wh \f}2 \int_{-\infty}^{\infty} \!\! \de h_0 \, e^{h_0 -\Bi \redv( h_0  )}   \moy{ \redv''(h(a) ) + \redv'(h(a))}_h
	\ .
 \end{array} \right.
\eeq
Of course, the saddle-point equation~\eqref{eqC3:dyn1} is the same as the one obtained with our Gaussian approximation in Eq.~\eqref{eq:dynA}, as it can be self-consistently checked by using the different rescalings~\eqref{eqC3:fricscal}-\eqref{eqC3:fricscal2} and~\eqref{eq:defalpha}.
Consequently, it also provides a closure relation for the variance ${\alpha(a,b)}$, the rescaled counterpart of Eq.~\eqref{eq:dynA-closure-relation}, which we will not write explicitly.
We emphasise that the bare brackets ${\moy{\bullet}}$ denote the `true' dynamical average for which ${y(a)}$ and ${h(a)}$ are Gaussian random variables in the infinite-dimensional limit, whereas $\moy{ \bullet }_y$ and $\moy{ \bullet }_h$ denote the dynamical average over two effective non-Gaussian stochastic processes.
In particular, the relations derived in section~\ref{secC3:largedM-path-integral-effective-action} for the statistical properties of $y$, which hold on the measure $\moy{\bullet}$, do not hold for the measure $\moy{\bullet}_y$.
For example, ${\moy{y^2(t)}_y \neq 2\D_r(t)}$ while ${\moy{y^2(t)} = 2\D_r(t)}$.

Explicitly, using Eq.~\eqref{eqC3:dyn1} we can substitute ${(2\a)^{-1}(a,b)}$ with its expression in terms of derivatives and memory functions in the Gaussian measure for $y$,
and therefore the stochastic process of $y$ becomes a standard Langevin process with white plus colored noise with memory ${\MM(a,b)}$,
and an additional spring constant ${\k(a)}$.
Then, plugging this effective dynamics in Eq.~\eqref{eqC3:dyn2bis} gives a closed equation for the superkernels ${\MM(a,b)}$ and ${\k(a)}$.

At last, coming back to the definition of ${\alpha(a,b)}$ in Eq.~\eqref{eq:defalpha}, we can distinguish its scalar and the Grassmann variables components and recognise them as encoding both the rescaled correlation and response functions.
As in Sec.~\ref{sec:path-Gaussian-ansatz}, causality implies that the SUSY correlators have no `${\ththbar{a}\ththbar{b}}$' component, hence they have the form
\beq
\label{eq:SUSYnoneq-bis}
\begin{split}
 \a (a,b)
 	&= \CC(t_a,t_b) + \th_a\bth_a \cR(t_b,t_a) + \th_b\bth_b \cR(t_a,t_b)
 \ , \\
 \MM(a,b) &= \MM_C(t_a,t_b) + \ththbar{a} \, \MM_R(t_b,t_a) + \ththbar{b} \, \MM_R(t_a ,t_b)
 \ , \\
 {\k}(a) &=\k(t)
 \ ,
\end{split}
\eeq
where we have defined the rescaled correlation and response:
\begin{equation}
\label{eq:CRscal}
 \CC(t,s)
 	=\frac{d^2}{\ell^2} C(t,s)
 	=\frac{d}{\ell^2N}\sum_{i=1}^N\ \moy{ \uu_i(t)\cdot\uu_i(s) }
 \ , \quad 
 \cR(t,s)
 	=\frac{d^2}{\ell^2} R(t,s)
 	=\frac{d}{\ell^2N}\sum_{i,\m}\left.\frac{\d\la u_{i\m}(t)\ra}{\d \l_{i\m}(s)}\right\vert_{\{\bm{\l}_i\}
 	=\{\bm0\}}
\end{equation}
consistently with their non-rescaled counterparts in Eq.~\eqref{eq:SUSYnoneq}.

\subsubsection{Effective stochastic process for the longitudinal motion}
\label{secC3:largedM-path-integral-eff-stoch-process}

Unraveling the effective stochastic process for ${y(t)}$ from the SUSY formulation in Eq.~\eqref{eq:defGauss},
we obtain a process with memory ${\lbrace \MM_C, \MM_R \rbrace }$, a spring constant $\k(t)$ and a potential ${\redv(h_0 + y(t) + \CC(t,t))}$.
The rescaled correlation at equal times ${\CC(t,t)}$ is nothing but the rescaled MSD 
first defined in  Eq.~\eqref{eq-proj-term1}, ${\CC(t,t) = \Delta_r(t)=\moy{\uu(t)^2} d/\ell^2}$.
The explicit dynamics for ${y(t)}$ is therefore given by the effective process~\eqref{eqC3:effproch}, first derived in Sec.~\ref{sec:effprocgap},
and conversely one can check that the SUSY path-integral representation of Eq.~\eqref{eqC3:effproch} precisely coincides
with $\moy{\bullet}_y$ as defined in Eq.~\eqref{eq:defGauss}.
The corresponding process for ${h(t)}$ is given in Eq.~\eqref{eq:yeff}.
From Eq.~\eqref{eq:defGauss-gaps} one can extract the equations for the memory kernels,
which also coincide with the results of Sec.~\ref{sec:Langlarged}, Eq.~\eqref{eq:yeff-kernels-trio}.

This concludes our alternative derivation of the infinite-dimensional limit of the effective dynamics obtained in Sec.~\ref{sec:cavity} and Sec.~\ref{sec:path}.
The assumptions that we had to make
--~namely the Gaussian ansatz in Sec.~\ref{sec:path-Gaussian-ansatz} and the decomposition between the longitudinal and transverse motions with the resulting fluctuations of the gap ${h(t) = d (|\rr_0 + \vv(t)|/\ell -1)}$ in Sec.~\ref{sec:Langlarged}~-
are now exact features of the ${d \to \infty}$ saddle point of the dynamical MSRDDJ action.
All that remains in order to close this dynamics is to determine the evolution of the MSD  ${\Delta_r (t)}$, and more generally the dynamical equations for ${\CC(t,s)}$ and ${\cR(t,s)}$.

\section{Dynamical equations for the correlation and response functions}
\label{sec:dynamical-equations-correlation-response}

In order to complete our derivation of the mean-field effective dynamics, we derive here the explicit expressions of the dynamical equations for correlation and response functions, which are the main dynamical observables:
\begin{itemize}

\item first of the correlation and response functions ${\lbrace C(t,t'),R(t,t') \rbrace}$ in Sec.~\ref{sec:dynamical-equations-correlation-response-section1},

\item secondly of the MSD functions ${\DE (t,t')}$ and ${\DE_r(t)=\DE(t,0)}$ in Sec.~\ref{sec:dynamical-equations-correlation-response-section2-MSD},

\item and thirdly of their infinite-dimensional counterparts ${\lbrace \CC(t,t'), \cR(t,t') \rbrace}$ and ${\lbrace \Delta(t,t'),\Delta_r(t) \rbrace}$ in Sec.~\ref{sec:dynamical-equations-correlation-response-section3-high-dim}; this actually amounts to a simple rescaling of all the correlators.

\end{itemize}
Thereafter we present a derivation from the effective stochastic process on the individual displacements ${\uu_i(t)=\xx_i(t)-\xx_i(0)}$, taking as a starting point the definitions~\eqref{eq:SUSYnoneq}.
These equations can also be obtained, in the SUSY path-integral formulation, from the closure relation~\eqref{eq:dynA-closure-relation}, 
using the causal ansatz Eq.~\eqref{eq:SUSYnoneq} for the superfields ${\lbrace A(a,b), M(a,b) \rbrace}$, leading to the same result.
As in previous sections, we set here $m=0$, $\G_R=0$.

Note that in the vectorial formulation, the equations~\eqref{eqC3:twop}-\eqref{eqC3:Mself} for the memory kernels are closed, and from their solutions one can then deduce the correlation and response.
Instead, in the infinite-dimensional scalar formulation, the equations~\eqref{eq:yeff}-\eqref{eq:yeff-kernels-trio} for the memory kernels involve $\D_r(t)$, and they are therefore coupled to the ones for correlations.
This is a complication of the scalar formulation that is important to keep in mind.

\subsection{Correlation and response}
\label{sec:dynamical-equations-correlation-response-section1}

In order to write a closed set of equations for the correlation and response, we start from the effective Langevin dynamics for ${\uu_i(t)}$ in Eq.~\eqref{eqC3:effproc}.
We need the two following relations, which are easy to prove by expressing the dynamical average through the MSRDDJ path integral (an explicit proof can be found in Ref.~\cite[Sec. 4.3]{CC05})
since particle and dimension indices are decoupled:
\begin{equation}
\label{eq:relCR}
 \begin{split}
  R(t,t')
  	& =\la \frac{\d  u_{i\m}(t)}{\d\left(\sqrt2\,\Xi_{i\m}(t')\right)}\ra
 \ ,\\
 \la \sqrt2\,\Xi_{i\m}(t)  u_{i\m}(t')\ra
 	&=\int_0^\io \!\! \de s\, \la \sqrt2\,\Xi_{i\m}(t) \sqrt2\,\Xi_{i\m}(s)  \ra R(t',s)
 \ ,
 \end{split}
\end{equation}
the last being valid due to the Gaussianity of the noise $\bm{\Xi}_i$.
One uses the first equation to get the evolution of the response by differentiating Eq.~\eqref{eqC3:effproc} and averaging over the noise.
The evolution of the correlation ${C(t,t')}$ is instead obtained by multiplying Eq.~\eqref{eqC3:effproc} by ${u_{i\m}(t')}$ and averaging, using the second line in~Eq.~\eqref{eq:relCR}.
We get the system of coupled dynamical equations:
\begin{equation}
\label{eq:CRfinite}
\begin{split}
 \z\frac{\partial }{\partial t} C(t,t')
  	&=2T\z R(t',t)-k(t)C(t,t')+\int_0^{t} \de s\,M_R(t,s)C(s,t')+\int_0^{t'}\de s\,\left[\G_C(t,s)+M_C(t,s)\right]R(t',s)
 \ ,\\
 \z\frac{\partial }{\partial t} R(t,t')
 	&=\d(t-t')-k(t)R(t,t')+\int_{t'}^{t} \de s\,M_R(t,s)R(s,t')
 \ .
\end{split}
\end{equation}
Here, some integration intervals have been truncated owing to the causality of response functions.

\subsection{Mean-square displacement}
\label{sec:dynamical-equations-correlation-response-section2-MSD}

A related quantity of interest in the dynamics is the MSD of a particle, related to the correlation function of the positions
\begin{equation}
\label{eq:DEdyndef}
\begin{split}
 \DE(t,t')
  	&=\frac{1}{d}\la \left[\xx_i(t)-\xx_i(t')\right]^2 \ra
  	=\frac{1}{d}\la \left[\uu_i(t)-\uu_i(t')\right]^2 \ra
	=C(t,t) + C(t',t') - 2 C(t,t') 
   \ , \\
 \DE_r(t)&= \DE(t,0)=C(t,t)
 \ ,\\
 C(t,t')
 	&=\frac12\left[ \DE_r(t) +  \DE_r(t')- \DE(t,t')\right]
 \ .
\end{split}
\end{equation}
The best procedure to obtain $\DE(t,t')$ is likely to be that of solving Eqs.~\eqref{eq:CRfinite}, and then deducing $\DE(t,t')$ from
$C(t,t')$ via its definition~\eqref{eq:DEdyndef}.
Yet, it is also interesting to write a closed dynamical equation for the MSD, thus eliminating $C(t,t')$. 
This can be done by first considering the time derivative
\begin{equation}
 \dot{\DE}_r(t)=\frac{\de}{\de t}C(t,t)
 	=\left.\left(\frac{\partial}{\partial t_1}+\frac{\partial}{\partial t_2}\right)C(t_1,t_2)\right\vert_{t_1=t_2=t}
 \ ,
\end{equation}
and expressing both partial derivatives with the help of the equation for the correlation evolution~\eqref{eq:CRfinite}.
Setting ${t_1=t}$ and ${t_2=t^-}$ this procedure yields
\begin{equation}
 \frac\z2\dot{\DE}_r(t)
 	=-k(t)\DE_r(t)+T\z R(t,t^-)+\int_0^t\de s\,M_R(t,s)C(s,t)+\int_0^t\de s\,\left[\G_C(t,s)+M_C(t,s)\right]R(t,s)
 \ .
\end{equation}
The short-time value ${R(t,t^-)}$ can be computed exactly by ignoring the potential%
\footnote{As in Ref.~\cite{MKZ16}, or by noting that the contribution to position increments of the potential is of ${\mathcal{O}(\de t)}$ while the noise is of ${\mathcal{O} \left(\sqrt{\de t}\right)}$.}.
In the free particle case one has%
\footnote{Note that this derivation holds in the overdamped limit; if inertia or the friction kernel are included, the properties of the response function at equal times change, and additional terms arise. We will consider a more general setting in Sec.~\ref{sec:summary-results}.}
\begin{equation}
\label{eq:Requaltimes}
 \z\dot{\bm x}=\bm \xi+\bm \l\hskip10pt \Rightarrow
 \hskip10pt  
 \la \bm x(t)-\bm x(0)\ra=\frac1\z \int_0^t\de t'\, \bm \l(t')
 \hskip10pt \Rightarrow
 \hskip10pt  
 \left.\frac{\d\la x^\m(t)- x^\m(0)\ra}{\d \l_\m(t')}\right\vert_{\bm\l=\bm 0}= \frac1\z \th(t-t')= R_{\rm free}(t,t')
 \ .
\end{equation}
So we can conclude that ${R(t,t^-)=1/\z}$ and:
\begin{equation}
\label{eq:DR}
 \frac\z2\dot{\DE}_r(t)=T-k(t)\DE_r(t)+\int_0^t\de s\,M_R(t,s)\frac12\left[ \DE_r(t) +  \DE_r(s)- \DE(s,t)\right]+\int_0^t\de s\,\left[\G_C(t,s)+M_C(t,s)\right]R(t,s)
 \ .
\end{equation}
From Eq.~\eqref{eq:DEdyndef}, we moreover have that ${\partial_t \DE(t,t') = \dot \DE_r(t) - 2 \partial_t C(t,t')}$.
So combining Eqs.~\eqref{eq:CRfinite} and~\eqref{eq:DR}, and recalling the definition ${\DE_r(t) = \DE(t,0)}$, 
we obtain an equation for the MSD that can replace the equation for $C(t,t')$ in Eqs.~\eqref{eq:CRfinite}:
\beq
\label{eq:Dclosed}
\begin{split}
 \frac{\z}2 \frac{\partial }{\partial t} \DE(t,t')
 	&= T + \frac{k(t)}2 [ \DE_r(t') - \DE_r(t) - \DE(t,t')] + \int_0^t \de s \, M_R(t,s) \frac12 [\DE_r(t) - \DE_r(t') - \DE(s,t) + \DE(s,t')]
 \\
	&+ \int_0^{\max(t,t')} \!\!\!\! \de s \, \left[\G_C(t,s)+M_C(t,s)\right] \left[R(t,s) - R(t',s)\right] - 2 T \z R(t',t)
 \ ,
\end{split}
\eeq
and reduces to Eq.~\eqref{eq:DR} for ${t'=0}$, as a self-consistency check.

\subsection{High-dimensional limit}
\label{sec:dynamical-equations-correlation-response-section3-high-dim}

The dynamical equations in the ${d\to\io}$ limit can be obtained directly from last section results, specifically from Eqs.~\eqref{eq:CRfinite}, \eqref{eq:DR} and~\eqref{eq:Dclosed}.
We recall the definition of the rescaled correlation and response in Eq.~\eqref{eq:CRscal},
and we similarly define the rescaled MSDs
\begin{equation}
\label{eq-def-MSD-rescaled}
 \D(t,t')
 	= \frac{d^2}{\ell^2} \DE(t,t')
 \ ,\qquad
 \D_r(t)=\frac{d^2}{\ell^2}  \DE_r(t)
\ .
\end{equation}
This last expression is consistent with the definition of ${\D_r(t)}$ already given in Eq.~\eqref{eq-proj-term1}, which was needed when defining the effective stochastic processes in the ${d \to \infty}$ limit, throughout Sec.~\ref{secC3:largedM}.

Rescaling all the quantities using Eq.~\eqref{eqC3:fricscal} and \eqref{eqC3:fricscal2}, 
we thus obtain:
\begin{equation}
\label{eq:scaldyneq}
\begin{split}
 \wh\z\frac{\partial }{\partial t}\CC(t,t')
  	=& 2T\wh\z\cR(t',t)-\k(t)\CC(t,t')+\int_0^{t}\de s\,\MM_R(t,s)\CC(s,t')
 		 +\int_0^{t'}\de s\,\left[\GG_C(t,s)+\MM_C(t,s)\right]\cR(t',s)
 \ ,\\
 \wh \z\frac{\partial }{\partial t}\cR(t,t')
 	=& \frac{\d(t-t')}{2}-\k(t)\cR(t,t')+\int_{t'}^{t}\de s\,\MM_R(t,s)\cR(s,t')
 \ ,\\
 \frac{\wh\z}2 \frac{\partial }{\partial t} \D(t,t')
 	=& \frac{T}2 + \frac{\k(t)}2 [ \D_r(t') - \D_r(t) - \D(t,t')] + \int_0^t \de s \, \MM_R(t,s) \frac12 [\D_r(t) - \D_r(t') - \D(s,t) + \D(s,t')]
 \\
		&+ \int_0^{\max(t,t')} \!\!\!\! \de s \, \left[\GG_C(t,s)+\MM_C(t,s)\right] \left[\cR(t,s) - \cR(t',s)\right] - 2 T \wh\z \cR(t',t)
 \\
 \frac{\wh\z}2 \frac{\partial }{\partial t} \D_r(t)
 	=& \frac{T}2 - \k(t) \D_r(t) + \int_0^t \de s \, \MM_R(t,s) \frac12 [\D_r(t) + \D_r(s) - \D(s,t) ]
 		+ \int_0^t\de s\,\left[\GG_C(t,s)+\MM_C(t,s)\right] \cR(t,s)
 \ .
 \end{split}
\end{equation}

\section{Summary of the results: infinite-dimensional mean-field dynamics}
\label{sec:summary-results}

The different results obtained until now,
first through the dynamical cavity in Sec.~\ref{sec:cavity},
secondly through the SUSY path-integral approach in Sec.~\ref{sec:path},
and thirdly in the infinite-dimensional limit in Sec.~\ref{secC3:largedM}
are gathered thereafter in a compact way.
We emphasise that this summary section is intended as a possible shortcut between the presentation of the general setting in Sec.~\ref{sec:setting} and the applications of our results that will be discussed in Sec.~\ref{sec:limit-cases}.
For the sake of completeness, we have restored all the single-particle terms in the initial Langevin dynamics~\eqref{eqC3:GENLang}, namely a finite mass~$m$ and non-local friction kernel~${\Gamma_R(t,s)}$ (see Eq.~\eqref{eq:defSUSY-part2} and following Ref.~\cite{ABUZ18}).
As mentioned in Sec.~\ref{sec:setting}, including a finite fluid velocity 
${\bm v_f(\xx,t)}$ requires some additional discussion, and this will be presented in the companion paper~\cite{AMZ19bis}.
Physically, all our derivations rely essentially on a few key features of high-dimensional physics:
small displacements of particle around their initial position (${\uu_i(t) = \mathcal{O}(1/d)}$),
uncorrelated numerous neighbours,
and a statistical isotropy and translational invariance of the system.

\subsection{High-dimensional vectorial formulation}
\label{sec:summary-results-high-dim-vectorial}

Our first result (derived in Sec.~\ref{sec:cavity} via the cavity approach and in Sec.~\ref{sec:path} via the path-integral approach) is that in the high-dimensional limit, the correlation functions of the ${N \, d}$-dimensional
many-body Langevin dynamics~\eqref{eqC3:GENLang} can be expressed in terms of an effective one-body (single-particle) $d$-dimensional stochastic process:
\begin{itemize}

\item
The individual displacements ${\uu_i(t)=\xx_i(t)-\RR_i}$, with the initial condition ${\uu_i(0)=\bm 0}$ for all particles and $\RR_i$ drawn from the Boltzmann distribution at temperature $T_0$,
are described by the effective stochastic process (identical for all particles)
given in~Eq.~\eqref{eqC3:effproc}:
\beq
\label{summary:uvec}
\begin{split}
 & m\ddot{\uu}(t) + \z  \dot \uu(t)+ \int_0^t \!\! \de s \, \G_R(t,s) \, \dot\uu(s)
 = - k(t) \uu(t) + \int_0^t \!\! \de s\, M_R(t,s) \, \uu(s) + \sqrt{2}\, \bm{\X}(t)
 \ , \\
 & \moy{\X_{\m}(t)}_{\bm{\Xi}} =0
 \ , \qquad
 \moy{\X_{\m}(t) \X_{\n}(t')}_{\bm{\Xi}}
 = \d_{\m\n} \left[ T \z \d(t-t') +\frac12 \G_C(t,t') +\frac12 M_C(t,t') \right]
 \ .
\end{split}
\eeq

\item 
The inter-particle distances
${\vv_{ij}(t) = \uu_{i}(t)-\uu_j(t)}$,
with the initial condition ${\vv_{ij}(0)=\bm 0}$ for all particles, satisfy the effective stochastic process (identical for all pairs)
given in~Eq.~\eqref{eqC3:twop}:
\beq
\label{summary:wvec}
 \begin{split}
 & \frac{m}{2} \ddot{\vv}(t)  	+ \frac\z2 \dot \vv(t) 
+ \frac12 \int_0^t \!\! \de s \, \G_R(t,s) \, \dot\vv(s)
 	=- \frac{k(t)}2 \vv(t) + \frac12 \int_0^t \!\! \de s\, M_R(t,s) \, \vv(s) - \nabla v(\rr_0 + \vv(t)) + \bm{\X}(t)
 	\ ,
\end{split}
\eeq
where the noise ${\bm\X(t)}$ has exactly the same statistics as in Eq.~\eqref{summary:uvec}.
Here ${\rr_0 = \RR_i - \RR_j}$ is a parameter representing the initial distance between the pair.

\item
The time-dependent kernels ${\arga{k(t),M_C(t,t'),M_R(t,s)}}$ that enter in the effective processes, and express in a mean-field way the interaction with the other particles,
are determined self-consistently by~Eq.~\eqref{eqC3:Mself} (via the cavity):
\beq
\label{summary:kernelsvec}
\begin{split}
 k(t)
 	&= \frac{\r}d \int \de\rr_0 \, e^{-\Bi v(\rr_0)} \moy{ \Lap v( \rr(t))}_{\vv}
 \ , \\
 M_C(t,t')
 	&=  \frac{\r}d \int \de \rr_0 \, e^{-\Bi v(\rr_0)} \moy{ \nabla v(\rr(t))  \cdot \nabla v (\rr(t')) }_{\vv}  
 \ , \\
 M_R(t,s)
 &=  \frac{\r}d \sum_{\m=1}^d \int \de \rr_0 \, e^{-\Bi v(\rr_0)}\,\left. \frac{\d \moy{ \nabla_{\m} v(\rr(t)) }_{\vv,\bm{P}}}{\d P_{\mu}(s)}\right\vert_{\bm P=\bm0}
 \ ,
\end{split}
\eeq
where ${\rr(t) = \rr_0 + \vv(t)}$,
the dynamical average is over the effective process~\eqref{summary:wvec} for ${\vv(t)}$, 
the response is computed by shifting ${\nabla v(\rr_0 + \vv(t)) \to \nabla v(\rr_0 + \vv(t)-\bm{P}(t))}$ in Eq.~\eqref{summary:wvec},
and ${g_{\text{eq}}(\valabs{\rr_0} | \Bi) =  e^{-\Bi v(\rr_0)}}$ is the (non-normalised) equilibrium distribution of the initial pair distances evaluated in the high-dimensional limit.
Note that the initial condition of the dynamics only enter in the equations via its pair correlation ${g_{\text{in}}(\valabs{\rr_0})}$; hence, more general initial conditions could be modeled via different choices of the radial function ${g_{\text{in}}(\valabs{\rr_0})}$ in the kernels of Eq.~\eqref{summary:kernelsvec}.

\item
One-time observables such as the pair correlation, the
potential energy and the stress tensor, defined in Eq.~\eqref{eq:ePi}, can be expressed as
\beq
\begin{split}
g(\rr,t) &= \int \de\rr_0 \, e^{-\Bi v(\rr_0)} \moy{ \d( \rr(t) - \rr)}_{\vv}
 \ , \\
 e(t)
 	&= \frac{\r}2 \int \de\rr_0 \, e^{-\Bi v(\rr_0)} \moy{ v( \rr(t))}_{\vv}
 \ , \\
 \Pi_{\m\n}(t)
 	 &= -\frac{\r^2}{2} \int \de\rr_0 \, e^{-\Bi v(\rr_0)}\moy{ 
  \frac{r_{\m}(t) r_{\n}(t)}{r(t)^2} \rr(t) \cdot \nabla v(\rr(t))}_{\vv}
 \ ,
\end{split}
\eeq
where in the expression of ${\Pi_{\m\n}(t)}$ we considered the overdamped limit ${m=0}$ for simplicity (the kinetic contribution is usually negligible in the glassy regime).
Note that in absence of any fluid flow (${\bm v_f=0}$) the off-diagonal components of ${\Pi_{\m\n}(t)}$ vanish by isotropy,
while the diagonal components give the pressure $P(t) = \Pi_{\m\m}(t)$.
Any other one-time observable which depends on particle gaps can be expressed in the same way.

\item
Two-time observables can be related to the memory kernels via integro-differential equations.
For example, defining the integro-differential operator
\beq
 \DD_2 f(t,t')
	\equiv m \frac{\partial^2}{\partial t^2} f(t,t')
			+ \z\frac{\partial }{\partial t} f(t,t')
			+ \int_0^t \de s \,\G_R(t,s)\frac{\partial }{\partial s} f(s,t')
 \ , \\
\eeq
we obtain
for the correlation and response defined in Eq.~\eqref{eq:CRdef}:
\begin{equation}
\label{eq:summaryCR}
\begin{split}
 \DD_2C(t,t')
  	&=2T\z R(t',t)-k(t)C(t,t')+\int_0^{t} \de s\,M_R(t,s)C(s,t')+\int_0^{t'}\de s\,\left[\G_C(t,s)+M_C(t,s)\right]R(t',s)
 \ ,\\
\DD_2 R(t,t')
 	&=\d(t-t')-k(t)R(t,t')+\int_{t'}^{t} \de s\,M_R(t,s)R(s,t')
 \ .
\end{split}
\end{equation}
\item From these one can derive the MSD, either via $\DE(t,t') = C(t,t) + C(t',t') - 2 C(t,t')$, or by writing a dynamical equation for it applying similar arguments%
\footnote{Most of the differences with respect to Eqs.~\eqref{eq:DR} and~\eqref{eq:Dclosed} come from the inertia. When $m\neq0$, the 
`equal-time' response vanishes, \ie $R(t,t^-)=0$, as can be checked from the Langevin equation~\eqref{eqC3:GENLang}, dropping all 
terms except inertial and frictional ones (and the response-generating field), which are the only relevant ones at very short times.} 
to the ones of Sec.~\ref{sec:dynamical-equations-correlation-response-section2-MSD}:
\begin{equation}
\label{eq:summaryMSD}
\begin{split}
 \frac12\DD_2 \DE(t,t')-&\frac12\int_0^{t} \de s\,\G_R(t,s)\frac{\partial}{\partial s}\DE(s,t)\\
  	=&-\frac{k(t)}{2}\left[\DE(t,t')+\DE_r(t)-\DE_r(t')\right]+\frac12\int_0^{t} \de s\,M_R(t,s)\left[\DE_r(t)-\DE_r(t')+\DE(s,t')-\DE(s,t)\right]\\
  	&-\frac m2 \partial_1\partial_2\DE(t,t) - 2T\z R(t',t)
  	+\int_0^{\max(t,t')}\de s\,\left[\G_C(t,s)+M_C(t,s)\right]\left[R(t,s)-R(t',s)\right]
 \ .
\end{split}
\end{equation}
For concision we noted $\partial_i$ the partial derivative with respect to the $i$-th argument of the function. 
The term ${-(m/2) \partial_1\partial_2\DE(t,t)=m\la \dot{\xx}_i^2(t) \ra/d}$ is twice the kinetic energy per particle and dimension,%
\footnote{As a consequence, at equilibrium this term reduces to ${-(m/2) \partial_1\partial_2\DE(t,t)=T_0=T}$.}  
measured at time $t$.
By definition, this equation evaluated at ${t'=0}$ gives the evolution of the `initial condition' MSD ${D_r(t)}$.

\end{itemize}

\subsection{Infinite-dimensional scalar formulation}
\label{sec:summary-results-infinite-dim-scalar}

The vectorial equations summarised in Sec.~\ref{sec:summary-results-high-dim-vectorial} are convenient because they have a clear physical interpretation, but their numerical solution is particularly difficult because one has to solve self-consistently a $d$-dimensional process.
As discussed throughout Sec.~\ref{secC3:largedM}, these equations can be reduced to one-dimensional
processes by essentially projecting on the longitudinal motion, \textit{i.e.}~$\vv(t)$ parallel to $\rr_0$. In this section we summarise the results of this procedure.
\begin{itemize}

\item
In order to obtain a non-trivial ${d\to\io}$ limit, the following rescalings have to be performed, with $\ell$ being a typical interaction scale of the pair potential, as discussed in~Eqs.~\eqref{eq:scalpotd}, \eqref{eqC3:fricscal}, and~\eqref{eqC3:fricscal2}:
\begin{eqnarray}
 && v(r) = \redv(h)
 \ , \qquad
 h = d(r/\ell -1)
 \quad \Leftrightarrow \quad
 r = \ell (1+h/d)
 \ ,
 \\
 && 
 \wh{m} = \frac{ \ell^2}{2d^2} m
 \ , \quad
  \wh\z = \frac{ \ell^2}{2d^2} \z 
 \ , \quad
 \GG_C(t,s) = \frac{ \ell^2}{2d^2} \G_C(t,s)
  \ , \quad
 \GG_R(t,s) = \frac{ \ell^2}{2d^2} \G_R(t,s)
 \ ,
 \\
 && \k(t) = \frac{ \ell^2}{2d^2} k(t)
 \ , \qquad
 \MM_C(t,t') = \frac{ \ell^2}{2d^2} M_C(t,t')
 \ , \qquad
 \MM_R(t,s) = \frac{ \ell^2}{2d^2} M_R(t,s)
 \ .
\end{eqnarray}
with the number density being rescaled into packing fraction ${\f=\r V_d (\ell/2)^d=\wh\f \, d/2^d}$
and ${\wh\f=\r V_d\ell^d/d}$, with ${V_d=\p^{d/2}/\G(d/2+1)}$ the volume of the unit radius sphere in dimension $d$, see Eq.~\eqref{eq:pf}.

\item 
One obtains an effective stochastic process for the longitudinal motion, 
\textit{i.e.}~the rescaled projection ${y(t) = \hat{\rr}_0 \cdot \vv(t) \, d/\ell }$, with ${y(0)=0}$ by definition, 
given by Eq.~\eqref{eqC3:effproch}:
\beq
\begin{split}
 & \wh{m} \ddot{y}(t) + \wh\z \dot y(t)+ \int_0^t \!\! \de s \, \GG_R(t,s) \, \dot{y}(s)
 	=	- \k(t) y(t)
 		+ \int_0^t \!\! \de s \, \MM_R(t,s) \, y(s)
 		- \redv'(h_0+y(t)+\D_r(t)) +  \X(t)
 \ , \\
 & \moy{ \X(t)}_{\X}=0
  	\ , \quad
 	\moy{ \X(t) \X(t')}_{\X}
 		=  2 T \wh\z \d(t-t') + \GG_C(t,t') + \MM_C(t,t')
 \ , \\
\end{split}
\eeq
with ${\Delta_r(t) = \moy{\uu(t)^2} d/\ell^2}$ the mean-square displacement with respect to the initial position.
This process can also be expressed in terms of the inter-particle gap ${h(t) = h_0+y(t)+\D_r(t)}$ as in 
Eq.~\eqref{eq:yeff}.

\item
The rescaled time-dependent kernels ${\arga{\k(t),\MM_C(t,t'),\MM_R(t,s)}}$ are obtained self-consistently
from Eq.~\eqref{eq:yeff-kernels-trio}:
\beq
\begin{split}
 \k(t)
 	&= \frac{\wh \f}2 \int^{\infty}_{-\infty} \, \de h_0 \, e^{h_0 -\Bi \redv( h_0  )}   \la \redv''(h(t)) + \redv'(h(t)) \ra_{h}
 \ , \\
 \MM_C(t,t')
 	&=  \frac{\wh\f}2 \int^{\infty}_{-\infty}  \de h_0 \, e^{h_0 -\Bi  \redv(h_0)}  
 		\langle \redv'(h(t)) \redv'(h(t')) \rangle_{h} 
 \ , \\
 \MM_R(t,s)
 	&=  \frac{\wh\f}2 \int^{\infty}_{-\infty}  \de h_0 \, e^{h_0 -\Bi  \redv(h_0)}  
		\left. \frac{\d \langle \redv'(h(t))  \rangle_{h,\PP}}{\d \PP(s)}\right\vert_{\PP=0}
 \ ,
\end{split}
\eeq
where the perturbation ${\PP(t)}$ acts in ${\redv'(h_0+y(t)+\D_r(t)) \to \redv'(h_0+y(t)+\D_r(t) - \PP(t))}$.
These definitions remain again valid if we assume a statistical isotropic initial condition, provided that we replace the equilibrium function ${g_{\text{eq}}(h_0)=e^{-\Bi \bar{v}(h_0)}}$ by the corresponding function ${g_{\text{in}}(h_0)}$.

\item
The scaled energy and pressure become
\beq
\begin{split}
 \wh e(t)
 	&=\frac{e(t)}d
 	= \frac{ \wh \f}2 \int^{\infty}_{-\infty}  \, \de h_0 \, e^{h_0 -\Bi \redv( h_0  )}   \la \redv(h(t)) \ra_{h}
 \ , \\
\wh p(t) &= \frac{ \b P(t) }{d \, \r} =
\frac{\b\Tr \hat \Pi(t)}{d^2\, \r}
 	= - \frac{ \wh \f}2 \int^{\infty}_{-\infty}  \, \de h_0 \, e^{h_0 -\Bi \redv( h_0  )} 
  \la \b\redv'(h(t)) \ra_{h}
 \ .
\end{split}
\eeq

\item
The correlation and response functions have also to be rescaled as
\beq
 \CC(t,t')
 	=\frac{d^2}{\ell^2} C(t,t')
 \ ,
\quad \cR(t,t')
 	=\frac{d^2}{\ell^2} R(t,t')
	\ , \quad
  \D(t,t')
 	= \frac{d^2}{\ell^2} \DE(t,t')
  \ ,
  \quad \D_r(t) = \D(t,0)
 \ .
\end{equation}
They satisfy a rescaled version of Eq.~\eqref{eq:summaryCR}:
\begin{equation}
\begin{split}
\wh \DD_2 \CC(t,t')
  	=& 2T\wh\z\cR(t',t)-\k(t)\CC(t,t')+\int_0^{t}\de s\,\MM_R(t,s)\CC(s,t')
 		 +\int_0^{t'}\de s\,\left[\GG_C(t,s)+\MM_C(t,s)\right]\cR(t',s)
 \ ,\\
 \wh \DD_2\cR(t,t')
 	=& \frac{\d(t-t')}{2}-\k(t)\cR(t,t')+\int_{t'}^{t}\de s\,\MM_R(t,s)\cR(s,t')
 \ ,
 \end{split}
\end{equation}
with the operator ${\wh\DD_2 = \frac{ \ell^2}{2d^2}  \DD_2}$.
Likewise, the rescaled MSD can be deduced by $\D(t,t') = \CC(t,t) + \CC(t',t') - 2 \CC(t,t')$ or by
an independent dynamical equation which is a rescaled version of Eq.~\eqref{eq:summaryMSD}: 
\begin{equation}
\begin{split}
 \frac12\wh\DD_2 \D(t,t')-&\frac12\int_0^{t} \de s\,\GG_R(t,s)\frac{\partial}{\partial s}\D(s,t)\\
  	=&-\frac{\k(t)}{2}\left[\D(t,t')+\D_r(t)-\D_r(t')\right]+\frac12\int_0^{t} \de s\,\MM_R(t,s)\left[\D_r(t)-\D_r(t')+\D(s,t')-\D(s,t)\right]\\
  	&-\frac{\wh m}{2} \partial_1\partial_2\D(t,t) - 2T\wh\z \cR(t',t)
  	+\int_0^{\max(t,t')}\de s\,\left[\GG_C(t,s)+\MM_C(t,s)\right]\left[\cR(t,s)-\cR(t',s)\right]
 \ . \\
\end{split}
\end{equation}

\end{itemize}

\section{Limit cases and applications}
\label{sec:limit-cases}

We now show how our dynamical equations
--~for the vectorial effective stochastic processes ${\uu(t)=\xx(t)-\RR}$ and ${\vv(t)=\rr(t)-\rr_0}$ in high dimension~--
reproduce previously known results, in particular concerning the equilibrium dynamics~\cite{MKZ16,KMZ16,Sz17,CKPUZ17} in Sec.~\ref{sec:limit-cases-equilibrium}
and the so-called `state-following' protocol~\cite{RUYZ15} in Sec.~\ref{sec:limit-cases-state-following}.

\subsection{Equilibrium dynamics}
\label{sec:limit-cases-equilibrium}

As a first check, we consider in Eq.~\eqref{eqC3:GENLang} the white-noise equilibrium case, \textit{i.e.}~where ${\G_C=0}$ and ${T=\Ti}$ (in addition to ${m =0}$, ${\Gamma_R=0}$) in order to align with the setting of Refs.~\cite{MKZ16,KMZ16,Sz17}. 
In this case, the two memory kernels ${M_{C,R}(t,s) = M_{C,R}(t-s)}$ are time-translational invariant (TTI) and related by the FDT relation 
${M_R(t-s) = \b \th(t-s) \partial_s M_C(t-s)}$~\cite{Ku66,Cu02,Ha97,ZBCK05,MSVW13}.
Furthermore, being in a steady state, the mean divergence of local forces is a constant, ${k(t) = k}$, given by its ${t=0}$ value in Eq.~\eqref{summary:kernelsvec}.
Therefore the effective stochastic process for $\vv(t)$, Eq.~\eqref{summary:wvec} 
can be simplified by using these properties.
It becomes, via an integration by parts and changing variable via Eq.~\eqref{eq-def-wrt-initial-condition} to the absolute inter-particle distance 
${\rr(t) = \rr_0 + \vv(t)}$:
\beq
 \frac\z2 \dot \rr(t)
 	= - \frac12 \argp{k - \b M_C(0)} \argp{\rr(t)-\rr_0} - \frac\b2 \int_0^t \de s\, M_C(t-s) \, \dot{\rr}(s) - \nabla v(\rr(t)) + \bm{\X}(t) \ .
\eeq
From Eqs.~\eqref{summary:kernelsvec} with ${T=\Ti}$, we have
\beq
\label{eq:IBPkM}
 k - \b M_C(0)
 	\propto \int \de \rr \, e^{-\b v(\rr)}\argp{ \b \left| \nabla v(\rr) \right|^2 - \Lap v( \rr)}
	= 0
 \ ,
\eeq
where the last equality can be proven by recognising that
$e^{-\b v(\rr)} \, \b \nabla v(\rr) = -\nabla e^{-\b v(\rr)}$ and integrating
by parts.
Therefore, we obtain
\beq
\label{eq-check-equ-white-noise}
 \frac\z2 \dot \rr(t) + \frac\b2 \int_0^t \de s\, M_C(t-s) \, \dot{\rr}(s)
  = -\nabla v(\rr(t)) + \bm{\X}(t)
 \ ,
\eeq
which coincides with the equilibrium equation obtained in Ref.~\cite{Sz17}.
The same argument holds for the single-particle effective process ${\uu(t)}$.
Finally, recalling that ${\bm{\X}(t)}$ is here a Gaussian noise of variance
${\moy{\X_{\m}(t) \X_{\n}(s)}_{\bm{\Xi}}
 	= \d_{\m\n} \left[ T \z \d(t-s) + \frac12 M_C(t,s) \right]}$,
this stochastic process describes indeed an equilibrium dynamics in a potential ${v(\rr)}$,
hence with the equilibrium probability distribution
${P_{\text{eq}}(\rr) \propto e^{-\beta v(\rr)}}$.

Note that for standard pair potentials $v(\rr)$ which decay to zero at large ${|\rr|}$, ${P_{\rm eq}(\rr)}$ is not normalisable (even if one can always introduce an artificial
cutoff at large distances).
The fact that ${v(\rr)}$ is not confining and ${P_{\rm eq}(\rr)}$ is not normalisable has an important physical meaning.
In fact, as we shall show below, if the memory kernel ${M_C(t)}$ decays to zero sufficiently fast
at large-time separation $t$, then on large timescales,
Eq.~\eqref{eq-check-equ-white-noise} becomes essentially a Brownian motion.
This implies a diffusive behaviour stemming from the decorrelated motion at large distances.

This statement can naturally be extended to the more general equilibrium case with colored-noise and retarded-friction kernels, in other words reinstated non-zero ${\Gamma_R}$ and ${\Gamma_C}$.
In this setting, Eq.~\eqref{eq:IBPkM} remains valid and the stochastic process~\eqref{eq-check-equ-white-noise} becomes:
\beq
\label{eq-check-equ-generic-friction-noise}
\begin{split}
 & \frac\z2 \dot \rr(t) + \frac12 \int_0^t \de s\, \argc{\b M_C(t-s)+ \Gamma_R(t-s)}\, \dot{\rr}(s)
  = -\nabla v(\rr(t)) + \bm{\X}(t)
 \ ,
 \\
 & \moy{\X_{\m}(t)}_{\bm{\Xi}}=0
 \, , \quad
 \moy{\X_{\m}(t) \X_{\n}(s)}_{\bm{\Xi}}
 	= \d_{\m\n} \left[ T \z \d(t-s) + \frac12 \G_C(t,s) + \frac12 M_C(t,s) \right]
\end{split}
\eeq
so we can reach the equilibrium probability distribution ${P_{\text{eq}}(\rr) \propto e^{-\beta v(\rr)}}$ if and only if the friction and noise kernels satisfy the 
FDT relation (`of the second kind')
${\Gamma_R(t)=\beta \theta(t) \Gamma_C(t)}$~\cite{Ku66,Cu02,Ha97,ZBCK05,MSVW13}.

Finally, we discuss the behaviour of correlation and response functions, specialising the discussion of Sec.~\ref{sec:dynamical-equations-correlation-response} to equilibrium,
where ${R(t,t') = R(t-t')}$ and ${\DE(t,t') = \DE(t-t')}$ are time-translationally invariant (TTI).
This property does not hold for the correlation ${C(t,t')}$, as \eg the MSD ${C(t,t)=\DE_r(t)}$ is not stationary. 
It is then more convenient to replace it with ${\DE(t-t')}$.
In this regime, correlations and responses are related by the FDT
\beq
 R(t)
 	=\frac{\b}2 \th(t) \dot \DE(t) \ , \qquad M_R(t)
 	= - \b \th(t) \dot M_C(t)
 \ .
\eeq
The unusual ${(-1/2)}$ factor in the first relation comes from the fact that ${\DE(t-t')}$ is the average of a squared displacement,
see \textit{e.g.}~Ref.~\cite{BB02}. The FDT relation for the MSD is also called the Einstein relation.
Using TTI and FDT in Eq.~\eqref{eq:Dclosed}, and recalling Eq.~\eqref{eq:IBPkM}, we obtain the equilibrium equation for the MSD (respectively in its rescaled form):
\beq
\label{eq:DEeqeq}
 \frac\z2 \dot \DE(t)
 	= T - \frac{\b}2 \int_0^t \de s\, M_C(t-s) \dot \DE(s)
 \ ,\qquad \qquad
 \wh\z\dot\D(t)
 	= T -\b \int_{0}^{t} \de s\, \MM_C(t-s)\dot \D(s)
 \ ,
\end{equation}
which coincides with the equilibrium dynamical equation for the MSD obtained in Refs.~\cite{MKZ16,KMZ16}.
As a consequence of this and of FDT, one has in equilibrium that ${\BB_{\text{MSD}}(t)=T}$ in~Eq.~\eqref{eq:yeff}.
The stochastic process in Eq.~\eqref{eq:yeff} at equilibrium then reduces to the one derived in Refs.~\cite{MKZ16,KMZ16}.
Note that from Eqs.~\eqref{eq:DEeqeq} one can explicitly check the following: if ${M_C(t-s)}$ decays to zero sufficiently fast 
so that it is integrable for ${|t-s|\to\io}$,
then ${\DE(t) \propto t}$ at large times~\cite{MKZ16},
\textit{i.e.}~the long-time behaviour is diffusive.

\subsection{Following glassy states under quasistatic perturbations}
\label{sec:limit-cases-state-following}

We have recalled in the equilibrium case that, when the memory kernels decay sufficiently fast at large times, the dynamics is diffusive.
The same result remains true out of equilibrium.
At high density and low temperatures, and in absence of energy injection, the dynamics can become arrested due to the formation of a glass phase~\cite{FP95,Go09,RUYZ15,CKPUZ17}.
In this situation, the memory kernels do not decay to zero, and the MSD reaches a finite plateau, while diffusion is arrested.
One can then write self-consistent equations for the plateau values, which do not require a full solution of the dynamics~\cite{FP95,Go09,MKZ16,Sz17} and are equivalently obtained via the replica method in the so-called Franz-Parisi or state-following formulation~\cite{FP95,RUYZ15}.
This trapped dynamics within a metastable state is a standard feature of out-of-equilibrium mean-field glassy models~\cite{CK93,CK94,FM94,BBM96,CLD96,CK08}.
We now discuss this construction, following the discussion in Refs.~\cite{MKZ16,KMZ16,Sz17}.

We first consider in Sec.~\ref{sec:limit-cases-state-following-ansatz} the simplest case, in which we prepare the system at some initial temperature $\Ti$ and then we instantaneously quench it to a different temperature $T$, in presence of a white noise with ${\G_C = \G_R=0}$.
We assume that both $\Ti$ and $T$ fall in the glass regime, in which there is no diffusion, and we want to establish dynamically the equations for the long-time limit of the MSD.
Note that in this context the temperature $\Ti$ is interpreted as the last temperature that the system can visit in equilibrium before being trapped in a glass state, which is usually identified with the glass transition temperature, \textit{i.e.}~${\Ti = T_g}$~\cite{RUYZ15}, and we recall that our derivation is valid only for $\Ti > T_{\rm K}$ (Kauzmann temperature).
We derive in Sec.~\ref{sec:limit-cases-state-following-long-time-MSD} the corresponding set of equations for the large-time limit of the MSDs.

Then, in the rest of this section we generalise the result
in Sec.~\ref{sec:limit-cases-state-following-tuning-persistence-time} \textit{(i)}~to non-zero `equilibrium' friction and noise kernels, \textit{i.e.}~satisfying ${\Gamma_R(t-s)=\beta \theta(t-s) \Gamma_C(t-s)}$, as we have just seen in Sec.~\ref{sec:limit-cases-equilibrium} for standard equilibrium,
and \textit{(ii)}~to constant random forces with ${\Gamma_R=0}$ but ${\Gamma_C(t-s)=f_0^2}$.

\subsubsection{Restricted equilibrium ansatz}
\label{sec:limit-cases-state-following-ansatz}

In a glassy regime,
the dynamics can be decomposed into a `fast' part --~corresponding to the initial transient regime where the system moves from $\Ti$ to $T$~-- and a `slow' part --~corresponding to the long-time correlations between the initial and final states since the system is assumed to be in the glass phase~\cite{CK93,Cu02,CK08}.
Besides, the fast part of the dynamics reaches a restricted equilibrium in the glass state at temperature $T$, in which TTI and FDT hold.
It is then more natural to work with the fluctuation ${\vv(t)=\rr(t)-\rr_0}$, since we might be confined to the metastable state picked up by the initial condition.

In practice, at long times ${t,s \to\io}$, we can consider the following ansatz to decompose the memory kernels and the noise of the Langevin equation~\eqref{summary:wvec}:
\beq
\label{eq-restricted-equilib-ansatz}
\begin{split}
 & M_C(t,s)
 	\underset{(t,s\to\io)}{\longrightarrow} M_f(t-s) + M_\io 
 \ , \\
 & M_R(t,s)
	 \underset{(t,s\to\io)}{\longrightarrow} \b \th(t-s)\partial_s M_f(t-s)
 \ , \\
 & \bm{\X}(t)
 	\underset{(t\to\io)}{\longrightarrow}  \bm{\X}^f(t) + \bm{\X}^\io 
  \quad \text{with} \quad \left\lbrace \begin{array}{l}
 	\moy{ \X^f_{\m}(t)}_{\text{req}, \bm{\X}^\infty} = 0
 	\, , \quad
 	\moy{ \X^f_{\m}(t) \X^f_{\n}(s) }_{\text{req}, \bm{\X}^\infty}
 		= \d_{\m\n} \left[ T \z \d(t-s) + \frac12 M_f(t-s) \right]
 	\ , \\ \\
 	\overline{ \X^\io_{\m}}=0
 	\, , \quad
 	\overline{ \X^\io_{\m} \X^\io_{\n}}
	 	= \d_{\m\n}  \frac12 M_\io 
 	\ ,
 \end{array} \right.
 \\
 & k(t)
	 \underset{(t\to\io)}{\longrightarrow} k_{\infty}
 \ .
\end{split}
\eeq
Here,
$M_\io$ is the long-time plateau of the memory function,
and ${M_f(t-s)}$ is the fast part that decays quickly to zero for ${|t-s|\to\io}$ and is related by FDT to the response ${M_R(t-s)}$ (which has no slow part).
The Gaussian noise ${\bm{\X}(t)}$ is also decomposed into two Gaussian components, with their corresponding statistical averages: ${\moy{\bullet}_{\text{req}, \bm{\X}^\infty}}$ is an equilibrium average restricted to a given value ${\bm{\X}^\infty}$, and ${\overline{\, \bullet \,}}$ denotes the `disorder average' over ${\bm{\X}^\io}$ (which characterises the glassy metabasin in which the system is trapped).
As for the kernel ${k(t)}$, it is a single-time quantity and as such it simply goes to its long-time limit value $k_{\infty}$.

Plugging this ansatz into the stochastic process for ${\vv(t)}$ given by Eq.~\eqref{summary:wvec}, we obtain for long times $t$:
\beq
 \frac\z2 \dot \vv(t)
	= 	- \frac{k_\io}2 \vv(t)
		+\frac\b2 M_f(0) \vv(t)
		- \frac\b2 \int_0^t \de s \, M_f(t-s) \dot\vv(s)
		-\nabla v (\rr_0 + \vv(t))
		+ \bm{\X}^f(t) + \bm{\X}^\io
\eeq
that we can rewrite, by regrouping the dissipative terms \textit{versus} the terms deriving from a potential and introducing ${k_{\rm eff} = k_\io - \b M_f(0)}$, as
\beq
\label{eq-state-following-rewriting-vv}
 \frac\z2 \dot \vv(t) + \frac\b2 \int_0^t \de s \, M_f(t-s) \dot\vv(s)
 = - \argc{ \nabla v (\rr_0 + \vv(t)) + \frac12 k_{\rm eff} \vv(t) - \bm{\X}^\io} + \bm{\X}^f(t)
 \ .
\eeq
This equation describes an equilibrium dynamics in the modified potential
\beq
\label{eq:veffdef}
 v_{\text{eff}}(\vv)
 	= v(\rr_0 + \vv) + \frac{k_{\rm eff}}{4} \argp{\vv - 2 \frac{\bm{\X}^\io }{ k_{\rm eff}}}^2  \ ,
\eeq
and has thus the following (normalised) probability distribution:
\beq
\label{eq:Peqr}
 P_{\rm req}(\vv;\rr_0,k_{\rm eff}, \bm{\X}^\io)
 	=  \argp{ \frac{\b k_{\rm eff}}{4\pi} }^{d/2} \frac{ e^{-\b v(\rr_0 + \vv) - \frac{\b k_{\rm eff}}4 (\vv - 2 \bm{\X}^\io/k_{\rm eff})^2 } }
{ e^{\frac{1}{\b k_{\rm eff}}\Lap   } e^{-\b v(\rr_0 + 2  \bm{\X}^\io/k_{\rm eff})}}
\ .
\eeq
In the denominator we have used the following identity to express in a compact way the $d$-dimensional convolution with a Gaussian measure of variance $A$ as a differential operator, similarly to Eq.~\eqref{eq:idGA}:
\beq
\label{eq:idG}
 \int \de \xx \, \frac{e^{- \frac{\xx^2}{2 A}}}{(2\pi A)^{d/2}} f(\rr + \xx)
 	= e^{\frac{A}2 \Lap} f(\rr)
\ , 
\eeq
where $\Lap$ is the Laplacian operator and ${f(\rr)}$ an arbitrary regular function.

Physically, ${P_{\rm req}(\vv;\rr_0,k_{\rm eff}, \bm{\X}^\io)}$ is the restricted equilibrium distribution of the glass state reached by the dynamics, selected by ${(k_{\rm eff}, \bm{\X}^\io)}$ acting effectively as 
quenched variables, and with a given initial condition~$\rr_0$.
This glass state is encoded in the effective potential~${v_{\text{eff}}(\vv)}$ as a quadratic well --~literally a metabasin
~-- of curvature ${k_{\rm eff} = k_\io - \b M_f(0)}$ and shifted by ${2\bm{\X}^\io/k_{\rm eff}}$.
Note that the ansatz~\eqref{eq-restricted-equilib-ansatz}, where there is a single timescale and 
${\bm{\X}^\io}$ is a Gaussian variable, corresponds to a replica-symmetric (RS) assumption on the metabasin structure of the free-energy landscape~\cite{nature,CKPUZ17}.
Further steps of replica-symmetry breaking would require the introduction of more well-separated time sectors~\cite{CK94}.

So within the RS assumption on the equilibrium initial condition, the restricted-equilibrium probability reached at large times depends on three unknowns, namely ${k_\io}$, ${M_f(0)}$ and $M_\io$.
Self-consistent equations for these quantities are derived by taking the ${t\to\io}$ limit of the self-consistent equations for the kernels, Eqs.~\eqref{summary:kernelsvec}.
We obtain
\beq
\label{eq:statefol1}
\begin{split}
 k^\io 
 	&	= \lim_{t\to\io} k(t)
 		= \frac{\r}d \int \de\rr_0 \, e^{-\Bi v(\rr_0)} \, \overline{\moy{ \Lap v(\rr_0 +\vv) }_{\rm req}}
 \ , \\
 M_\io 
 	&	= \lim_{t\to\io} \lim_{|t-s|\to\io} M_C(t,s)
 		=  \frac{\r}d \int \de \rr_0 \, e^{-\Bi v(\rr_0)} \overline{  | \la \nabla v(\rr_0 +\vv)  \ra_{\rm req} |^2}
 \ , \\
 M_f(0)
 	&	= \lim_{t\to\io} [ M_C(t,t) - M_\io ]
 		=  \frac{\r}d \int \de \rr_0 \, e^{-\Bi v(\rr_0)} \argc{
			\overline{  \moy{ | \nabla v(\rr_0 +\vv) |^2 }_{\rm req}} 
				- \overline{  | \la \nabla v(\rr_0 +\vv)  \ra_{\rm req} |^2} }
\ ,
\end{split}
\eeq
where one should first take the average over $P_{\rm req}$,
then an average over the Gaussian noise $\bm{\X}^\io$,
and finally integrate over the initial condition $\rr_0$.
These relations provide closed equations for $k_{\rm eff}$ and $M_\io$.
Note that using Eqs.~\eqref{eq:Peqr} and \eqref{eq:idG}, the averages that enter in these equations can be explicitly expressed as
\beq
\label{eq-moy-equilib-disorder-state-following}
 \overline{\moy{ f(\rr_0 +\vv) }_{\rm req}^{\alpha}}
	=\overline{
 		\argc{\frac{ e^{\frac{1}{\b k_{\rm eff}} \Lap  } e^{-\b v(\rr_0 + 2  \bm{\X}^\io/k_{\rm eff})} f(\rr_0 +  2  \bm{\X}^\io/k_{\rm eff}) }
		{ e^{\frac{1}{\b k_{\rm eff}}\Lap   } e^{-\b v(\rr_0 + 2  \bm{\X}^\io/k_{\rm eff})}}  
	}^\alpha}
	= e^{\frac{M_\io}{k_{\rm eff}^2}\Lap } \arga{\argc{
 \frac{ e^{\frac{1}{\b k_{\rm eff}}\Lap   } e^{-\b v(\rr_0 )} f(\rr_0 ) }
{ e^{\frac{1}{\b k_{\rm eff}}\Lap   } e^{-\b v(\rr_0 )}}  }^\alpha}
 \ .
\eeq
In order to be more explicit regarding the kernels given in Eq.~\eqref{eq:statefol1}, one would need to specify the pairwise potential ${v(\rr)}$.

\subsubsection{Order parameters: long-time mean-square displacements}
\label{sec:limit-cases-state-following-long-time-MSD}

The self-consistent equations in Eq.~\eqref{eq:statefol1} can be equivalently expressed in terms of two more physically transparent order parameters, corresponding to long-time limits of the MSDs.
For this, we first plug our ansatz~\eqref{eq-restricted-equilib-ansatz} into the single-particle effective process for ${\uu(t)}$
given by Eq.~\eqref{summary:uvec}, and obtain for long time $t$:
\beq
 \z \dot \uu(t) + \int_0^t \de s \, \b M_f(t-s) \, \dot\uu(s)
 	= - \argc{k_{\rm eff} \uu(t) - \sqrt{2} \bm{\X}^\io} + \sqrt{2} \bm{\X}^f(t)
 \ .
\eeq
Once again, as for Eq.~\eqref{eq-state-following-rewriting-vv}, this is the equilibrium dynamics associated to the restricted equilibrium distribution
\beq
\label{eq:Peqr-single-particle}
 P_{\rm req}(\uu; k_{\rm eff},\bm{\X}^\io)
 	=  \left(\frac{\b k_{\rm eff}}{2\pi} \right)^{d/2}   e^{- \frac\b2 k_{\rm eff} (\uu - \sqrt{2}\, \bm{\X}^\io/k_{\rm eff})^2  }  
 \ .
\eeq
From this Gaussian distribution we can easily compute the long-time limits of the two distinct MSDs:
\beq\label{eq:MSDslongtime}
\begin{split}
 \DE_r 
 	&= \frac1d \lim_{t\to\io} \la | \xx(t) - \xx(0) |^2 \ra 
 	= \frac1d \overline{  \la | \uu |^2 \ra_{\rm req}}
 	= \frac{1}{\b k_{\rm eff}} + \frac{M_\io}{k_{\rm eff}^2} 
 \ , \\
 \DE 
	&= \frac1d \lim_{t\to\io} \lim_{|t-s|\to\io}  \la | \xx(t) - \xx(s) |^2 \ra 
	= \frac2d \overline{ [  \la | \uu |^2 \ra_{\rm req} - | \la \uu \ra_{\rm req} |^2 ] }
	=  \frac{2}{\b k_{\rm eff}}
 \ .
\end{split}
\eeq
Introducing for convenience
${\AE = \overline{| \la \uu \ra_{\rm req} |^2} = 2 \DE_r - \DE = 2 M_\io/k_{\rm eff}^2}$, 
one can therefore express the self-consistent equations as follows:
\beq
\label{eq-state-following-equa1}
\begin{split}
 \frac1{\DE} - \frac{\AE}{\DE^2}
 	= \frac{\b}2 ( k_{\rm eff} - \b M_\io )
 	& \stackrel{\eqref{eq:statefol1}}{=}
 		\frac{\b \r}{2d} \int \de\rr_0 \, e^{-\Bi v(\rr_0)}
		\overline{\la \Lap v(\rr_0 +\vv) - \b | \nabla v(\rr_0 +\vv) |^2 \ra_{\rm req}}
\\
	& \stackrel{\eqref{eq-moy-equilib-disorder-state-following}}{=}
		 \frac{\b \r}{2d} \int \de\rr_0 \, e^{-\Bi v(\rr_0)} e^{\frac{\AE}{2}\Lap   }
	\left[ \frac{ e^{\frac{\DE}{2}\Lap   } e^{-\b v(\rr_0 )} [\Lap v(\rr_0) - \b | \nabla v(\rr_0) |^2] }
{ e^{\frac{\DE}{2}\Lap   } e^{-\b v(\rr_0 )}}  \right]
\\
	& = -\frac{ \r}{d} \int \de\rr_0 \, e^{-\Bi v(\rr_0)}  e^{\frac{\AE}{2}\Lap }
	\left[ \frac{ \frac{\Lap}2 e^{\frac{\DE}{2}\Lap   }  e^{-\b v(\rr_0 )} }
{ e^{\frac{\DE}{2}\Lap   } e^{-\b v(\rr_0 )}}  \right]
\ ,
\end{split}
\eeq
and
\beq
\label{eq-state-following-equa2}
\begin{split}
 \frac1{\DE}
	=	\frac{\b k_{\rm eff}}2
	& \stackrel{\eqref{eq:statefol1}}{=}
		\frac{\b \r}{2d} \int \de\rr_0 \, e^{-\Bi v(\rr_0)}  \overline{\la \Lap v(\rr_0 +\vv) - \b | \nabla v(\rr_0 +\vv) |^2 \ra_{\rm req} + \b | \la \nabla v(\rr_0 +\vv) | \ra_{\rm req} |^2}
 \\
	& \stackrel{\eqref{eq-moy-equilib-disorder-state-following}}{=}
		 \frac{\b \r}{2d} \int \de\rr_0 \, e^{-\Bi v(\rr_0)}  e^{\frac{\AE}{2}\Lap   }
		 \argc{
		 \frac{ e^{\frac{\DE}{2}\Lap   } e^{-\b v(\rr_0 )} [\Lap v(\rr_0) - \b | \nabla v(\rr_0) |^2] }
		{ e^{\frac{\DE}{2}\Lap   } e^{-\b v(\rr_0 )}}  
		+ \b \left|  \frac{ e^{\frac{\DE}{2}\Lap   } e^{-\b v(\rr_0 )}  \nabla v(\rr_0)  }
		{ e^{\frac{\DE}{2}\Lap   } e^{-\b v(\rr_0 )}}   \right|^2
}
 \\
	& = -\frac{\r}{d} \int \de\rr_0 \, e^{-\Bi v(\rr_0)} e^{\frac{\AE}{2}\Lap   } \frac{\Lap}2\log
			\argc{ e^{\frac{\DE}2 \Lap} e^{-\b v(\rr_0)} }
 \ .
\end{split}
\eeq
From these two last equations one can access ${\lbrace \DE,\AE \rbrace}$ and thus the long-time limit MSDs ${\lbrace \DE, \DE_r \rbrace}$ or,
 equivalently, ${\lbrace k_{\rm eff},M_\infty \rbrace}$.
Physically, we recall that
$\DE_r$ is the MSD of each particle with respect to its initial position,
$\DE$ the MSD of the inter-particle distance fluctuation,
and $M_\infty$ the long-time force-force correlator (and characterises the metabasins distribution).

Eqs.~\eqref{eq-state-following-equa1}-\eqref{eq-state-following-equa2} 
may also be obtained by taking the derivatives with respect to $\AE$ and $\DE$ of the glass free energy
\beq
\label{eq-state-following-free-energy}
 -\b {\rm f}_g
 	= \frac{d}2 \left[ 1 + \log(\pi \DE) + \frac{\AE}{\DE} \right] + \frac\r2 \int \de \rr_0 \, e^{-\Bi v(\rr_0)} e^{\frac{\AE}2 \Lap} \log \left[ e^{\frac{\DE}2 \Lap} e^{-\b v(\rr_0)} \right]
\ .
\eeq
In the ${d\to\io}$ limit, this free energy and the corresponding equations
are precisely equivalent to the ones obtained from the state-following replica method~\cite{RUYZ15}; the vectorial derivation in the replica context will be presented in Ref.~\cite{SZ2666}.
Note that when ${T=\Ti}$, \textit{i.e.}~when the system is prepared and let evolve in the same equilibrium, one can check from Eq.~\eqref{eq-state-following-equa1} that ${\DE = \DE_r = \AE}$ and one obtains a single equation for either $\DE$ or $M_\io$.
This equation provides the mean-field dynamical glass transition of the $d$-dimensional equilibrium liquid, and it is equivalent to the one obtained by the replica method in Ref.~\cite{PZ10} and from the dynamics by Szamel~\cite{Sz17}.
The transition occurs when by tuning the temperature, a non-zero solution for $M_\io$ or $\DE$ appears discontinuously.
Note that correspondingly, one has ${ k_{\rm eff} = k_\io - \b M_f(0) \neq 0}$, which implies that the effective potential~\eqref{eq:veffdef} becomes confining,
while for ${M_\io=0}$ one has ${ k_{\rm eff} =0}$, the potential is not confining and the dynamics is diffusive (${\DE=\io}$), \textit{i.e.}~there is no well-formed metabasin.
States prepared at $\Ti$ such that the dynamics is arrested can be followed at ${T\neq \Ti}$ while keeping a non-zero solution for ${\lbrace k_{\rm eff},M_\infty \rbrace}$ and a confining potential, but at too high $T$ again the solution disappears discontinuously giving rise to a glass spinodal~\cite{RUYZ15}.

\subsubsection{Colored `equilibrium' noise and constant random forces}
\label{sec:limit-cases-state-following-tuning-persistence-time}

In the general setting in Sec.~\ref{sec:setting}, we allow for a colored noise (${\Gamma_C \neq 0}$) and a non-local friction kernel (${\Gamma_R \neq 0}$).
There are in fact two cases to which the above state-following results can be generalised straightforwardly:
on the one hand, the case of non-zero `equilibrium' friction and noise kernels, \textit{i.e.}~satisfying ${\Gamma_R^{\rm{eq}}(t-s)=\beta \theta(t-s) \Gamma_C^{\rm{eq}}(t-s)}$ 
and decaying to zero on the restricted-equilibration timescale, similarly to the previous behaviour of the pair of transient memory kernels $M_{f,R}$.
On the other hand, one can discuss the case of constant random forces (\textit{i.e.}~infinite persistence time) with the time-independent noise kernel ${\Gamma_C(t,s) = f_0^2}$.

We can treat these two cases conjointly by considering
\beq
\label{eq-state-following-generalised-kernels}
 \Gamma_C(t-s) = \Gamma_C^{\rm{eq}}(t-s) + f_0^2
 \, , \qquad
 \Gamma_R(t-s) = \Gamma_R^{\rm{eq}}(t-s)
 \ .
\eeq
Indeed, coming back to the generic long-time ansatz in Eq.~\eqref{eq-restricted-equilib-ansatz}, we can simply modify the noise distributions:
\beq
\label{eq-restricted-equilib-ansatz-bis}
\left\lbrace \begin{array}{l}
 	\moy{ \X^f_{\m}(t)}_{\text{req}, \bm{\X}^\infty} = 0
 	\, , \quad
 	\moy{ \X^f_{\m}(t) \X^f_{\n}(s) }_{\text{req}, \bm{\X}^\infty}
 		= \d_{\m\n} \left[ T \z \d(t-s) + \frac12 M_f(t-s) + \frac12 \Gamma_C^{\rm{eq}}(t-s) \right]
 	\ , \\ \\
 	\overline{ \X^\io_{\m}}=0
 	\, , \quad
 	\overline{ \X^\io_{\m} \X^\io_{\n}}
	 	= \d_{\m\n}  \frac12 \argp{M_\io + f_0^2}
 	\ ,
 \end{array} \right.
\eeq
which modifies the long-time stochastic process for $\vv(t)$ as follows:
\beq
 \frac\z2 \dot \vv(t) + \frac{\b}2 \int_0^t \de s \, \argc{M_f(t-s) + \Gamma_C^{\rm{eq}}(t-s) } \dot\vv(s)
 = - \argc{ \nabla v (\rr_0 + \vv(t)) + \frac12 \argp{k_{\infty}- \b M_f(0)} \vv(t) - \bm{\X}^\io} + \bm{\X}^f(t)
 \ .
\eeq
This process has exactly the same (normalised) probability distribution ${P_{\rm req}(\vv;\rr_0,k_{\rm eff}, \bm{\X}^\io)}$ as in Eq.~\eqref{eq:Peqr} for a given ${\bm{\X}^\io}$,
with the same ${k_{\rm eff} = k_\io - \b M_f(0)}$, and similarly for the single-particle stochastic process in Eq.~\eqref{eq:Peqr-single-particle}.
In other words, our colored `equilibrium' noise does not modify the restricted equilibrium distribution of the glass state reached by the dynamics (as expected, by definition).
However, the metabasins characterised by $\bm{\X}^\io$ have a modified distribution because of the constant random forces, so the following combined average is slightly adjusted with respect to Eq.~\eqref{eq-moy-equilib-disorder-state-following}, by replacing 
${M_\infty \to M_\infty + f_0^2}$:
\beq
\label{eq-moy-equilib-disorder-state-following-bis}
 \overline{\moy{ f(\rr_0 +\vv) }_{\rm req}^\alpha}
	= e^{\frac{M_\io + f_0^2}{k_{\rm eff}^2}\Lap } \arga{\argc{
 \frac{ e^{\frac{1}{\b k_{\rm eff}}\Lap   } e^{-\b v(\rr_0 )} f(\rr_0 ) }
{ e^{\frac{1}{\b k_{\rm eff}}\Lap   } e^{-\b v(\rr_0 )}}  }^\alpha}
 \ .
\eeq
The self-consistent equations for ${\lbrace k^\io, M_\io, M_f(0)\rbrace}$ are the same as in Eq.~\eqref{eq:statefol1}, though they now depend implicitly on ${M_\infty + f_0^2}$ via the statistical average over $\bm{\Xi}^{\infty}$ (denoted by the overline).

As a result the long-time limits of the MSDs, Eq.~\eqref{eq:MSDslongtime}, are modified but only $\DE_r$ is affected:
\beq
\label{eq-state-following-equa0-bis}
\DE_r = \frac{1}{\b k_{\rm eff}} + \frac{M_\io + f_0^2}{k_{\rm eff}^2}
 \, , \qquad
 \DE = \frac{2}{\b k_{\rm eff}}
 \, , \qquad
 \AE = \overline{| \la \uu \ra_{\rm req} |^2} = 2 \DE_r - \DE \ ,
\eeq
and correspondingly their self-consistent coupled equation~\eqref{eq-state-following-equa1} is slightly altered as follows:
\beq
\label{eq-state-following-equa1-constant-forces}
 \frac1{\DE} - \frac{\AE}{\DE^2}
 	= \frac{\b}2 ( k_{\rm eff} - \b M_\io - \b f_0^2)
	= - \frac{\beta^2}{2} f_0^2
	-\frac{ \r}{d} \int \de\rr_0 \, e^{-\Bi v(\rr_0)}  e^{\frac{\AE}{2}\Lap }
	\left[ \frac{ \frac{\Lap}2 e^{\frac{\DE}{2}\Lap   }  e^{-\b v(\rr_0 )} }
{ e^{\frac{\DE}{2}\Lap   } e^{-\b v(\rr_0 )}}  \right]
\ ,
\eeq
while Eq.~\eqref{eq-state-following-equa2} for ${1/\DE}$ remains unchanged.

For a given set of inputs ${\lbrace \rho, \beta_0, v(\rr) \rbrace}$, the values of the long-time MSDs ${\lbrace \DE, \DE_r  \rbrace}$ (or ${\lbrace \DE, \AE  \rbrace}$) are thus fixed by almost the same equations as for a white `equilibrium' noise, the only modifications being the addition of $f_0^2$ in Eqs.~\eqref{eq-state-following-equa0-bis}-\eqref{eq-state-following-equa1-constant-forces}.
So, provided that the ansatz given by Eqs.~\eqref{eq-restricted-equilib-ansatz} and~\eqref{eq-restricted-equilib-ansatz-bis} for the short- \textit{versus} long-time decomposition of the kernels is valid, then the system self-consistently fixes the MSDs,
and consequently the variance of long-time forces
${\overline{\bm{\Xi}^{\infty} \cdot \bm{\Xi}^{\infty}}=\frac{d}{2} (M_{\infty} + f_0^2)}$
and the metabasin curvature ${k_{\rm eff} = k_\io - \b M_f(0)}$.
One can in particular start by a glass state at $\Ti$ with finite ${\lbrace \DE, \DE_r  \rbrace}$ and progressively increase the applied random force $f_0$.
For small $f_0$, the solution for ${\lbrace \DE, \DE_r  \rbrace}$ remains finite and describes the elastic response to the solid to the applied force~\cite{RUYZ15}. Upon increasing $f_0$, at some point an instability occurs signaling the yielding of the solid~\cite{RUYZ15}.
Beyond that value of the force, the dynamics does not admit anymore a confined solution, and diffusion appears together with ${M_\io=0}$.
The investigation of the dynamics around this yielding transition is a very interesting potential application of our dynamical equations.

\section{Conclusions}
\label{sec-conclusion}

In this paper, we have obtained the effective mean-field dynamics that describes exactly the generalised Langevin dynamics of a many-body system of interacting particles, in the joint infinite-dimensional and thermodynamic limit.
We have derived these Dynamical Mean-Field Equations (DMFE) in a very general setting including generic friction and noise kernels.
The generalisation to a finite fluid velocity (describing \textit{e.g.}~an applied strain) is straightforward but requires some additional technicalities since statistical isotropy is then broken, that will be presented in the companion paper~\cite{AMZ19bis}.
As such, our DMFE can be applied to model several out-of-equilibrium situations,
ranging from active matter to the mechanical properties of amorphous solids or the micro-rheology of yield stress fluids.
They thus provide a versatile toolbox for addressing the infinite-dimensional exact benchmark of a given model, defined by the set ${\lbrace v(\rr),m,\zeta,T,\Gamma_R(t,s),\Gamma_C(t,s) \rbrace}$, as summarised by Sec.~\ref{sec:setting} and~\ref{sec:summary-results}.


We have used two complementary derivations, via a dynamical cavity method and via a path-integral approach in a supersymmetric form.
The former relied on a few assumptions made on high-dimensional physics, whereas the latter properly justified them as exact features of the ${d\to\infty}$ path-integral saddle point.
Physically, these key features can be summarised as follows.
In a dense regime, the particles remain essentially `close' to their initial positions, namely at a distance of ${\mathcal{O}(1/d)}$.
Consequently, the fluctuations of the inter-particle distances $\vv_{ij}(t)$ happen mostly in the `longitudinal' direction of ${\rr_{ij}(0)}$, and the `transverse' motion is diffusive.
In this context, the interactions between particles can be encoded in three independent kernels, which in turn define effective stochastic processes for the absolute and relative fluctuations of the particle positions:
${k(t)}$ is simply the mean divergence of local forces,
${M_C(t,t')}$ the force-force correlator,
and ${M_R(t,t')}$ the mean linear response of the local force.
These kernels are self-consistently defined as statistical averages over the initial condition
and over the dynamical trajectories of the effective stochastic processes.
Last, all parameters and observables must be properly rescaled with respect to the dimension, in order to have a proper high-dimensional limit.


As a first application of our results, we have checked that we recover the equilibrium dynamics of Refs.~\cite{MKZ16,KMZ16,Sz17}.
Moreover we have provided a dynamical derivation of the `state-following' equations describing the response of a glass to quasistatic perturbations
--~otherwise obtained by static approaches using replicas \cite{RUYZ15,SZ2666}~--
including the cases of quasistatic random forces.
However, this dynamical derivation strongly relies on an ansatz assuming a timescale separation between the long- \textit{versus} short-time components of the kernels, the latter contributing to a colored `equilibrium' noise satisfying a FDT relation.
Being able to go beyond this quasistatic assumption and to characterise a system truly out of equilibrium, as well as to compute the full time-dependence of the self-consistent kernels, will be the main challenge following our work.
Nevertheless, everything is at hand to numerically study our DMFE, for each specific setting of interest.


Two cases will be particularly interesting in that respect.
The first is the pure active matter case, trying to assess for instance if and how an active noise with a finite persistence time could modify the nature and temperature of the glass transition --~if there is such a transition~-- similarly to the study in Ref.~\cite{BK13} on the driven $p$-spin glass model.
The second is the constant random forces case,
which is conjectured to be equivalent to shear in infinite dimension.
Recently~\cite{LX18} it has been shown numerically that random forces of constant amplitude reproduce many features of shear in a two-dimensional harmonic particle system, regarding specifically the vicinity of the zero-temperature jamming transition;
a systematic investigation of our infinite-dimensional DMFE could provide an exact analytical benchmark to discuss for instance the slight discrepancies of the critical exponents measured in Ref.~\cite{LX18}, with respect to standard jamming in a shear flow.


Finally, as mentioned in the introduction, MCT is to date the unique microscopic theory which could account for a variety of out-of-equilibrium properties of strongly interacting particle systems at finite dimension.
With our results, we have now a reliable reference theory in the infinite-dimensional limit, 
so one important issue to address will be a more systematic exploration of the connections between MCT and our DFME, and further the understanding of the finite-dimensional corrections to our mean-field description.

\section*{Acknowledgments}

We would like to thank Giulio Biroli, Matthias Fuchs, Jorge Kurchan, Alessandro Manacorda, Grzegorz Szamel and Pierfrancesco Urbani
for fruitful discussions related to this work.
This project has received funding from the European Research Council (ERC) under the European Union Horizon 2020 research and innovation programme (grant agreement n. 723955 - GlassUniversality).





\begin{thebibliography}{106}
\expandafter\ifx\csname natexlab\endcsname\relax\def\natexlab#1{#1}\fi
\expandafter\ifx\csname bibnamefont\endcsname\relax
  \def\bibnamefont#1{#1}\fi
\expandafter\ifx\csname bibfnamefont\endcsname\relax
  \def\bibfnamefont#1{#1}\fi
\expandafter\ifx\csname citenamefont\endcsname\relax
  \def\citenamefont#1{#1}\fi
\expandafter\ifx\csname url\endcsname\relax
  \def\url#1{\texttt{#1}}\fi
\expandafter\ifx\csname urlprefix\endcsname\relax\def\urlprefix{URL }\fi
\providecommand{\bibinfo}[2]{#2}
\providecommand{\eprint}[2][]{\url{#2}}

\bibitem[{\citenamefont{Hansen and McDonald}(2013)}]{hansen}
\bibinfo{author}{\bibfnamefont{J.-P.} \bibnamefont{Hansen}} \bibnamefont{and}
  \bibinfo{author}{\bibfnamefont{I.~R.} \bibnamefont{McDonald}},
  \emph{\bibinfo{title}{Theory of Simple Liquids}}
  (\bibinfo{publisher}{Elsevier}, \bibinfo{year}{2013}).

\bibitem[{\citenamefont{Bonn et~al.}(2017)\citenamefont{Bonn, Denn, Berthier,
  Divoux, and Manneville}}]{BDBDM17}
\bibinfo{author}{\bibfnamefont{D.}~\bibnamefont{Bonn}},
  \bibinfo{author}{\bibfnamefont{M.~M.} \bibnamefont{Denn}},
  \bibinfo{author}{\bibfnamefont{L.}~\bibnamefont{Berthier}},
  \bibinfo{author}{\bibfnamefont{T.}~\bibnamefont{Divoux}}, \bibnamefont{and}
  \bibinfo{author}{\bibfnamefont{S.}~\bibnamefont{Manneville}},
  \bibinfo{journal}{Reviews of Modern Physics} \textbf{\bibinfo{volume}{89}},
  \bibinfo{pages}{035005} (\bibinfo{year}{2017}).

\bibitem[{\citenamefont{Marchetti et~al.}(2013)\citenamefont{Marchetti, Joanny,
  Ramaswamy, Liverpool, Prost, Rao, and Simha}}]{MJRL13}
\bibinfo{author}{\bibfnamefont{M.~C.} \bibnamefont{Marchetti}},
  \bibinfo{author}{\bibfnamefont{J.-F.} \bibnamefont{Joanny}},
  \bibinfo{author}{\bibfnamefont{S.}~\bibnamefont{Ramaswamy}},
  \bibinfo{author}{\bibfnamefont{T.~B.} \bibnamefont{Liverpool}},
  \bibinfo{author}{\bibfnamefont{J.}~\bibnamefont{Prost}},
  \bibinfo{author}{\bibfnamefont{M.}~\bibnamefont{Rao}}, \bibnamefont{and}
  \bibinfo{author}{\bibfnamefont{R.~A.} \bibnamefont{Simha}},
  \bibinfo{journal}{Reviews of Modern Physics} \textbf{\bibinfo{volume}{85}},
  \bibinfo{pages}{1143} (\bibinfo{year}{2013}).

\bibitem[{\citenamefont{Rodney et~al.}(2011)\citenamefont{Rodney, Tanguy, and
  Vandembroucq}}]{RTV11}
\bibinfo{author}{\bibfnamefont{D.}~\bibnamefont{Rodney}},
  \bibinfo{author}{\bibfnamefont{A.}~\bibnamefont{Tanguy}}, \bibnamefont{and}
  \bibinfo{author}{\bibfnamefont{D.}~\bibnamefont{Vandembroucq}},
  \bibinfo{journal}{Modelling and Simulation in Materials Science and
  Engineering} \textbf{\bibinfo{volume}{19}}, \bibinfo{pages}{083001}
  (\bibinfo{year}{2011}).

\bibitem[{\citenamefont{G{\"o}tze}(2009)}]{Go09}
\bibinfo{author}{\bibfnamefont{W.}~\bibnamefont{G{\"o}tze}},
  \emph{\bibinfo{title}{Complex dynamics of glass-forming liquids: A
  mode-coupling theory}}, vol. \bibinfo{volume}{143} (\bibinfo{publisher}{OUP,
  USA}, \bibinfo{year}{2009}).

\bibitem[{\citenamefont{Reichman and Charbonneau}(2005)}]{reichman2005mode}
\bibinfo{author}{\bibfnamefont{D.~R.} \bibnamefont{Reichman}} \bibnamefont{and}
  \bibinfo{author}{\bibfnamefont{P.}~\bibnamefont{Charbonneau}},
  \bibinfo{journal}{Journal of Statistical Mechanics: Theory and Experiment}
  \textbf{\bibinfo{volume}{2005}}, \bibinfo{pages}{P05013}
  (\bibinfo{year}{2005}).

\bibitem[{\citenamefont{Janssen}(2018)}]{janssen2018mode}
\bibinfo{author}{\bibfnamefont{L.}~\bibnamefont{Janssen}},
  \bibinfo{journal}{arXiv:1806.01369}  (\bibinfo{year}{2018}).

\bibitem[{\citenamefont{Cavagna}(2009)}]{Ca09}
\bibinfo{author}{\bibfnamefont{A.}~\bibnamefont{Cavagna}},
  \bibinfo{journal}{Physics Reports} \textbf{\bibinfo{volume}{476}},
  \bibinfo{pages}{51} (\bibinfo{year}{2009}).

\bibitem[{\citenamefont{G\"{o}tze}(1999)}]{Go99}
\bibinfo{author}{\bibfnamefont{W.}~\bibnamefont{G\"{o}tze}},
  \bibinfo{journal}{Journal of Physics: Condensed Matter}
  \textbf{\bibinfo{volume}{11}}, \bibinfo{pages}{A1} (\bibinfo{year}{1999}).

\bibitem[{\citenamefont{Kob and Andersen}(1995{\natexlab{a}})}]{KA95a}
\bibinfo{author}{\bibfnamefont{W.}~\bibnamefont{Kob}} \bibnamefont{and}
  \bibinfo{author}{\bibfnamefont{H.~C.} \bibnamefont{Andersen}},
  \bibinfo{journal}{Physical Review E} \textbf{\bibinfo{volume}{51}},
  \bibinfo{pages}{4626} (\bibinfo{year}{1995}{\natexlab{a}}).

\bibitem[{\citenamefont{Kob and Andersen}(1995{\natexlab{b}})}]{KA95b}
\bibinfo{author}{\bibfnamefont{W.}~\bibnamefont{Kob}} \bibnamefont{and}
  \bibinfo{author}{\bibfnamefont{H.~C.} \bibnamefont{Andersen}},
  \bibinfo{journal}{Physical Review E} \textbf{\bibinfo{volume}{52}},
  \bibinfo{pages}{4134} (\bibinfo{year}{1995}{\natexlab{b}}).

\bibitem[{\citenamefont{Fabbian et~al.}(1999)\citenamefont{Fabbian, G{\"o}tze,
  Sciortino, Tartaglia, and Thiery}}]{fabbian1999}
\bibinfo{author}{\bibfnamefont{L.}~\bibnamefont{Fabbian}},
  \bibinfo{author}{\bibfnamefont{W.}~\bibnamefont{G{\"o}tze}},
  \bibinfo{author}{\bibfnamefont{F.}~\bibnamefont{Sciortino}},
  \bibinfo{author}{\bibfnamefont{P.}~\bibnamefont{Tartaglia}},
  \bibnamefont{and} \bibinfo{author}{\bibfnamefont{F.}~\bibnamefont{Thiery}},
  \bibinfo{journal}{Physical Review E} \textbf{\bibinfo{volume}{59}},
  \bibinfo{pages}{R1347} (\bibinfo{year}{1999}).

\bibitem[{\citenamefont{Dawson et~al.}(2000)\citenamefont{Dawson, Foffi, Fuchs,
  G{\"o}tze, Sciortino, Sperl, Tartaglia, Voigtmann, and
  Zaccarelli}}]{dawson2000}
\bibinfo{author}{\bibfnamefont{K.}~\bibnamefont{Dawson}},
  \bibinfo{author}{\bibfnamefont{G.}~\bibnamefont{Foffi}},
  \bibinfo{author}{\bibfnamefont{M.}~\bibnamefont{Fuchs}},
  \bibinfo{author}{\bibfnamefont{W.}~\bibnamefont{G{\"o}tze}},
  \bibinfo{author}{\bibfnamefont{F.}~\bibnamefont{Sciortino}},
  \bibinfo{author}{\bibfnamefont{M.}~\bibnamefont{Sperl}},
  \bibinfo{author}{\bibfnamefont{P.}~\bibnamefont{Tartaglia}},
  \bibinfo{author}{\bibfnamefont{T.}~\bibnamefont{Voigtmann}},
  \bibnamefont{and}
  \bibinfo{author}{\bibfnamefont{E.}~\bibnamefont{Zaccarelli}},
  \bibinfo{journal}{Physical Review E} \textbf{\bibinfo{volume}{63}},
  \bibinfo{pages}{011401} (\bibinfo{year}{2000}).

\bibitem[{\citenamefont{Bergenholtz and Fuchs}(1999)}]{bergenholtz1999}
\bibinfo{author}{\bibfnamefont{J.}~\bibnamefont{Bergenholtz}} \bibnamefont{and}
  \bibinfo{author}{\bibfnamefont{M.}~\bibnamefont{Fuchs}},
  \bibinfo{journal}{Physical Review E} \textbf{\bibinfo{volume}{59}},
  \bibinfo{pages}{5706} (\bibinfo{year}{1999}).

\bibitem[{\citenamefont{Miyazaki and Reichman}(2002)}]{miyazaki2002molecular}
\bibinfo{author}{\bibfnamefont{K.}~\bibnamefont{Miyazaki}} \bibnamefont{and}
  \bibinfo{author}{\bibfnamefont{D.~R.} \bibnamefont{Reichman}},
  \bibinfo{journal}{Physical Review E} \textbf{\bibinfo{volume}{66}},
  \bibinfo{pages}{050501} (\bibinfo{year}{2002}).

\bibitem[{\citenamefont{Fuchs and Cates}(2002)}]{FC02}
\bibinfo{author}{\bibfnamefont{M.}~\bibnamefont{Fuchs}} \bibnamefont{and}
  \bibinfo{author}{\bibfnamefont{M.~E.} \bibnamefont{Cates}},
  \bibinfo{journal}{Physical Review Letters} \textbf{\bibinfo{volume}{89}},
  \bibinfo{pages}{248304} (\bibinfo{year}{2002}).

\bibitem[{\citenamefont{Miyazaki et~al.}(2006)\citenamefont{Miyazaki, Wyss,
  Weitz, and Reichman}}]{miyazaki2006nonlinear}
\bibinfo{author}{\bibfnamefont{K.}~\bibnamefont{Miyazaki}},
  \bibinfo{author}{\bibfnamefont{H.~M.} \bibnamefont{Wyss}},
  \bibinfo{author}{\bibfnamefont{D.~A.} \bibnamefont{Weitz}}, \bibnamefont{and}
  \bibinfo{author}{\bibfnamefont{D.~R.} \bibnamefont{Reichman}},
  \bibinfo{journal}{EPL (Europhysics Letters)} \textbf{\bibinfo{volume}{75}},
  \bibinfo{pages}{915} (\bibinfo{year}{2006}).

\bibitem[{\citenamefont{Brader et~al.}(2009)\citenamefont{Brader, Voigtmann,
  Fuchs, Larson, and Cates}}]{Br09}
\bibinfo{author}{\bibfnamefont{J.~M.} \bibnamefont{Brader}},
  \bibinfo{author}{\bibfnamefont{T.}~\bibnamefont{Voigtmann}},
  \bibinfo{author}{\bibfnamefont{M.}~\bibnamefont{Fuchs}},
  \bibinfo{author}{\bibfnamefont{R.~G.} \bibnamefont{Larson}},
  \bibnamefont{and} \bibinfo{author}{\bibfnamefont{M.~E.} \bibnamefont{Cates}},
  \bibinfo{journal}{Proceedings of the National Academy of Sciences} \textbf{\bibinfo{volume}{106}},
  \bibinfo{pages}{15186} (\bibinfo{year}{2009}).

\bibitem[{\citenamefont{Brader et~al.}(2012)\citenamefont{Brader, Cates, and
  Fuchs}}]{BCF12}
\bibinfo{author}{\bibfnamefont{J.~M.} \bibnamefont{Brader}},
  \bibinfo{author}{\bibfnamefont{M.~E.} \bibnamefont{Cates}}, \bibnamefont{and}
  \bibinfo{author}{\bibfnamefont{M.}~\bibnamefont{Fuchs}},
  \bibinfo{journal}{Physical Review E} \textbf{\bibinfo{volume}{86}},
  \bibinfo{pages}{021403} (\bibinfo{year}{2012}).

\bibitem[{\citenamefont{Szamel et~al.}(2015)\citenamefont{Szamel, Flenner, and
  Berthier}}]{SFB15}
\bibinfo{author}{\bibfnamefont{G.}~\bibnamefont{Szamel}},
  \bibinfo{author}{\bibfnamefont{E.}~\bibnamefont{Flenner}}, \bibnamefont{and}
  \bibinfo{author}{\bibfnamefont{L.}~\bibnamefont{Berthier}},
  \bibinfo{journal}{Physical Review E} \textbf{\bibinfo{volume}{91}},
  \bibinfo{pages}{062304} (\bibinfo{year}{2015}).
  
\bibitem[{\citenamefont{Nandi and Gov}(2017)}]{NG17}
\bibinfo{author}{\bibfnamefont{S.~K.} \bibnamefont{Nandi}} \bibnamefont{and}
  \bibinfo{author}{\bibfnamefont{N.~S.} \bibnamefont{Gov}},
  \bibinfo{journal}{Soft Matter} \textbf{\bibinfo{volume}{13}},
  \bibinfo{pages}{7609} (\bibinfo{year}{2017}).

\bibitem[{\citenamefont{Ikeda and Berthier}(2013)}]{IB13}
\bibinfo{author}{\bibfnamefont{A.}~\bibnamefont{Ikeda}} \bibnamefont{and}
  \bibinfo{author}{\bibfnamefont{L.}~\bibnamefont{Berthier}},
  \bibinfo{journal}{Physical Review E} \textbf{\bibinfo{volume}{88}},
  \bibinfo{pages}{052305} (\bibinfo{year}{2013}).

\bibitem[{\citenamefont{Schweizer}(1989)}]{schweizer1989microscopic}
\bibinfo{author}{\bibfnamefont{K.~S.} \bibnamefont{Schweizer}},
  \bibinfo{journal}{Journal of Chemical Physics}
  \textbf{\bibinfo{volume}{91}}, \bibinfo{pages}{5802} (\bibinfo{year}{1989}).

\bibitem[{\citenamefont{Schweizer and Saltzman}(2003)}]{schweizer2003entropic}
\bibinfo{author}{\bibfnamefont{K.~S.} \bibnamefont{Schweizer}}
  \bibnamefont{and} \bibinfo{author}{\bibfnamefont{E.~J.}
  \bibnamefont{Saltzman}}, \bibinfo{journal}{Journal of Chemical Physics}
  \textbf{\bibinfo{volume}{119}}, \bibinfo{pages}{1181} (\bibinfo{year}{2003}).

\bibitem[{\citenamefont{Bhattacharyya et~al.}(2008)\citenamefont{Bhattacharyya,
  Bagchi, and Wolynes}}]{BBW08}
\bibinfo{author}{\bibfnamefont{S.~M.} \bibnamefont{Bhattacharyya}},
  \bibinfo{author}{\bibfnamefont{B.}~\bibnamefont{Bagchi}}, \bibnamefont{and}
  \bibinfo{author}{\bibfnamefont{P.~G.} \bibnamefont{Wolynes}},
  \bibinfo{journal}{Proceedings of the National Academy of Sciences}
  \textbf{\bibinfo{volume}{105}}, \bibinfo{pages}{16077}
  (\bibinfo{year}{2008}).

\bibitem[{\citenamefont{Mari and Kurchan}(2011)}]{MK11}
\bibinfo{author}{\bibfnamefont{R.}~\bibnamefont{Mari}} \bibnamefont{and}
  \bibinfo{author}{\bibfnamefont{J.}~\bibnamefont{Kurchan}},
  \bibinfo{journal}{Journal of Chemical Physics} \textbf{\bibinfo{volume}{135}},
  \bibinfo{pages}{124504} (\bibinfo{year}{2011}).

\bibitem[{\citenamefont{Szamel}(2013)}]{szamel2013mode}
\bibinfo{author}{\bibfnamefont{G.}~\bibnamefont{Szamel}},
  \bibinfo{journal}{Progress of Theoretical and Experimental Physics}
  \textbf{\bibinfo{volume}{2013}} (\bibinfo{year}{2013}).

\bibitem[{\citenamefont{Kim et~al.}(2014)\citenamefont{Kim, Kawasaki, Jacquin,
  and van Wijland}}]{kim2014equilibrium}
\bibinfo{author}{\bibfnamefont{B.}~\bibnamefont{Kim}},
  \bibinfo{author}{\bibfnamefont{K.}~\bibnamefont{Kawasaki}},
  \bibinfo{author}{\bibfnamefont{H.}~\bibnamefont{Jacquin}}, \bibnamefont{and}
  \bibinfo{author}{\bibfnamefont{F.}~\bibnamefont{van Wijland}},
  \bibinfo{journal}{Physical Review E} \textbf{\bibinfo{volume}{89}},
  \bibinfo{pages}{012150} (\bibinfo{year}{2014}).

\bibitem[{\citenamefont{Rizzo and Voigtmann}(2015)}]{rizzo2015qualitative}
\bibinfo{author}{\bibfnamefont{T.}~\bibnamefont{Rizzo}} \bibnamefont{and}
  \bibinfo{author}{\bibfnamefont{T.}~\bibnamefont{Voigtmann}},
  \bibinfo{journal}{EPL (Europhysics Letters)} \textbf{\bibinfo{volume}{111}},
  \bibinfo{pages}{56008} (\bibinfo{year}{2015}).

\bibitem[{\citenamefont{Rizzo}(2016)}]{rizzo2016dynamical}
\bibinfo{author}{\bibfnamefont{T.}~\bibnamefont{Rizzo}},
  \bibinfo{journal}{Physical Review B} \textbf{\bibinfo{volume}{94}},
  \bibinfo{pages}{014202} (\bibinfo{year}{2016}).

\bibitem[{\citenamefont{Janssen and Reichman}(2015)}]{janssen2015microscopic}
\bibinfo{author}{\bibfnamefont{L.~M.} \bibnamefont{Janssen}} \bibnamefont{and}
  \bibinfo{author}{\bibfnamefont{D.~R.} \bibnamefont{Reichman}},
  \bibinfo{journal}{Physical Review Letters} \textbf{\bibinfo{volume}{115}},
  \bibinfo{pages}{205701} (\bibinfo{year}{2015}).

\bibitem[{\citenamefont{Leutheusser}(1984)}]{L84}
\bibinfo{author}{\bibfnamefont{E.}~\bibnamefont{Leutheusser}},
  \bibinfo{journal}{Physical Review A} \textbf{\bibinfo{volume}{29}},
  \bibinfo{pages}{2765} (\bibinfo{year}{1984}).

\bibitem[{\citenamefont{Bengtzelius et~al.}(1984)\citenamefont{Bengtzelius,
  Gotze, and Sjolander}}]{BGS84}
\bibinfo{author}{\bibfnamefont{U.}~\bibnamefont{Bengtzelius}},
  \bibinfo{author}{\bibfnamefont{W.}~\bibnamefont{Gotze}}, \bibnamefont{and}
  \bibinfo{author}{\bibfnamefont{A.}~\bibnamefont{Sjolander}},
  \bibinfo{journal}{Journal of Physics C: Solid State Physics}
  \textbf{\bibinfo{volume}{17}}, \bibinfo{pages}{5915} (\bibinfo{year}{1984}).

\bibitem[{\citenamefont{Kirkpatrick and Thirumalai}(1987{\natexlab{a}})}]{KT87}
\bibinfo{author}{\bibfnamefont{T.~R.} \bibnamefont{Kirkpatrick}}
  \bibnamefont{and}
  \bibinfo{author}{\bibfnamefont{D.}~\bibnamefont{Thirumalai}},
  \bibinfo{journal}{Physical Review Letters} \textbf{\bibinfo{volume}{58}},
  \bibinfo{pages}{2091} (\bibinfo{year}{1987}{\natexlab{a}}).

\bibitem[{\citenamefont{Kirkpatrick and
  Thirumalai}(1987{\natexlab{b}})}]{KT87b}
\bibinfo{author}{\bibfnamefont{T.~R.} \bibnamefont{Kirkpatrick}}
  \bibnamefont{and}
  \bibinfo{author}{\bibfnamefont{D.}~\bibnamefont{Thirumalai}},
  \bibinfo{journal}{Physical Review B} \textbf{\bibinfo{volume}{36}},
  \bibinfo{pages}{5388} (\bibinfo{year}{1987}{\natexlab{b}}).

\bibitem[{\citenamefont{Bouchaud et~al.}(1996)\citenamefont{Bouchaud,
  Cugliandolo, Kurchan, and M{\'e}zard}}]{BCKM96}
\bibinfo{author}{\bibfnamefont{J.-P.} \bibnamefont{Bouchaud}},
  \bibinfo{author}{\bibfnamefont{L.}~\bibnamefont{Cugliandolo}},
  \bibinfo{author}{\bibfnamefont{J.}~\bibnamefont{Kurchan}}, \bibnamefont{and}
  \bibinfo{author}{\bibfnamefont{M.}~\bibnamefont{M{\'e}zard}},
  \bibinfo{journal}{Physica A: Statistical Mechanics and its Applications}
  \textbf{\bibinfo{volume}{226}}, \bibinfo{pages}{243} (\bibinfo{year}{1996}).

\bibitem[{\citenamefont{Kirkpatrick et~al.}(1989)\citenamefont{Kirkpatrick,
  Thirumalai, and Wolynes}}]{KTW89}
\bibinfo{author}{\bibfnamefont{T.~R.} \bibnamefont{Kirkpatrick}},
  \bibinfo{author}{\bibfnamefont{D.}~\bibnamefont{Thirumalai}},
  \bibnamefont{and} \bibinfo{author}{\bibfnamefont{P.~G.}
  \bibnamefont{Wolynes}}, \bibinfo{journal}{Physical Review A}
  \textbf{\bibinfo{volume}{40}}, \bibinfo{pages}{1045} (\bibinfo{year}{1989}).

\bibitem[{\citenamefont{Lubchenko and Wolynes}(2007)}]{LW07}
\bibinfo{author}{\bibfnamefont{V.}~\bibnamefont{Lubchenko}} \bibnamefont{and}
  \bibinfo{author}{\bibfnamefont{P.~G.} \bibnamefont{Wolynes}},
  \bibinfo{journal}{Annual Review of Physical Chemistry}
  \textbf{\bibinfo{volume}{58}}, \bibinfo{pages}{235} (\bibinfo{year}{2007}).

\bibitem[{\citenamefont{Wolynes and Lubchenko}(2012)}]{WL12}
\bibinfo{editor}{\bibfnamefont{P.}~\bibnamefont{Wolynes}} \bibnamefont{and}
  \bibinfo{editor}{\bibfnamefont{V.}~\bibnamefont{Lubchenko}}, eds.,
  \emph{\bibinfo{title}{Structural Glasses and Supercooled Liquids: Theory,
  Experiment, and Applications}} (\bibinfo{publisher}{Wiley},
  \bibinfo{year}{2012}).

\bibitem[{\citenamefont{Berthier et~al.}(2011)\citenamefont{Berthier, Biroli,
  Bouchaud, Cipelletti, and van Saarloos}}]{BBBCS11}
\bibinfo{author}{\bibfnamefont{L.}~\bibnamefont{Berthier}},
  \bibinfo{author}{\bibfnamefont{G.}~\bibnamefont{Biroli}},
  \bibinfo{author}{\bibfnamefont{J.-P.} \bibnamefont{Bouchaud}},
  \bibinfo{author}{\bibfnamefont{L.}~\bibnamefont{Cipelletti}},
  \bibnamefont{and} \bibinfo{author}{\bibfnamefont{W.}~\bibnamefont{van
  Saarloos}}, \emph{\bibinfo{title}{Dynamical Heterogeneities and Glasses}}
  (\bibinfo{publisher}{Oxford University Press}, \bibinfo{year}{2011}).

\bibitem[{\citenamefont{Cugliandolo and Kurchan}(1993)}]{CK93}
\bibinfo{author}{\bibfnamefont{L.~F.} \bibnamefont{Cugliandolo}}
  \bibnamefont{and} \bibinfo{author}{\bibfnamefont{J.}~\bibnamefont{Kurchan}},
  \bibinfo{journal}{Physical Review Letters} \textbf{\bibinfo{volume}{71}},
  \bibinfo{pages}{173} (\bibinfo{year}{1993}).

\bibitem[{\citenamefont{Berthier et~al.}(2000)\citenamefont{Berthier, Barrat,
  and Kurchan}}]{BBK00}
\bibinfo{author}{\bibfnamefont{L.}~\bibnamefont{Berthier}},
  \bibinfo{author}{\bibfnamefont{J.-L.} \bibnamefont{Barrat}},
  \bibnamefont{and} \bibinfo{author}{\bibfnamefont{J.}~\bibnamefont{Kurchan}},
  \bibinfo{journal}{Physical Review E} \textbf{\bibinfo{volume}{61}},
  \bibinfo{pages}{5464} (\bibinfo{year}{2000}).

\bibitem[{\citenamefont{Berthier and Kurchan}(2013)}]{BK13}
\bibinfo{author}{\bibfnamefont{L.}~\bibnamefont{Berthier}} \bibnamefont{and}
  \bibinfo{author}{\bibfnamefont{J.}~\bibnamefont{Kurchan}},
  \bibinfo{journal}{Nature Physics} \textbf{\bibinfo{volume}{9}},
  \bibinfo{pages}{310} (\bibinfo{year}{2013}).

\bibitem[{\citenamefont{Kirkpatrick and Wolynes}(1987)}]{KW87}
\bibinfo{author}{\bibfnamefont{T.~R.} \bibnamefont{Kirkpatrick}}
  \bibnamefont{and} \bibinfo{author}{\bibfnamefont{P.~G.}
  \bibnamefont{Wolynes}}, \bibinfo{journal}{Physical Review A}
  \textbf{\bibinfo{volume}{35}}, \bibinfo{pages}{3072} (\bibinfo{year}{1987}).

\bibitem[{\citenamefont{Monasson}(1995)}]{Mo95}
\bibinfo{author}{\bibfnamefont{R.}~\bibnamefont{Monasson}},
  \bibinfo{journal}{Physical Review Letters} \textbf{\bibinfo{volume}{75}},
  \bibinfo{pages}{2847} (\bibinfo{year}{1995}).

\bibitem[{\citenamefont{Franz and Parisi}(1995)}]{FP95}
\bibinfo{author}{\bibfnamefont{S.}~\bibnamefont{Franz}} \bibnamefont{and}
  \bibinfo{author}{\bibfnamefont{G.}~\bibnamefont{Parisi}},
  \bibinfo{journal}{Journal de Physique I} \textbf{\bibinfo{volume}{5}},
  \bibinfo{pages}{1401} (\bibinfo{year}{1995}).

\bibitem[{\citenamefont{Barrat et~al.}(1996)\citenamefont{Barrat, Burioni, and
  M{\'e}zard}}]{BBM96}
\bibinfo{author}{\bibfnamefont{A.}~\bibnamefont{Barrat}},
  \bibinfo{author}{\bibfnamefont{R.}~\bibnamefont{Burioni}}, \bibnamefont{and}
  \bibinfo{author}{\bibfnamefont{M.}~\bibnamefont{M{\'e}zard}},
  \bibinfo{journal}{Journal of Physics A: Mathematical and General}
  \textbf{\bibinfo{volume}{29}}, \bibinfo{pages}{L81} (\bibinfo{year}{1996}).

\bibitem[{\citenamefont{Rainone et~al.}(2015)\citenamefont{Rainone, Urbani,
  Yoshino, and Zamponi}}]{RUYZ15}
\bibinfo{author}{\bibfnamefont{C.}~\bibnamefont{Rainone}},
  \bibinfo{author}{\bibfnamefont{P.}~\bibnamefont{Urbani}},
  \bibinfo{author}{\bibfnamefont{H.}~\bibnamefont{Yoshino}}, \bibnamefont{and}
  \bibinfo{author}{\bibfnamefont{F.}~\bibnamefont{Zamponi}},
  \bibinfo{journal}{Physical Review Letters} \textbf{\bibinfo{volume}{114}},
  \bibinfo{pages}{015701} (\bibinfo{year}{2015}).

\bibitem[{\citenamefont{Krzakala and Zdeborov{\'a}}(2010)}]{KZ10}
\bibinfo{author}{\bibfnamefont{F.}~\bibnamefont{Krzakala}} \bibnamefont{and}
  \bibinfo{author}{\bibfnamefont{L.}~\bibnamefont{Zdeborov{\'a}}},
  \bibinfo{journal}{EPL (Europhysics Letters)} \textbf{\bibinfo{volume}{90}}, \bibinfo{pages}{66002}
  (\bibinfo{year}{2010}).

\bibitem[{\citenamefont{Zdeborov\'a and Krzakala}(2010)}]{KZ10b}
\bibinfo{author}{\bibfnamefont{L.}~\bibnamefont{Zdeborov\'a}} \bibnamefont{and}
  \bibinfo{author}{\bibfnamefont{F.}~\bibnamefont{Krzakala}},
  \bibinfo{journal}{Physical Review B} \textbf{\bibinfo{volume}{81}},
  \bibinfo{pages}{224205} (\bibinfo{year}{2010}).

\bibitem[{\citenamefont{Ikeda and Miyazaki}(2010)}]{IM10}
\bibinfo{author}{\bibfnamefont{A.}~\bibnamefont{Ikeda}} \bibnamefont{and}
  \bibinfo{author}{\bibfnamefont{K.}~\bibnamefont{Miyazaki}},
  \bibinfo{journal}{Physical Review Letters} \textbf{\bibinfo{volume}{104}},
  \bibinfo{pages}{255704} (\bibinfo{year}{2010}).

\bibitem[{\citenamefont{Schmid and Schilling}(2010)}]{SS10}
\bibinfo{author}{\bibfnamefont{B.}~\bibnamefont{Schmid}} \bibnamefont{and}
  \bibinfo{author}{\bibfnamefont{R.}~\bibnamefont{Schilling}},
  \bibinfo{journal}{Physical Review E} \textbf{\bibinfo{volume}{81}},
  \bibinfo{pages}{041502} (\bibinfo{year}{2010}).

\bibitem[{\citenamefont{Charbonneau et~al.}(2011)\citenamefont{Charbonneau,
  Ikeda, Parisi, and Zamponi}}]{CIPZ11}
\bibinfo{author}{\bibfnamefont{P.}~\bibnamefont{Charbonneau}},
  \bibinfo{author}{\bibfnamefont{A.}~\bibnamefont{Ikeda}},
  \bibinfo{author}{\bibfnamefont{G.}~\bibnamefont{Parisi}}, \bibnamefont{and}
  \bibinfo{author}{\bibfnamefont{F.}~\bibnamefont{Zamponi}},
  \bibinfo{journal}{Physical Review Letters} \textbf{\bibinfo{volume}{107}},
  \bibinfo{pages}{185702} (\bibinfo{year}{2011}).

\bibitem[{\citenamefont{Parisi and Zamponi}(2006)}]{PZ06a}
\bibinfo{author}{\bibfnamefont{G.}~\bibnamefont{Parisi}} \bibnamefont{and}
  \bibinfo{author}{\bibfnamefont{F.}~\bibnamefont{Zamponi}},
  \bibinfo{journal}{Journal of Statistical Mechanics: Theory and Experiment}
  \textbf{\bibinfo{volume}{2006}}, \bibinfo{pages}{P03017}
  (\bibinfo{year}{2006}).

\bibitem[{\citenamefont{Parisi and Zamponi}(2010)}]{PZ10}
\bibinfo{author}{\bibfnamefont{G.}~\bibnamefont{Parisi}} \bibnamefont{and}
  \bibinfo{author}{\bibfnamefont{F.}~\bibnamefont{Zamponi}},
  \bibinfo{journal}{Review Modern Physics} \textbf{\bibinfo{volume}{82}},
  \bibinfo{pages}{789} (\bibinfo{year}{2010}).

\bibitem[{\citenamefont{Kurchan et~al.}(2012)\citenamefont{Kurchan, Parisi, and
  Zamponi}}]{KPZ12}
\bibinfo{author}{\bibfnamefont{J.}~\bibnamefont{Kurchan}},
  \bibinfo{author}{\bibfnamefont{G.}~\bibnamefont{Parisi}}, \bibnamefont{and}
  \bibinfo{author}{\bibfnamefont{F.}~\bibnamefont{Zamponi}},
  \bibinfo{journal}{Journal of Statistical Mechanics: Theory and Experiment}
  \textbf{\bibinfo{volume}{2012}}, \bibinfo{pages}{P10012}
  (\bibinfo{year}{2012}).

\bibitem[{\citenamefont{Kurchan et~al.}(2013)\citenamefont{Kurchan, Parisi,
  Urbani, and Zamponi}}]{KPUZ13}
\bibinfo{author}{\bibfnamefont{J.}~\bibnamefont{Kurchan}},
  \bibinfo{author}{\bibfnamefont{G.}~\bibnamefont{Parisi}},
  \bibinfo{author}{\bibfnamefont{P.}~\bibnamefont{Urbani}}, \bibnamefont{and}
  \bibinfo{author}{\bibfnamefont{F.}~\bibnamefont{Zamponi}},
  \bibinfo{journal}{Journal of Physical Chemistry B} \textbf{\bibinfo{volume}{117}},
  \bibinfo{pages}{12979} (\bibinfo{year}{2013}).

\bibitem[{\citenamefont{Charbonneau et~al.}(2017)\citenamefont{Charbonneau,
  Kurchan, Parisi, Urbani, and Zamponi}}]{CKPUZ17}
\bibinfo{author}{\bibfnamefont{P.}~\bibnamefont{Charbonneau}},
  \bibinfo{author}{\bibfnamefont{J.}~\bibnamefont{Kurchan}},
  \bibinfo{author}{\bibfnamefont{G.}~\bibnamefont{Parisi}},
  \bibinfo{author}{\bibfnamefont{P.}~\bibnamefont{Urbani}}, \bibnamefont{and}
  \bibinfo{author}{\bibfnamefont{F.}~\bibnamefont{Zamponi}},
  \bibinfo{journal}{Annual Review of Condensed Matter Physics}
  \textbf{\bibinfo{volume}{8}}, \bibinfo{pages}{265} (\bibinfo{year}{2017}).

\bibitem[{\citenamefont{Kurchan}(1992)}]{Ku92}
\bibinfo{author}{\bibfnamefont{J.}~\bibnamefont{Kurchan}},
  \bibinfo{journal}{Journal de Physique I} \textbf{\bibinfo{volume}{2}},
  \bibinfo{pages}{1333} (\bibinfo{year}{1992}).

\bibitem[{\citenamefont{Kurchan}(2003)}]{Ku03}
\bibinfo{author}{\bibfnamefont{J.}~\bibnamefont{Kurchan}},
  \bibinfo{journal}{Markov Processes and Related Fields}
  \textbf{\bibinfo{volume}{9}}, \bibinfo{pages}{243} (\bibinfo{year}{2003}).

\bibitem[{\citenamefont{Parisi and Rizzo}(2013)}]{PR12}
\bibinfo{author}{\bibfnamefont{G.}~\bibnamefont{Parisi}} \bibnamefont{and}
  \bibinfo{author}{\bibfnamefont{T.}~\bibnamefont{Rizzo}},
  \bibinfo{journal}{Physical Review E} \textbf{\bibinfo{volume}{87}},
  \bibinfo{pages}{012101} (\bibinfo{year}{2013}).

\bibitem[{\citenamefont{Caltagirone et~al.}(2012)\citenamefont{Caltagirone,
  Ferrari, Leuzzi, Parisi, Ricci-Tersenghi, and Rizzo}}]{CFLPRR12}
\bibinfo{author}{\bibfnamefont{F.}~\bibnamefont{Caltagirone}},
  \bibinfo{author}{\bibfnamefont{U.}~\bibnamefont{Ferrari}},
  \bibinfo{author}{\bibfnamefont{L.}~\bibnamefont{Leuzzi}},
  \bibinfo{author}{\bibfnamefont{G.}~\bibnamefont{Parisi}},
  \bibinfo{author}{\bibfnamefont{F.}~\bibnamefont{Ricci-Tersenghi}},
  \bibnamefont{and} \bibinfo{author}{\bibfnamefont{T.}~\bibnamefont{Rizzo}},
  \bibinfo{journal}{Physical Review Letters} \textbf{\bibinfo{volume}{108}},
  \bibinfo{pages}{085702} (\bibinfo{year}{2012}).

\bibitem[{\citenamefont{Kurchan et~al.}(2016)\citenamefont{Kurchan, Maimbourg,
  and Zamponi}}]{KMZ16}
\bibinfo{author}{\bibfnamefont{J.}~\bibnamefont{Kurchan}},
  \bibinfo{author}{\bibfnamefont{T.}~\bibnamefont{Maimbourg}},
  \bibnamefont{and} \bibinfo{author}{\bibfnamefont{F.}~\bibnamefont{Zamponi}},
  \bibinfo{journal}{Journal of Statistical Mechanics: Theory and Experiment}
  \textbf{\bibinfo{volume}{2016}}, \bibinfo{pages}{033210}
  (\bibinfo{year}{2016}), \eprint{arXiv:1512.02186}.

\bibitem[{\citenamefont{Maimbourg et~al.}(2016)\citenamefont{Maimbourg,
  Kurchan, and Zamponi}}]{MKZ16}
\bibinfo{author}{\bibfnamefont{T.}~\bibnamefont{Maimbourg}},
  \bibinfo{author}{\bibfnamefont{J.}~\bibnamefont{Kurchan}}, \bibnamefont{and}
  \bibinfo{author}{\bibfnamefont{F.}~\bibnamefont{Zamponi}},
  \bibinfo{journal}{Physical Review Letters} \textbf{\bibinfo{volume}{116}},
  \bibinfo{pages}{015902} (\bibinfo{year}{2016}).

\bibitem[{\citenamefont{Szamel}(2017)}]{Sz17}
\bibinfo{author}{\bibfnamefont{G.}~\bibnamefont{Szamel}},
  \bibinfo{journal}{Physical Review Letters} \textbf{\bibinfo{volume}{119}},
  \bibinfo{pages}{155502} (\bibinfo{year}{2017}).

\bibitem[{\citenamefont{Georges et~al.}(1996)\citenamefont{Georges, Kotliar,
  Krauth, and Rozenberg}}]{GKKR96}
\bibinfo{author}{\bibfnamefont{A.}~\bibnamefont{Georges}},
  \bibinfo{author}{\bibfnamefont{G.}~\bibnamefont{Kotliar}},
  \bibinfo{author}{\bibfnamefont{W.}~\bibnamefont{Krauth}}, \bibnamefont{and}
  \bibinfo{author}{\bibfnamefont{M.~J.} \bibnamefont{Rozenberg}},
  \bibinfo{journal}{Reviews of Modern Physics} \textbf{\bibinfo{volume}{68}},
  \bibinfo{pages}{13} (\bibinfo{year}{1996}).

\bibitem[{\citenamefont{Urbani and Zamponi}(2017)}]{UZ17}
\bibinfo{author}{\bibfnamefont{P.}~\bibnamefont{Urbani}} \bibnamefont{and}
  \bibinfo{author}{\bibfnamefont{F.}~\bibnamefont{Zamponi}},
  \bibinfo{journal}{Physical Review Letters} \textbf{\bibinfo{volume}{118}},
  \bibinfo{pages}{038001} (\bibinfo{year}{2017}).

\bibitem[{\citenamefont{Sellitto and
  Zamponi}(2013)}]{sellitto2013thermodynamic}
\bibinfo{author}{\bibfnamefont{M.}~\bibnamefont{Sellitto}} \bibnamefont{and}
  \bibinfo{author}{\bibfnamefont{F.}~\bibnamefont{Zamponi}},
  \bibinfo{journal}{EPL (Europhysics Letters)} \textbf{\bibinfo{volume}{103}},
  \bibinfo{pages}{46005} (\bibinfo{year}{2013}).

\bibitem[{\citenamefont{Altieri et~al.}(2018)\citenamefont{Altieri, Urbani, and
  Zamponi}}]{altieri2018microscopic}
\bibinfo{author}{\bibfnamefont{A.}~\bibnamefont{Altieri}},
  \bibinfo{author}{\bibfnamefont{P.}~\bibnamefont{Urbani}}, \bibnamefont{and}
  \bibinfo{author}{\bibfnamefont{F.}~\bibnamefont{Zamponi}},
  \bibinfo{journal}{Physical Review Letters} \textbf{\bibinfo{volume}{121}},
  \bibinfo{pages}{185503} (\bibinfo{year}{2018}).

\bibitem[{\citenamefont{Charbonneau et~al.}(2014)\citenamefont{Charbonneau,
  Kurchan, Parisi, Urbani, and Zamponi}}]{nature}
\bibinfo{author}{\bibfnamefont{P.}~\bibnamefont{Charbonneau}},
  \bibinfo{author}{\bibfnamefont{J.}~\bibnamefont{Kurchan}},
  \bibinfo{author}{\bibfnamefont{G.}~\bibnamefont{Parisi}},
  \bibinfo{author}{\bibfnamefont{P.}~\bibnamefont{Urbani}}, \bibnamefont{and}
  \bibinfo{author}{\bibfnamefont{F.}~\bibnamefont{Zamponi}},
  \bibinfo{journal}{Nature Communications} \textbf{\bibinfo{volume}{5}},
  \bibinfo{pages}{3725} (\bibinfo{year}{2014}).

\bibitem[{\citenamefont{Mangeat and Zamponi}(2016)}]{MZ16}
\bibinfo{author}{\bibfnamefont{M.}~\bibnamefont{Mangeat}} \bibnamefont{and}
  \bibinfo{author}{\bibfnamefont{F.}~\bibnamefont{Zamponi}},
  \bibinfo{journal}{Physical Review E} \textbf{\bibinfo{volume}{93}},
  \bibinfo{pages}{012609} (\bibinfo{year}{2016}).

\bibitem[{\citenamefont{Charbonneau et~al.}(2012)\citenamefont{Charbonneau,
  Ikeda, Parisi, and Zamponi}}]{CIPZ12}
\bibinfo{author}{\bibfnamefont{P.}~\bibnamefont{Charbonneau}},
  \bibinfo{author}{\bibfnamefont{A.}~\bibnamefont{Ikeda}},
  \bibinfo{author}{\bibfnamefont{G.}~\bibnamefont{Parisi}}, \bibnamefont{and}
  \bibinfo{author}{\bibfnamefont{F.}~\bibnamefont{Zamponi}},
  \bibinfo{journal}{Proceedings of the National Academy of Sciences}
  \textbf{\bibinfo{volume}{109}}, \bibinfo{pages}{13939}
  (\bibinfo{year}{2012}).

\bibitem[{\citenamefont{Charbonneau et~al.}(2015)\citenamefont{Charbonneau,
  Corwin, Parisi, and Zamponi}}]{CCPZ15}
\bibinfo{author}{\bibfnamefont{P.}~\bibnamefont{Charbonneau}},
  \bibinfo{author}{\bibfnamefont{E.~I.} \bibnamefont{Corwin}},
  \bibinfo{author}{\bibfnamefont{G.}~\bibnamefont{Parisi}}, \bibnamefont{and}
  \bibinfo{author}{\bibfnamefont{F.}~\bibnamefont{Zamponi}},
  \bibinfo{journal}{Physical Review Letters} \textbf{\bibinfo{volume}{114}},
  \bibinfo{pages}{125504} (\bibinfo{year}{2015}).


\bibitem[{\citenamefont{Charbonneau et~al.}(2018)\citenamefont{Charbonneau,
  Charbonneau, and Szamel}}]{CCS18}
\bibinfo{author}{\bibfnamefont{B.}~\bibnamefont{Charbonneau}},
  \bibinfo{author}{\bibfnamefont{P.}~\bibnamefont{Charbonneau}},
  \bibnamefont{and} \bibinfo{author}{\bibfnamefont{G.}~\bibnamefont{Szamel}},
  \bibinfo{journal}{The Journal of Chemical Physics}
  \textbf{\bibinfo{volume}{148}}, \bibinfo{pages}{224503}
  (\bibinfo{year}{2018}).


\bibitem[{\citenamefont{Agoritsas et~al.}(2018)\citenamefont{Agoritsas, Biroli,
  Urbani, and Zamponi}}]{ABUZ18}
\bibinfo{author}{\bibfnamefont{E.}~\bibnamefont{Agoritsas}},
  \bibinfo{author}{\bibfnamefont{G.}~\bibnamefont{Biroli}},
  \bibinfo{author}{\bibfnamefont{P.}~\bibnamefont{Urbani}}, \bibnamefont{and}
  \bibinfo{author}{\bibfnamefont{F.}~\bibnamefont{Zamponi}},
  \bibinfo{journal}{Journal of Physics A: Mathematical and Theoretical}
  \textbf{\bibinfo{volume}{51}}, \bibinfo{pages}{085002}
  (\bibinfo{year}{2018}).

\bibitem[{\citenamefont{Berthier and Biroli}(2011)}]{BB11}
\bibinfo{author}{\bibfnamefont{L.}~\bibnamefont{Berthier}} \bibnamefont{and}
  \bibinfo{author}{\bibfnamefont{G.}~\bibnamefont{Biroli}},
  \bibinfo{journal}{Review Modern Physics} \textbf{\bibinfo{volume}{83}},
  \bibinfo{pages}{587} (\bibinfo{year}{2011}).

\bibitem[{\citenamefont{Kubo}(1966)}]{Ku66}
\bibinfo{author}{\bibfnamefont{R.}~\bibnamefont{Kubo}},
  \bibinfo{journal}{Reports on Progress in Physics}
  \textbf{\bibinfo{volume}{29}}, \bibinfo{pages}{255} (\bibinfo{year}{1966}).

\bibitem[{\citenamefont{Cugliandolo}(2003)}]{Cu02}
\bibinfo{author}{\bibfnamefont{L.~F.} \bibnamefont{Cugliandolo}}, in
  \emph{\bibinfo{booktitle}{Slow Relaxations and nonequilibrium dynamics in
  condensed matter}} (\bibinfo{publisher}{Springer}, \bibinfo{year}{2003}),
  \eprint{{\tt arXiv:cond-mat/0210312}}.

\bibitem[{\citenamefont{H{\"a}nggi}(1997)}]{Ha97}
\bibinfo{author}{\bibfnamefont{P.}~\bibnamefont{H{\"a}nggi}}, in
  \emph{\bibinfo{booktitle}{Stochastic dynamics}}
  (\bibinfo{publisher}{Springer}, \bibinfo{year}{1997}), pp.
  \bibinfo{pages}{15--22}.

\bibitem[{\citenamefont{Zamponi et~al.}(2005)\citenamefont{Zamponi, Bonetto,
  Cugliandolo, and Kurchan}}]{ZBCK05}
\bibinfo{author}{\bibfnamefont{F.}~\bibnamefont{Zamponi}},
  \bibinfo{author}{\bibfnamefont{F.}~\bibnamefont{Bonetto}},
  \bibinfo{author}{\bibfnamefont{L.~F.} \bibnamefont{Cugliandolo}},
  \bibnamefont{and} \bibinfo{author}{\bibfnamefont{J.}~\bibnamefont{Kurchan}},
  \bibinfo{journal}{Journal of Statistical Mechanics: Theory and Experiment}
  \textbf{\bibinfo{volume}{2005}}, \bibinfo{pages}{P09013}
  (\bibinfo{year}{2005}).

\bibitem[{\citenamefont{Maes et~al.}(2013)\citenamefont{Maes, Safaverdi, Visco,
  and van Wijland}}]{MSVW13}
\bibinfo{author}{\bibfnamefont{C.}~\bibnamefont{Maes}},
  \bibinfo{author}{\bibfnamefont{S.}~\bibnamefont{Safaverdi}},
  \bibinfo{author}{\bibfnamefont{P.}~\bibnamefont{Visco}}, \bibnamefont{and}
  \bibinfo{author}{\bibfnamefont{F.}~\bibnamefont{van Wijland}},
  \bibinfo{journal}{Physical Review E} \textbf{\bibinfo{volume}{87}}
  (\bibinfo{year}{2013}).

\bibitem[{\citenamefont{Barrat et~al.}(1997)\citenamefont{Barrat, Franz, and
  Parisi}}]{BFP97}
\bibinfo{author}{\bibfnamefont{A.}~\bibnamefont{Barrat}},
  \bibinfo{author}{\bibfnamefont{S.}~\bibnamefont{Franz}}, \bibnamefont{and}
  \bibinfo{author}{\bibfnamefont{G.}~\bibnamefont{Parisi}},
  \bibinfo{journal}{Journal of Physics A: Mathematical and General}
  \textbf{\bibinfo{volume}{30}}, \bibinfo{pages}{5593} (\bibinfo{year}{1997}).

\bibitem[{\citenamefont{Berthier et~al.}(2017)\citenamefont{Berthier, Flenner,
  and Szamel}}]{BFS17}
\bibinfo{author}{\bibfnamefont{L.}~\bibnamefont{Berthier}},
  \bibinfo{author}{\bibfnamefont{E.}~\bibnamefont{Flenner}}, \bibnamefont{and}
  \bibinfo{author}{\bibfnamefont{G.}~\bibnamefont{Szamel}},
  \bibinfo{journal}{New Journal of Physics} \textbf{\bibinfo{volume}{19}},
  \bibinfo{pages}{125006} (\bibinfo{year}{2017}).

\bibitem[{\citenamefont{Fily and Marchetti}(2012)}]{FM12}
\bibinfo{author}{\bibfnamefont{Y.}~\bibnamefont{Fily}} \bibnamefont{and}
  \bibinfo{author}{\bibfnamefont{M.~C.} \bibnamefont{Marchetti}},
  \bibinfo{journal}{Physical Review Letters} \textbf{\bibinfo{volume}{108}},
  \bibinfo{pages}{235702} (\bibinfo{year}{2012}).

\bibitem[{\citenamefont{Yang et~al.}(2014)\citenamefont{Yang, Manning, and
  Marchetti}}]{YMM14}
\bibinfo{author}{\bibfnamefont{X.}~\bibnamefont{Yang}},
  \bibinfo{author}{\bibfnamefont{M.~L.} \bibnamefont{Manning}},
  \bibnamefont{and} \bibinfo{author}{\bibfnamefont{M.~C.}
  \bibnamefont{Marchetti}}, \bibinfo{journal}{Soft Matter}
  \textbf{\bibinfo{volume}{10}}, \bibinfo{pages}{6477} (\bibinfo{year}{2014}).

\bibitem[{\citenamefont{Ikeda et~al.}(2013)\citenamefont{Ikeda, Berthier, and
  Sollich}}]{IBS13}
\bibinfo{author}{\bibfnamefont{A.}~\bibnamefont{Ikeda}},
  \bibinfo{author}{\bibfnamefont{L.}~\bibnamefont{Berthier}}, \bibnamefont{and}
  \bibinfo{author}{\bibfnamefont{P.}~\bibnamefont{Sollich}},
  \bibinfo{journal}{Soft Matter} \textbf{\bibinfo{volume}{9}},
  \bibinfo{pages}{7669} (\bibinfo{year}{2013}).

\bibitem[{\citenamefont{Kawasaki et~al.}(2015)\citenamefont{Kawasaki,
  Coslovich, Ikeda, and Berthier}}]{KCIB15}
\bibinfo{author}{\bibfnamefont{T.}~\bibnamefont{Kawasaki}},
  \bibinfo{author}{\bibfnamefont{D.}~\bibnamefont{Coslovich}},
  \bibinfo{author}{\bibfnamefont{A.}~\bibnamefont{Ikeda}}, \bibnamefont{and}
  \bibinfo{author}{\bibfnamefont{L.}~\bibnamefont{Berthier}},
  \bibinfo{journal}{Physical Review E} \textbf{\bibinfo{volume}{91}},
  \bibinfo{pages}{012203} (\bibinfo{year}{2015}).

\bibitem[{\citenamefont{Gazuz et~al.}(2009)\citenamefont{Gazuz, Puertas,
  Voigtmann, and Fuchs}}]{GPVF09}
\bibinfo{author}{\bibfnamefont{I.}~\bibnamefont{Gazuz}},
  \bibinfo{author}{\bibfnamefont{A.~M.} \bibnamefont{Puertas}},
  \bibinfo{author}{\bibfnamefont{T.}~\bibnamefont{Voigtmann}},
  \bibnamefont{and} \bibinfo{author}{\bibfnamefont{M.}~\bibnamefont{Fuchs}},
  \bibinfo{journal}{Physical Review Letters} \textbf{\bibinfo{volume}{102}},
  \bibinfo{pages}{248302} (\bibinfo{year}{2009}).

\bibitem[{\citenamefont{Liao and Xu}(2018)}]{LX18}
\bibinfo{author}{\bibfnamefont{Q.}~\bibnamefont{Liao}} \bibnamefont{and}
  \bibinfo{author}{\bibfnamefont{N.}~\bibnamefont{Xu}}, \bibinfo{journal}{Soft
  Matter} \textbf{\bibinfo{volume}{14}}, \bibinfo{pages}{853}
  (\bibinfo{year}{2018}).

  
\bibitem{AMZ19bis}
\bibinfo{author}{\bibfnamefont{E.}~\bibnamefont{Agoritsas}},
  \bibinfo{author}{\bibfnamefont{T.}~\bibnamefont{Maimbourg}}, \bibnamefont{and}
  \bibinfo{author}{\bibfnamefont{F.}~\bibnamefont{Zamponi}},
  \bibinfo{journal}{in preparation (2019)}.

\bibitem[{\citenamefont{M\'ezard et~al.}(1987)\citenamefont{M\'ezard, Parisi,
  and Virasoro}}]{MPV87}
\bibinfo{author}{\bibfnamefont{M.}~\bibnamefont{M\'ezard}},
  \bibinfo{author}{\bibfnamefont{G.}~\bibnamefont{Parisi}}, \bibnamefont{and}
  \bibinfo{author}{\bibfnamefont{M.~A.} \bibnamefont{Virasoro}},
  \emph{\bibinfo{title}{Spin glass theory and beyond}}
  (\bibinfo{publisher}{World Scientific}, \bibinfo{address}{Singapore},
  \bibinfo{year}{1987}).

\bibitem[{\citenamefont{Frisch et~al.}(1985)\citenamefont{Frisch, Rivier, and
  Wyler}}]{FRW85}
\bibinfo{author}{\bibfnamefont{H.~L.} \bibnamefont{Frisch}},
  \bibinfo{author}{\bibfnamefont{N.}~\bibnamefont{Rivier}}, \bibnamefont{and}
  \bibinfo{author}{\bibfnamefont{D.}~\bibnamefont{Wyler}},
  \bibinfo{journal}{Physical Review Letters} \textbf{\bibinfo{volume}{54}},
  \bibinfo{pages}{2061} (\bibinfo{year}{1985}).

\bibitem[{\citenamefont{Wyler et~al.}(1987)\citenamefont{Wyler, Rivier, and
  Frisch}}]{WRF87}
\bibinfo{author}{\bibfnamefont{D.}~\bibnamefont{Wyler}},
  \bibinfo{author}{\bibfnamefont{N.}~\bibnamefont{Rivier}}, \bibnamefont{and}
  \bibinfo{author}{\bibfnamefont{H.~L.} \bibnamefont{Frisch}},
  \bibinfo{journal}{Physical Review A} \textbf{\bibinfo{volume}{36}},
  \bibinfo{pages}{2422} (\bibinfo{year}{1987}).

\bibitem[{\citenamefont{Frisch and Percus}(1999)}]{FP99}
\bibinfo{author}{\bibfnamefont{H.~L.} \bibnamefont{Frisch}} \bibnamefont{and}
  \bibinfo{author}{\bibfnamefont{J.~K.} \bibnamefont{Percus}},
  \bibinfo{journal}{Physical Review E} \textbf{\bibinfo{volume}{60}},
  \bibinfo{pages}{2942} (\bibinfo{year}{1999}).

\bibitem[{\citenamefont{Martin et~al.}(1973)\citenamefont{Martin, Siggia, and
  Rose}}]{MSR}
\bibinfo{author}{\bibfnamefont{P.~C.} \bibnamefont{Martin}},
  \bibinfo{author}{\bibfnamefont{E.~D.} \bibnamefont{Siggia}},
  \bibnamefont{and} \bibinfo{author}{\bibfnamefont{H.~A.} \bibnamefont{Rose}},
  \bibinfo{journal}{Physical Review A} \textbf{\bibinfo{volume}{8}},
  \bibinfo{pages}{423} (\bibinfo{year}{1973}).

\bibitem[{\citenamefont{Janssen}(1976)}]{Ja76}
\bibinfo{author}{\bibfnamefont{H.-K.} \bibnamefont{Janssen}},
  \bibinfo{journal}{Zeitschrift f{\"u}r Physik B Condensed Matter}
  \textbf{\bibinfo{volume}{23}}, \bibinfo{pages}{377} (\bibinfo{year}{1976}).

\bibitem[{\citenamefont{De~Dominicis}(1978)}]{DD78}
\bibinfo{author}{\bibfnamefont{C.}~\bibnamefont{De~Dominicis}},
  \bibinfo{journal}{Physical Review B} \textbf{\bibinfo{volume}{18}},
  \bibinfo{pages}{4913} (\bibinfo{year}{1978}).

\bibitem[{\citenamefont{Castellani and Cavagna}(2005)}]{CC05}
\bibinfo{author}{\bibfnamefont{T.}~\bibnamefont{Castellani}} \bibnamefont{and}
  \bibinfo{author}{\bibfnamefont{A.}~\bibnamefont{Cavagna}},
  \bibinfo{journal}{Journal of Statistical Mechanics: Theory and Experiment}
  \textbf{\bibinfo{volume}{2005}}, \bibinfo{pages}{P05012}
  (\bibinfo{year}{2005}).

\bibitem[{\citenamefont{Kamenev}(2009)}]{Kamenev}
\bibinfo{author}{\bibfnamefont{A.}~\bibnamefont{Kamenev}},
  \emph{\bibinfo{title}{Field Theory of Non-Equilibrium Systems}}
  (\bibinfo{publisher}{Cambridge University Press}, \bibinfo{year}{2009}).

\bibitem[{\citenamefont{Zinn-Justin}(2002)}]{Zinn-Justin}
\bibinfo{author}{\bibfnamefont{J.}~\bibnamefont{Zinn-Justin}},
  \emph{\bibinfo{title}{Quantum field theory and critical phenomena}}
  (\bibinfo{publisher}{Oxford University Press}, \bibinfo{year}{2002}).

\bibitem[{\citenamefont{Lechenault et~al.}(2010)\citenamefont{Lechenault,
  Candelier, Dauchot, Bouchaud, and Biroli}}]{LCDBB10}
\bibinfo{author}{\bibfnamefont{F.}~\bibnamefont{Lechenault}},
  \bibinfo{author}{\bibfnamefont{R.}~\bibnamefont{Candelier}},
  \bibinfo{author}{\bibfnamefont{O.}~\bibnamefont{Dauchot}},
  \bibinfo{author}{\bibfnamefont{J.~P.} \bibnamefont{Bouchaud}},
  \bibnamefont{and} \bibinfo{author}{\bibfnamefont{G.}~\bibnamefont{Biroli}},
  \bibinfo{journal}{Soft Matter} \textbf{\bibinfo{volume}{6}},
  \bibinfo{pages}{3059} (\bibinfo{year}{2010}).

\bibitem[{\citenamefont{Aron et~al.}(2010)\citenamefont{Aron, Biroli, and
  Cugliandolo}}]{ABC10}
\bibinfo{author}{\bibfnamefont{C.}~\bibnamefont{Aron}},
  \bibinfo{author}{\bibfnamefont{G.}~\bibnamefont{Biroli}}, \bibnamefont{and}
  \bibinfo{author}{\bibfnamefont{L.~F.} \bibnamefont{Cugliandolo}},
  \bibinfo{journal}{Journal of Statistical Mechanics: Theory and Experiment}
  \textbf{\bibinfo{volume}{2010}}, \bibinfo{pages}{P11018}
  (\bibinfo{year}{2010}).

\bibitem[{\citenamefont{Maimbourg and Kurchan}(2016)}]{MK16}
\bibinfo{author}{\bibfnamefont{T.}~\bibnamefont{Maimbourg}} \bibnamefont{and}
  \bibinfo{author}{\bibfnamefont{J.}~\bibnamefont{Kurchan}},
  \bibinfo{journal}{{EPL} (Europhysics Letters)}
  \textbf{\bibinfo{volume}{114}}, \bibinfo{pages}{60002}
  (\bibinfo{year}{2016}).

\bibitem[{\citenamefont{Berthier and Barrat}(2002)}]{BB02}
\bibinfo{author}{\bibfnamefont{L.}~\bibnamefont{Berthier}} \bibnamefont{and}
  \bibinfo{author}{\bibfnamefont{J.-L.} \bibnamefont{Barrat}},
  \bibinfo{journal}{Journal of Chemical Physics}
  \textbf{\bibinfo{volume}{116}}, \bibinfo{pages}{6228} (\bibinfo{year}{2002}).

\bibitem[{\citenamefont{Cugliandolo and Kurchan}(1994)}]{CK94}
\bibinfo{author}{\bibfnamefont{L.}~\bibnamefont{Cugliandolo}} \bibnamefont{and}
  \bibinfo{author}{\bibfnamefont{J.}~\bibnamefont{Kurchan}},
  \bibinfo{journal}{Journal of Physics A: Mathematical and General}
  \textbf{\bibinfo{volume}{27}}, \bibinfo{pages}{5749} (\bibinfo{year}{1994}).

\bibitem[{\citenamefont{Franz and M{\'{e}}zard}(1994)}]{FM94}
\bibinfo{author}{\bibfnamefont{S.}~\bibnamefont{Franz}} \bibnamefont{and}
  \bibinfo{author}{\bibfnamefont{M.}~\bibnamefont{M{\'{e}}zard}},
  \bibinfo{journal}{EPL (Europhysics Letters)} \textbf{\bibinfo{volume}{26}},
  \bibinfo{pages}{209} (\bibinfo{year}{1994}).

\bibitem[{\citenamefont{Cugliandolo and Doussal}(1996)}]{CLD96}
\bibinfo{author}{\bibfnamefont{L.~F.} \bibnamefont{Cugliandolo}}
  \bibnamefont{and} \bibinfo{author}{\bibfnamefont{P.~L.}
  \bibnamefont{Doussal}}, \bibinfo{journal}{Physical Review E}
  \textbf{\bibinfo{volume}{53}}, \bibinfo{pages}{1525} (\bibinfo{year}{1996}).

\bibitem[{\citenamefont{Cugliandolo and Kurchan}(2008)}]{CK08}
\bibinfo{author}{\bibfnamefont{L.~F.} \bibnamefont{Cugliandolo}}
  \bibnamefont{and} \bibinfo{author}{\bibfnamefont{J.}~\bibnamefont{Kurchan}},
  \bibinfo{journal}{Journal of Physics A: Mathematical and Theoretical}
  \textbf{\bibinfo{volume}{41}}, \bibinfo{pages}{324018}
  (\bibinfo{year}{2008}).

\bibitem[{\citenamefont{Scalliet and Zamponi}(2018)}]{SZ2666}
  \bibinfo{author}{\bibfnamefont{C.}~\bibnamefont{Scalliet}} \bibnamefont{and}
  \bibinfo{author}{\bibfnamefont{F.}~\bibnamefont{Zamponi}},
  \bibinfo{journal}{in preparation}  (\bibinfo{year}{2019}).

\end{thebibliography}

\end{document}